\documentclass[nohyperref]{article}

\usepackage{microtype}
\usepackage{graphicx}
\usepackage{subcaption} %  for subfigures environments 
\usepackage{makecell}
\usepackage{booktabs} % for professional tables
\usepackage{float}
\restylefloat{table}
\usepackage[compact]{titlesec}         % you need this package
\titlespacing{\section}{0pt}{0pt}{0pt} % this reduces space between (sub)sections to 0pt, for example
\AtBeginDocument{%                     % this will reduce spaces between parts (above and below) of texts within a (sub)section to 0pt, for example - like between an 'eqnarray' and text
  \setlength\abovedisplayskip{6pt}
  \setlength\belowdisplayskip{6pt}}

\usepackage{hyperref}

\usepackage[accepted]{icml2023}

\usepackage{amsmath}
\usepackage{amssymb}
\usepackage{mathtools}
\usepackage{amsthm}
\usepackage{booktabs}

\usepackage[capitalize,noabbrev]{cleveref}

\newtheoremstyle{exampstyle}
  {1.5em} % Space above
  {1.5em} % Space below
  {} % Body font
  {} % Indent amount
  {\bfseries} % Theorem head font
  {.} % Punctuation after theorem head
  {.5em} % Space after theorem head
  {} % Theorem head spec (can be left empty, meaning `normal')

\theoremstyle{plain}
\newtheorem{theorem}{Theorem}[section]

\theoremstyle{exampstyle}
\newtheorem{definition}[theorem]{Definition}

\theoremstyle{remark}

\newcommand{\E}{\mathbb{E}}
\renewcommand{\S}{\mathbb{S}}
\renewcommand{\H}{\mathbb{H}}

\newcommand{\R}{\mathbb{R}}

\usepackage[disable,textsize=tiny]{todonotes}
\icmltitlerunning{Product Manifold Representations for Learning on Biological Pathways}

\begin{document}

\twocolumn[
% \icmltitle{Mixed-Curvature Representation Learning for Biological Pathway Graphs - Journal Version}
\icmltitle{Product Manifold Representations for Learning on Biological Pathways}


\icmlsetsymbol{equal}{*}

\begin{icmlauthorlist}
\icmlauthor{Daniel McNeela}{uwcs,morg,uwbs}
\icmlauthor{Frederic Sala\textsuperscript{*}}{uwcs}
\icmlauthor{Anthony Gitter\textsuperscript{*}}{uwcs,morg,uwbs}
\end{icmlauthorlist}

\icmlaffiliation{uwbs}{Department of Biostatistics and Medical Informatics, University of Wisconsin-Madison, Madison, WI, USA}
\icmlaffiliation{uwcs}{Department of Computer Sciences, University of Wisconsin-Madison, Madison, WI, USA}
\icmlaffiliation{morg}{Morgridge Institute for Research, Madison, WI, USA}

\icmlcorrespondingauthor{Anthony Gitter}{gitter@biostat.wisc.edu}

\icmlkeywords{protein-protein interaction, edge prediction, graph representation learning, non-Euclidean, graph neural network}

\vskip 0.3in
]

\printAffiliationsAndNotice{\icmlEqualContribution} % otherwise use the standard text.

\begin{abstract}
Machine learning models that embed graphs in non-Euclidean spaces have shown substantial benefits in 
a variety of contexts, but their application has not been studied extensively in
the biological domain, particularly with respect to biological pathway graphs.
Such graphs 
exhibit a variety of complex 
network structures, presenting challenges to existing embedding approaches. Learning high-quality embeddings for biological pathway graphs is important for 
researchers looking to understand the underpinnings of disease and train high-quality predictive models on
these networks.
In this work, we investigate the effects of embedding pathway graphs in non-Euclidean mixed-curvature
spaces and compare against traditional Euclidean graph representation learning models. We then train a supervised model
using the learned node embeddings to predict missing protein-protein interactions 
in pathway graphs. We find large reductions in distortion and boosts on in-distribution edge prediction performance as a result of using mixed-curvature embeddings and their corresponding graph neural network
models. However, we find that mixed-curvature representations underperform existing baselines on out-of-distribution edge prediction performance suggesting that these representations may overfit to the training graph topology. We provide our Mixed-Curvature Product Graph Convolutional Network code at \href{https://github.com/mcneela/Mixed-Curvature-GCN}{https://github.com/mcneela/Mixed-Curvature-GCN} and our pathway analysis code at \href{https://github.com/mcneela/Mixed-Curvature-Pathways}{https://github.com/mcneela/Mixed-Curvature-Pathways}.
\end{abstract}

\section{Introduction}
\label{introduction}
Machine learning methods for embedding graphs \cite{wu_comprehensive_2021} enable learning on data ranging from social media networks,
to proteins and molecules, to phylogenies and knowledge graphs. These embeddings then enable useful
node classification and edge prediction models, which can perform tasks as diverse as predicting whether
a molecule is active against a given drug target or whether two users are likely to share a preference
for a given product.

While traditional graph learning methods employ Euclidean representations \citep{node2vec}, it is known
that for certain graphs Euclidean representations are unable to perfectly preserve graph structure, 
regardless of what algorithm is used \citep{bourgain1985}. As a result, recent works have studied 
whether lower graph distortion can be achieved by embedding in non-Euclidean spaces, such as 
hyperbolic space \citep{pmlr-v80-sala18a}. Generally speaking, lower distortion 
correlates with better downstream task performance.

The works of \citet{Gu2018LearningMR} and \citet{giovanni2022heterogeneous} have 
examined embeddings into mixed-curvature products of spaces and heterogeneous manifolds, 
respectively. By ``mixed-curvature products of spaces'', we mean a Cartesian
product of embedding spaces having constant positive, negative, or zero curvature.
These correspond to the model spherical, hyperbolic, and Euclidean spaces, 
respectively \citep{Lee2019IntroductionTR}.
While these methods have been evaluated on standard graph benchmarking datasets, their 
hypotheses have not been validated for specialized graphs found in biological pathways and networks, which may have properties and topologies that differ from general graphs in other domains. The product space approach is appealing because deciding a priori which space to embed into can be challenging.

{\it Biological pathways} are graphs that encode cellular processes. Typically, the nodes in
such graphs are entities such as genes, proteins, or metabolites, and the edges designate relationships 
between them. For example, an edge connecting nodes A and B might indicate that the presence 
of protein A controls the transcription of gene B. Pathways are an important object of study in network biology
as they can be used to infer subcellular relationships and understand the mechanisms underlying disease.

Embedding biological pathways is difficult because no canonical methods exist. Pathways exhibit
a high degree of complexity. Some, due to a lack of study, are sparse while others
exhibit high inter-connectivity.
Their complex network structures suggest that non-Euclidean 
representations might provide significant benefits.
However, no systematic
study of embedding methods applied to biological pathway graphs has been undertaken. Only Euclidean embedding methods have been applied
to pathway graphs \citep{path2vec, Pershad2020-ti}, and
because pathway graphs differ from standard graphs used
to benchmark non-Euclidean embedding models, it is unknown to what
extent these models would work for pathway graphs.

In this work, we study non-Euclidean embeddings of biological pathway graphs and their performance
relative to standard Euclidean embeddings.\footnote{This work extends our workshop paper \citet{mcneela_mixed-curvature_2023}.} We perform a large-scale test of a variety of embedding
methods to pathway graphs taken from PathBank, Reactome, HumanCyc, the NCI Pathway Interaction Database, and KEGG and embed into a number of different combinations of spaces.
For each pathway graph, we determine a best embedding space, as measured by the lowest graph distortion,
and learn the node embeddings for that graph. 
Although biological pathway databases are of high quality, they are incomplete and only capture a fraction of the knowledge about the relevant biological processes \cite{hanspers_pathway_2020}.
Therefore, we investigate the downstream performance of our graph embeddings by predicting potentially missing pathway edges.
Instead of only predicting held out pathway edges, we also examine the considerably more challenging task of predicting protein-protein interactions (PPIs) from an independent database that may participate in the pathway.
We find that mixed-curvature representations
outperform, as measured by distortion, Euclidean representations in all cases.
Furthermore, the positive impact of improved embeddings seems to generalize to downstream tasks such as edge prediction, where we find improvements in area under the receiver operating characteristic curve (AUROC) and average precision (AP) for held out in-distribution edges. However, for out-of-distribution STRING PPI edges, we find that the mixed-curvature representations underperform baselines, perhaps suggesting that these 
representations overfit to the in-distribution graph topology.

\section{Background and Related Work}

\subsection{Non-Euclidean Embeddings and Machine Learning}
Much of the research into non-Euclidean embeddings in machine learning originated in studies of graphs and networks, where they were originally
used to embed concept ontologies. For example, \citet{nickel2017poincare} developed a method for
embedding into the Poincar\'e model of hyperbolic space and used it to generate node embeddings for the
WORDNET ontology. \citet{pmlr-v80-sala18a} expanded on this work by determining the precision-dimensionality
tradeoffs inherent in hyperbolic embeddings. \citet{DBLP:journals/corr/abs-1805-09112} and \citet{chami2019hyperbolic} then produced generalized Hyperbolic Neural Networks and Hyperbolic Graph Convolutional Networks (GCNs) to perform prediction directly in hyperbolic space on data of various types. % Chami model was technically called `Hyperbolic graph convolutional neural networks' but it means the same thing and helps to use GCN consistently instead of GCN / GNN / GCNN
\citet{Gu2018LearningMR} extended graph representation learning to a Cartesian product 
of hyperbolic, spherical, and Euclidean spaces, while \citet{giovanni2022heterogeneous} further generalized
this to a product of manifolds of heterogeneous curvature. Our approach follows that of \citet{Gu2018LearningMR}. More recently, hyperbolic
and non-Euclidean geometries have been applied to
contrastive learning \citep{desai23a} and attention
blocks \citep{tseng2023coneheads}.

Hyperbolic modeling approaches have been applied to embed biological and chemical data 
exhibiting hierarchical structure.
Many methods focus on drug discovery, specifically drug repurposing \citep{Lau2023} or drug target prediction \cite{yue_flone_2023, poleksic_hyperbolic_2023, zahra_selection_2023, pogany_towards_2023}.
The hierarchical relationships in the Gene Ontology make it appealing for learning hyperbolic representations of its terms and their associated genes \cite{kim_hig2vec_2021, li_hyperbolic_2023}.
In transcriptomics, methods have been developed to visualize and estimate the curvature of gene expression samples \cite{zhou_hyperbolic_2021}, embed single-cell RNA-seq data \cite{klimovskaia_poincare_2020, ding_deep_2021}, and model differential expression signatures \cite{pogany_hyperbolic_2023}.
Finally, embedding biological sequences has enabled visualizing members of a protein family \cite{susmelj_poincare_2023} and Bayesian phylogenetic inference \cite{macaulay_fidelity_2023}.

\subsection{Pathway Graphs and Embeddings} 
Pathway graphs have been well-studied from a biological
perspective, but embedding them to facilitate downstream prediction tasks is relatively new. For example,
\citet{path2vec} developed a method called pathway2vec, which combines five different Euclidean
embedding methods to learn embeddings of biological pathways. Similarly, \citet{Pershad2020-ti} used node2vec
to generate embeddings of PPI networks and used the resulting embeddings
as one component of a method to predict response to psychiatric drugs. Pathway embeddings are
a crucial input to models that operate on pathway network structures to make predictions.
Euclidean GCNs have been broadly applied to biological pathways to predict cancer subtypes \citep{lee2020cancer, hayakawa2022pathway}, synthetic lethality \citep{lai2021predicting}, PPIs \citep{pham2021link}, cancer survival \citep{liang2022risk}, textual pathway descriptions \cite{yang2022pathway2text}, and subcellular localization \citep{magnano2023graph}.
However, there has been no systematic study investigating the use of non-Euclidean and mixed-curvature 
embedding models for pathway graphs. 

\subsection{PPI Prediction} 
PPIs can be predicted based on combinations of proteins' sequence, expression, functional, evolutionary, or 3D structural features \cite{durham2023predicting}.
One PPI prediction formulation is as an edge or link prediction task in a network of known PPIs \citep{li2022graph}. % this reference is okay, may look for something better
%Link prediction is a common task used to test graph 
%embedding and neural network models. We use the terms link prediction and edge prediction interchangeably. \tony{I feel like edge prediction is more common in the comp bio world, could we use that as the primary term and link prediction as the synonym?}
This can be accomplished using features from the original graph or by learning node embeddings as features for the edge prediction task.
For example, \citet{feng2020signaling} predict signaling cascades from 
PPI graphs by integrating transcriptomics and copy-number data into
a GCN. \citet{jiang2020biochem} use graph embeddings to predict links that indicate an
enzymatic reaction between pairs of molecules from the KEGG database. 
Finally, \citet{zhang2019ppi} and \citet{liu2020ppi} leverage a combination of sequence and network information
to predict PPIs.
\begin{table*}[htbp]
\centering
\resizebox{\textwidth}{!}{%
\begin{tabular}{|l|c|c|c|c|c|}
\hline
 &
  \multicolumn{1}{c|}{\textbf{PathBank}} &
  \multicolumn{1}{c|}{\textbf{Reactome}} &
  \multicolumn{1}{c|}{\textbf{KEGG}} &
  \multicolumn{1}{c|}{\textbf{HumanCyc}} &
  \multicolumn{1}{c|}{\textbf{NCI}} \\ \hline
Number of graphs                           & 875    & 1723    & 82     & 242    & 212    \\ \hline
Number of graphs with single best space    & 411    & 884     & 47     & 210    & 144    \\ \hline
Number of graphs with multiple best spaces & 464    & 839     & 35     & 32     & 68     \\ \hline
Mean number of nodes                    & 63.23  & 54.36   & 159.73 & 104.83 & 100.95 \\ \hline
Mean number of edges                    & 283.56 & 1091.79 & 657.39 & 289.62 & 531.98 \\ \hline
Median number of nodes                     & 53     & 24      & 152    & 73     & 73     \\ \hline
Median number of edges                     & 162    & 65      & 532    & 133    & 194    \\ \hline
\end{tabular}%
}
\caption{General statistics summarizing the five pathway databases.}
\label{tab:summary-stats}
\end{table*}

\section{Methods}
\subsection{Data Processing}
We downloaded five pathway datasets from Pathway Commons \citep{rodchenkov2020pathway} v12, namely,
PathBank \citep{wishart2020pathbank}, Reactome \citep{gillespie_reactome_2022}, HumanCyc \citep{romero_computational_2004}, KEGG \citep{kanehisa_kegg_2012}, and the NCI Pathway Interaction Database \cite{schaefer_pid_2009}. We used the {\tt .txt} files 
containing interaction participants, edge types, and associated metadata.
For each pathway in each dataset, we created a NetworkX \cite{hagberg_exploring_2008} graph object and ignored Pathway Commons edge types, treating
each edge as undirected with no additional annotations.
We then generated edge lists for the undirected graphs and learned embeddings using the mixed-curvature embedding Python library \citep{Gu2018LearningMR}.
All subsequent experiments were performed only for those pathway graphs for which
a unique ``best'' embedding space could be determined, defined as the space exhibiting
minimum distortion. Those pathways which did not exhibit a unique embedding space
of minimum distortion were discarded as there exists no canonical way to select a 
representative product space to compare embedding and edge prediction methods
against. We also discarded one pathway from the
Reactome dataset having 36,293 edges because its runtime
on the edge prediction task was prohibitive and
it is an extreme outlier in terms of
pathway size (Table~\ref{tab:summary-stats}).

For edge prediction, we downloaded Homo sapiens PPIs from STRING v12.0 \citep{szklarczyk2023string}. In STRING, PPIs are listed as pairs of ENSEMBL protein identifiers. 
For each gene or protein node name in the given pathway dataset, 
we map it to its UniProt SwissProt identifier using the \textit{MyGene} API \citep{xin}. For each ENSEMBL protein identifier, we
map it to the \texttt{Ensembl\_HGNC\_uniprot\_id} listed in the STRING aliases file. We then iterate over all STRING PPI edges. For each pathway, if there exists a STRING PPI
corresponding to two UniProt identifiers in the given pathway, % \todo{Insert `that are not already connected'?} % I actually don't think I checked whether they were already connected. So there could be some overlap with the validation set
we add an edge to the potential test set for that graph. 
We only included STRING edges with experimental evidence and discarded all edges with a score of less than 500 out of 1000. This
yields a dataset of reasonable size for edge prediction, although the score used to filter 
edges can be tuned. Filtering using higher experimental evidence scores would likely yield
less false positives in the list of candidate edges for each pathway graph; however, the 
relationship here is not strictly linear.
 
\subsection{Learning Embeddings}
We adopt the notation $\S^{s_i}_C$ for the spherical space having constant positive
curvature $C$ and dimension $s_i$, $\H^{h_j}_C$ for the hyperbolic space having constant negative curvature
$-C$ and dimension $h_j$, and $\E^{e_k}$ for the Euclidean space having zero curvature and dimension
$e_k$.
Given a collection of spherical, hyperbolic, and Euclidean spaces,
each having an associated dimension and curvature, we write their Cartesian product 
space as 
\begin{equation}\label{eq:sig}\mathcal{P} = \prod_{i=1}^m \S_{C_i}^{s_i} \times \prod_{j=1}^n\H_{C_{m+j}}^{h_j} \times \prod_{k=1}^p {\E}^{e_k},\end{equation}
Here the product space $\mathcal{P}$ consists of $m+n+p$ spaces and has total dimension $\sum_{i} s_i + \sum_{j} h_j + \sum_k e_k$.
Following \citet{Gu2018LearningMR}, we call each individual $\S^{s_i},\H^{h_i},\E^{e_i}$ a \emph{component} and 
the decomposition of the product space into components as in Eq.~\ref{eq:sig} the \emph{signature} of the 
space.

We learn an embedding function $f: \mathcal{G} \to \mathcal{P}$ where
$\mathcal{G}$ is the space of pathway graphs and $\mathcal{P}$ is a 
product manifold with a fixed, predefined signature. For embedding node 
$u$ of a graph $G$ on a product manifold $\mathcal{P}$ having $k$th hyperbolic
component $\H^{h_j}_{C_{j}}$ with curvature $-C_{j}$, we randomly initialize 
$f^{k}(u) = (p_0, p_1, \ldots, p_{h_j})$ to a point on the hyperbolic manifold,
parameterized using the {\it hyperboloid model}, i.e.
\begin{equation}
-{p_0}^2 + {p_1}^2 + \cdots + {p_{h_j}}^2 = -C_{j},
\end{equation} 
Similarly, for the $l$th spherical component with curvature $C_i$, 
we randomly initialize
$f^l(u) = (p_0, p_1, \ldots, p_{s_i})$ lying on the sphere
\begin{equation}
    {p_0}^2 + {p_1}^2 + \cdots + {p_{s_i}}^2 = C_i
\end{equation}
and for the $m$th Euclidean component, we randomly initialize 
$f^m(u) = (p_1 + \cdots p_{e_k})$. For the 
Euclidean components, $f$ is unconstrained. For the hyperbolic and spherical
components, we require one extra dimension because we embed hyperbolic and
spherical spaces of dimension $n$ in $\R^{n+1}$.
Squared distances in the product space decompose as sums of squared 
distances in the component spaces \citep{Lee2019IntroductionTR}, i.e.
\begin{equation}
    d_{\mathcal{P}}(u, v)^2 = \sum_i d_{M_i}(\pi_i(u), \pi_i(v))^2
\end{equation}
for $u, v \in \mathcal{P}$, where $\pi_i$ denotes projection onto the $i$th 
component. The learning of $f$ takes place via  optimization of the following 
loss function
\begin{equation}
\mathcal{L}(f) = \sum_{1 \leq u < v \leq n} \left| \left( \frac{d_{\mathcal{P}}(f(u), f(v))}{d_G(u, v)}\right)^2 - 1\right|
\end{equation}
where $d_G(u, v)$ denotes the graph 
distance between nodes $u$ and $v$, defined as the length of the 
shortest path connecting them in $G$, and
$d_{\mathcal{P}}(f(u), f(v))$ denotes the 
product manifold geodesic distance between the
learned embeddings for nodes $u$ and $v$.
To learn the embedding function, we initially randomize the
embeddings for all nodes, then train $f$ so as to
minimize $\mathcal{L}$ using the Riemannian stochastic gradient descent
optimization algorithm \citep{rsgd}, which 
extends stochastic gradient descent to arbitrary 
Riemannian manifolds.

The main metric used to evaluate our embeddings is the average graph distance {\it distortion}. We define the distortion $\mathcal{D}$ of a learned 
embedding $f$ relative to a graph $G = (V, E)$ to be
$$\mathcal{D}(f) = \frac{1}{|V|^2}\sum_{u, v \in V, u \neq v} 
\frac{|d_{\mathcal{P}}(f(u), f(v)) - d_G(u, v)|}
{d_G(u, v)}$$
%where $d_G(u, v)$ denotes the graph distance between nodes $u$ and $v$, defined as the length of the shortest path connecting them in $G$.

\subsection{Hyperparameters}
We create 252 embeddings for each pathway graph corresponding to different combinations of the 
number of hyperbolic, spherical, and Euclidean spaces; the learning rate; and the dimensionalities of each space (Appendix).
We test having different numbers of components of each space, where the number of components of each type
ranges from 0 to 3. We also constrain the total dimensionality of all spaces to sum to 100. This is
to keep the comparison across different combinations fair by ensuring they each have the same 
representational capacity as governed by the number of dimensions.

\begin{figure*}[hbtp]
\begin{subfigure}[t]{0.5\textwidth}
\includegraphics[width=\textwidth]{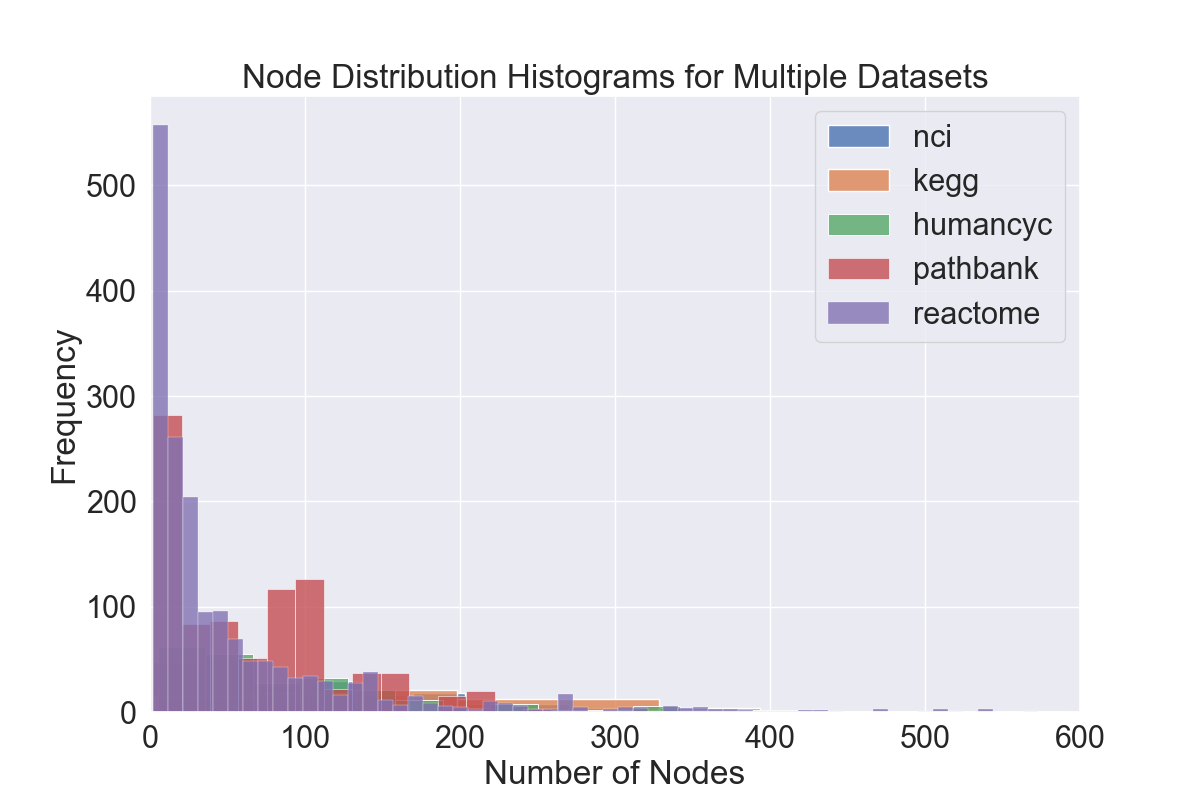}
\caption{Node Distributions}
\end{subfigure}
\hfill
\begin{subfigure}[t]{0.5\textwidth}
\includegraphics[width=\textwidth]{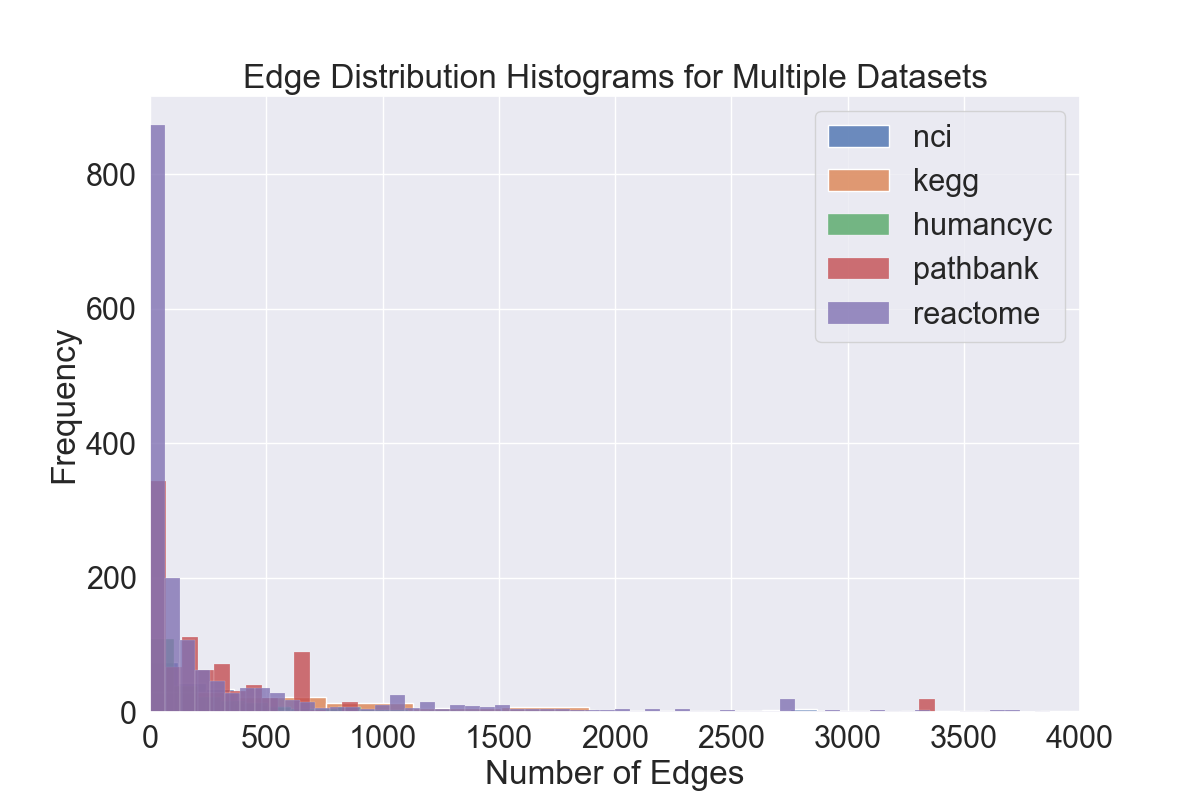}
\caption{Edge Distributions}
\end{subfigure}
\caption{Histograms of node and edge distributions for the
pathway databases studied. A few outliers were excluded.}
\label{fig:histograms}
\end{figure*}

\subsection{Mixed-Curvature Product GCN}
Similar to \citet{bachmann20a}, we extend the Hyperbolic GCN of \citet{chami2019hyperbolic}
to products of constant curvature manifolds. The following facts of
Riemannian geometry allow us to extend GCNs from single manifolds to
product manifolds. 

\begin{definition}
Given Riemannian manifolds, $(M_1, g_1), \ldots, (M_n, g_n)$, we can construct a \textit{Riemannian product manifold} $(M_1 \times M_2 \times \cdots \times M_n, g)$
with the \textit{product metric}, $g = g_1 \oplus \cdots \oplus g_n$, where the distance decomposes
as $$d_M(x, y)^2 = d_{M_1}(x_1, y_1)^2 + \cdots + 
d_{M_n}(x_n, y_n)^2$$
for any $x = (x_1, \ldots, x_n), y = (y_1, \ldots, y_n) \in M$.
\end{definition}

Thus, operations in the tangent spaces, $T_pM$, of the product manifold 
$M$ decompose as direct sums of operations over the constituent tangent
spaces $T_{p_i}M_i$, where $M = M_1 \times \cdots \times M_n$ and
$p = (p_1, \ldots, p_n)$.

On each pathway graph for which we perform edge
prediction, we match the signature of the product manifold in the GCN to the one used to learn
the pathway embedding. The operations in the Product GCN are defined analogously to the ones
in the hyperbolic GCN paper \citep{chami2019hyperbolic}.
Briefly put, pathway nodes are represented as a Cartesian product of points
on some combination of components of $\H^n$, $\S^n$, and $\E^n$ manifolds. At each layer in the Product GCN,
the points are projected into a tangent space to the manifold via the $\log$ map of the product space 
and a GCN operates in this Euclidean tangent space using the standard message-passing updates. 
The points are then projected back onto the manifold after the layer operation via the product
space $\exp$ map. Both the $\log$ map and the $\exp$ map decompose as sums over the constituent
manifolds in the product space. The same procedure is used for attention and activation layers,
with projection into the tangent space, followed by a standard layer operation, followed by projection back onto the manifold. For spherical (S) and hyperbolic
(H) manifolds having curvatures $C$ and $-C$, respectively, the $\exp$ and $\log$
maps are defined as follows:

\begin{align*}
    \text{(S)} \quad \exp^C_{\mathbf{x}}(\mathbf{v}) &= \cos\left(\frac{\mathbf{v}}{{\|\mathbf{v}\|}}\right)\mathbf{x} + C\sin \left( \frac{\mathbf{v}}{\|\mathbf{v}\|}\right)\mathbf{v} \\
\text{(H)} \quad \exp_{\mathbf{x}}^C(\mathbf{v}) &= 
\cosh\left(\frac{\|\mathbf{v}\|}{\sqrt{C}}\right) \mathbf{x} + 
\sqrt{C} \sinh\left(\frac{\|\mathbf{v}\|}{\sqrt{C}}\right) 
\frac{\mathbf{v}}{\|\mathbf{v}\|},
\end{align*}

\begin{align*}
\text{(S)} \quad \log_{\mathbf{x}}^C(\mathbf{y}) &= 
\cos^{-1}(\langle \mathbf{x}, \mathbf{y}\rangle_C)\frac{\mathbf{y}-\mathbf{x}-\langle \mathbf{x}, \mathbf{y-x}\rangle_C}{\|\mathbf{y}-\mathbf{x}-\langle \mathbf{x}, \mathbf{y-x}\rangle_C\|}\\
\text{(H)} \quad \log_{\mathbf{x}}^C(\mathbf{y}) &= \sqrt{C} \, \operatorname{arcosh}\left(-\langle \mathbf{x}, \mathbf{y} \rangle / C \right)
\frac{\mathbf{y} + \frac{1}{C} \langle \mathbf{x}, \mathbf{y} \rangle \mathbf{x}}
{\|\mathbf{y} + \frac{1}{C} \langle \mathbf{x}, \mathbf{y} \rangle \mathbf{x}\|}
\end{align*}

The Product GCN is trained using the Riemannian Adam 
optimization algorithm \citep{becigneul2018riemannian} on the edge prediction task described in Section~\ref{sec:edge_pred}.

\section{Results}
%\todo[inline]{Make sure every main text figure and supplementary figure is referenced from the main text.} % Will do this during revision
\subsection{Pathway Embeddings}
\label{sec:path_embed}

We first summarize each of the pathway databases in Table~\ref{tab:summary-stats} and
provide node and edge histograms in Figure~\ref{fig:histograms}. \todo{Preferred caption phrasing would have the number of outliers excluded. Capitalize database names the same way as in the text. In the next version, showing each database in a short long row on its own may be easier to see. KEGG has few graphs so it is covered up by the larger databases.}
There exist uniquely beneficial product spaces, as measured by distortion, for only roughly
half of the pathways.
Most pathways are small by graph machine learning standards
with the majority having less than 100 nodes and less than 200 edges. Thus, pathways exhibit
a unique opportunity to study the effects of popular graph learning techniques, including GCNs, on small graphs, as most graph machine learning benchmarks 
possess thousands, or even millions and billions, of nodes and edges.

\begin{figure*}[hbtp]
\begin{subfigure}[h]{0.5\textwidth}
\includegraphics[width=\textwidth]{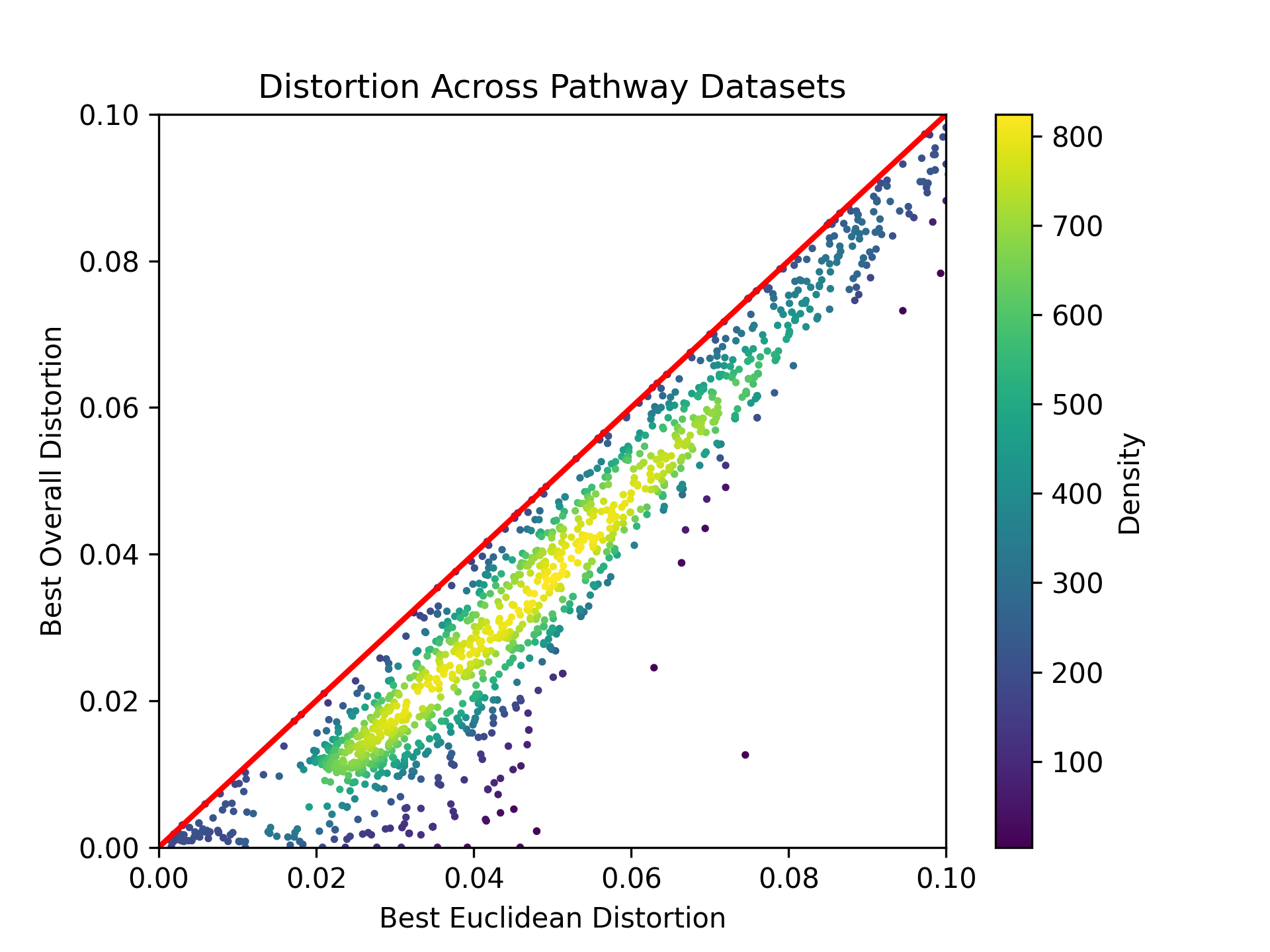}
\caption{Best overall versus Euclidean distortions}
\end{subfigure}
\hfill
\begin{subfigure}[h]{0.5\textwidth}
   \includegraphics[width=\textwidth]{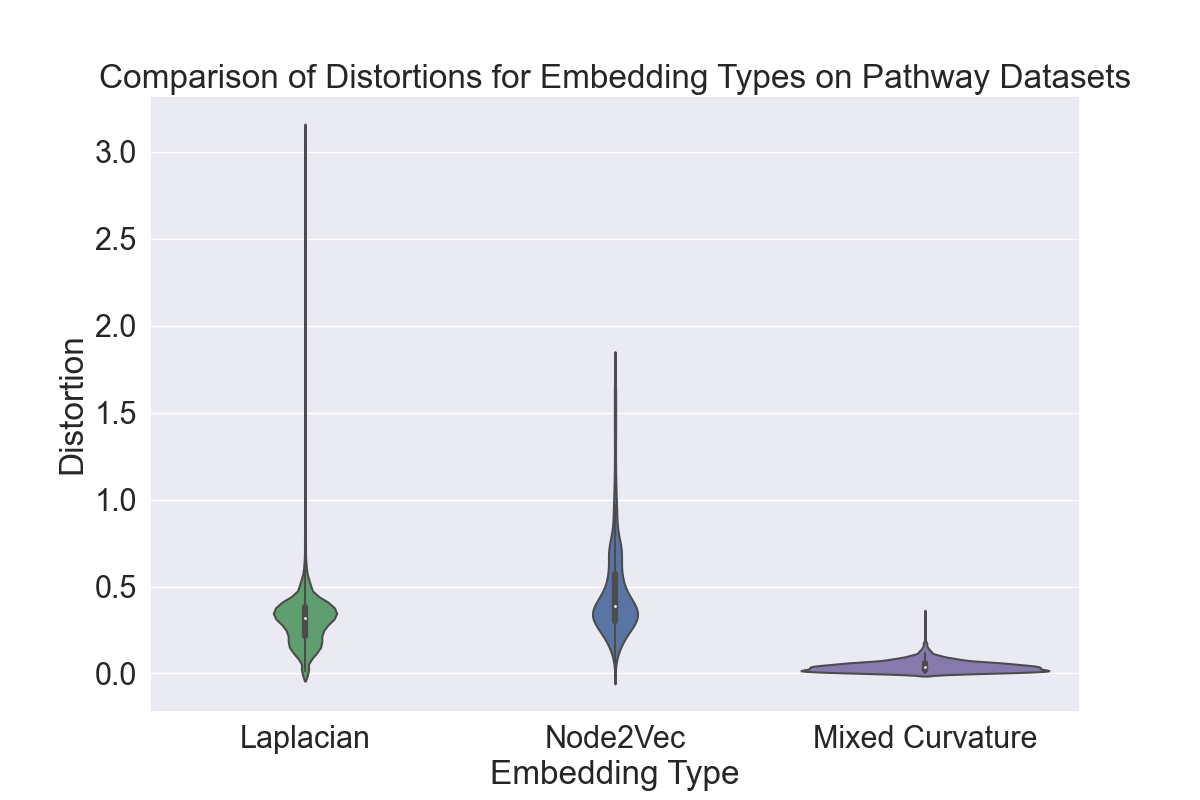}
   \caption{Pathway distortion distributions}
\end{subfigure}
\caption{Best overall mixed-curvature versus best Euclidean distortions across all pathway datasets (a). See Figure 
\ref{fig:distortion-scatter-apdx} for individual pathway datasets. Distortions across all datasets for the graph Laplacian, node2vec, and mixed-curvature embedding methods (b).}
\label{fig:dist-violin}
\end{figure*}

Next, for each pathway graph, we determine the best combination of hyperbolic, Euclidean, and 
spherical components as determined by lowest distortion.
Figure~\ref{fig:dist-violin}(a) demonstrates the
reductions in distortion gained from learning the non-Euclidean embeddings 
over all graphs in each of the five pathway datasets we studied. The red diagonal
indicates the position at which the distortions of the 
Euclidean embeddings and mixed-curvature product embeddings would be equal.
Points lying below this line indicate graphs for which a product representation 
yielded a reduction in distortion over a Euclidean representation.
Because a fully Euclidean embedding is a special case of the mixed-curvature product embedding, the best mixed-curvature product embedding should always have better distortion than the best Euclidean embedding.
The question is whether the improvement is marginal or substantial on biological pathways.
We observe that mixed-curvature product spaces provide marked 
reductions in distortion relative to the standard Euclidean 
embedding, with many graphs achieving a greater than 50\% reduction 
in distortion.

We further compare our mixed-curvature embeddings against two common graph embedding 
baselines. %\todo{Do we need short methods section introducing how the baselines learn embeddings?} % Leave for now, restructure during revisions
The first is node2vec \citep{node2vec}, a popular graph embedding method which uses an algorithm analogous to word2vec \citep{w2v} trained on
random walks taken from the graph. We also compare against a standard embedding method 
formed by taking eigenvectors of the graph Laplacian matrix \citep{Bonald_2019}.
For both methods, we embed the pathway graphs from all datasets and compute distortions for
each pathway embedding in each dataset. Because node2vec and the graph Laplacian embedding
do not directly minimize distortion in their loss functions, they can arbitrarily scale
embedding vectors, causing them to perform poorly on a naive calculation of the distortion metric.
Therefore, we perform a scaling optimization for the node2vec and graph Laplacian embeddings,
finding for each dataset and embedding type the constant $c$ which minimizes the below objective
$$\min_{c \in \R^+} \sum_{i < j} \left(\frac{c \cdot d(x_i, x_j)}{d_G(n_i, n_j)} - 1\right)$$
where $d(x_i, x_j)$ is the distance between the embeddings of nodes $i$ and $j$, and $d_G(n_i, n_j)$
is the shortest-path graph distance between nodes $i$ and $j$. %\todo{Scaling description better for Methods section} % Restructure during revisions
After calculating the optimal scaling factor, $c$, we scale the embeddings by 
this factor and recalculate the distortions for all pathways in the dataset.
We summarize the distortions in Figure~\ref{fig:dist-violin}(b).
The mixed-curvature embeddings exhibit significantly lower mean distortion and variance than the node2vec and Laplacian baselines.

\subsection{Edge Prediction}
\label{sec:edge_pred}
\subsubsection{Overview}
We train on individual pathway graphs to predict a set of held-out edges, then test our prediction models on the test set of experimentally validated edges from STRING. %\todo{We need to be more clear about how pathway edges are split, it is hard to follow because we don't have a methods section about that.}
For each pathway, we determine its optimal embedding space using the results of our 
hyperparameter sweep for the distortion task.
That is, for each pathway graph, we select the signature for 
the embedding space (number of hyperbolic, Euclidean, and spherical components, as well 
as their respective dimensionalities) that minimizes the distortion across the set of 
embeddings for that pathway graph. We then perform edge prediction on two different datasets:

\paragraph{In-Distribution}
An in-distribution validation set containing held-out edges from the original pathway graph. These edges are not seen by the model during training, but the validation set performance metrics, such as validation accuracy and validation loss, \textit{are} used to guide model training and hyperparameter selection. 95\% of the 
edges from each pathway were used for training, and 5\% were used
for validation.

\paragraph{Out-of-Distribution}
An out-of-distribution test set that includes edges from the STRING PPI database. This dataset is not seen by the model during training
and \textit{is not} used to guide model training or hyperparameter selection.
These edges are general PPIs and do not necessarily have the same biological context as the pathway.
However, pathway edge predictions supported by a general PPI are more plausible than those that are not.

We expect that the Product GCN, initialized with the
min-distortion product space embeddings, has a better inductive
bias in its pretrained representations that more preferentially 
match the topology of each graph than any of the other baselines.
Thus, we expect that the Product GCN should more accurately 
predict in-distribution edges held out from the original graph.
Although we are less certain what the effect of introducing 
non-Euclidean geometry should be to the prediction of 
out-of-distribution edges from an external database, we would hope
that the Product GCN would still outperform the other baselines
on this task.

\subsubsection{Baselines and Product GCN}
We compare four different methods on the edge prediction task:

\begin{figure*}[ht]
\begin{subfigure}[h]{.5\textwidth}
    \includegraphics[width=\textwidth]{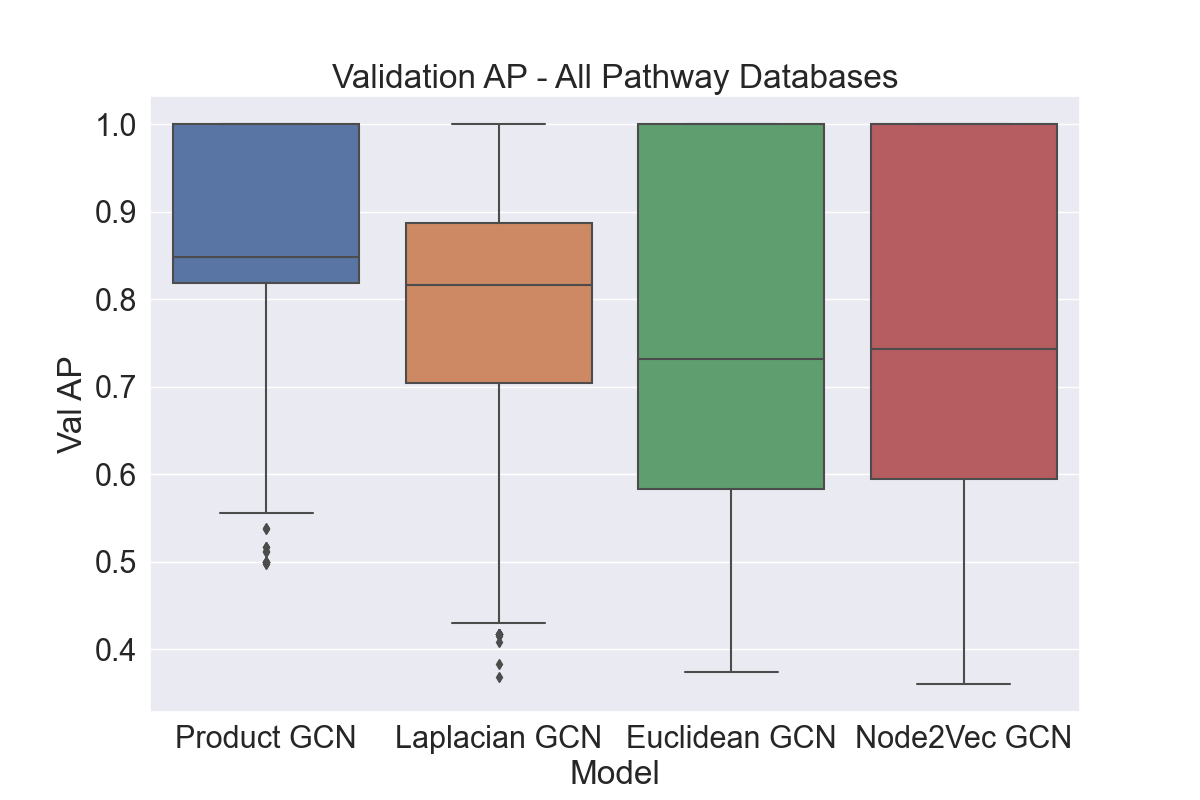}
\end{subfigure}%
\hfill
\begin{subfigure}[h]{.5\textwidth}
    \includegraphics[width=\textwidth]{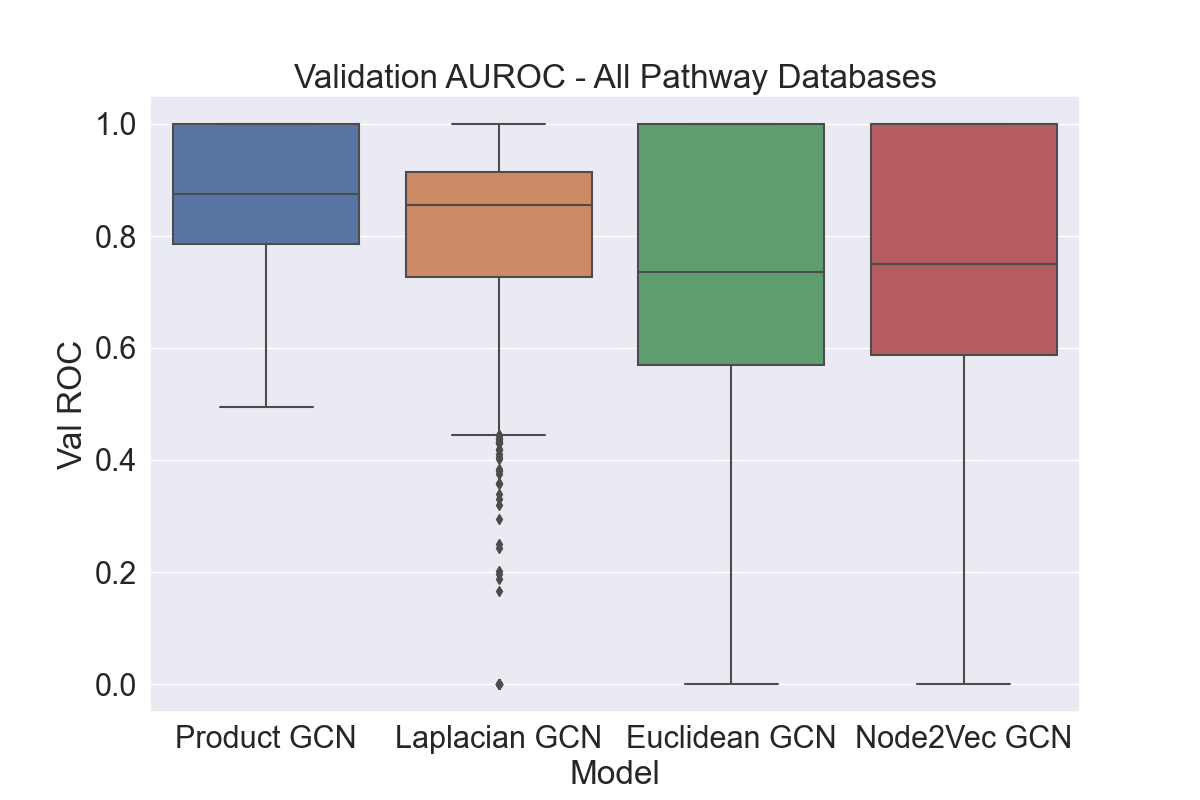}
\end{subfigure}
\begin{subfigure}[h]{.5\textwidth}
    \includegraphics[width=\textwidth]{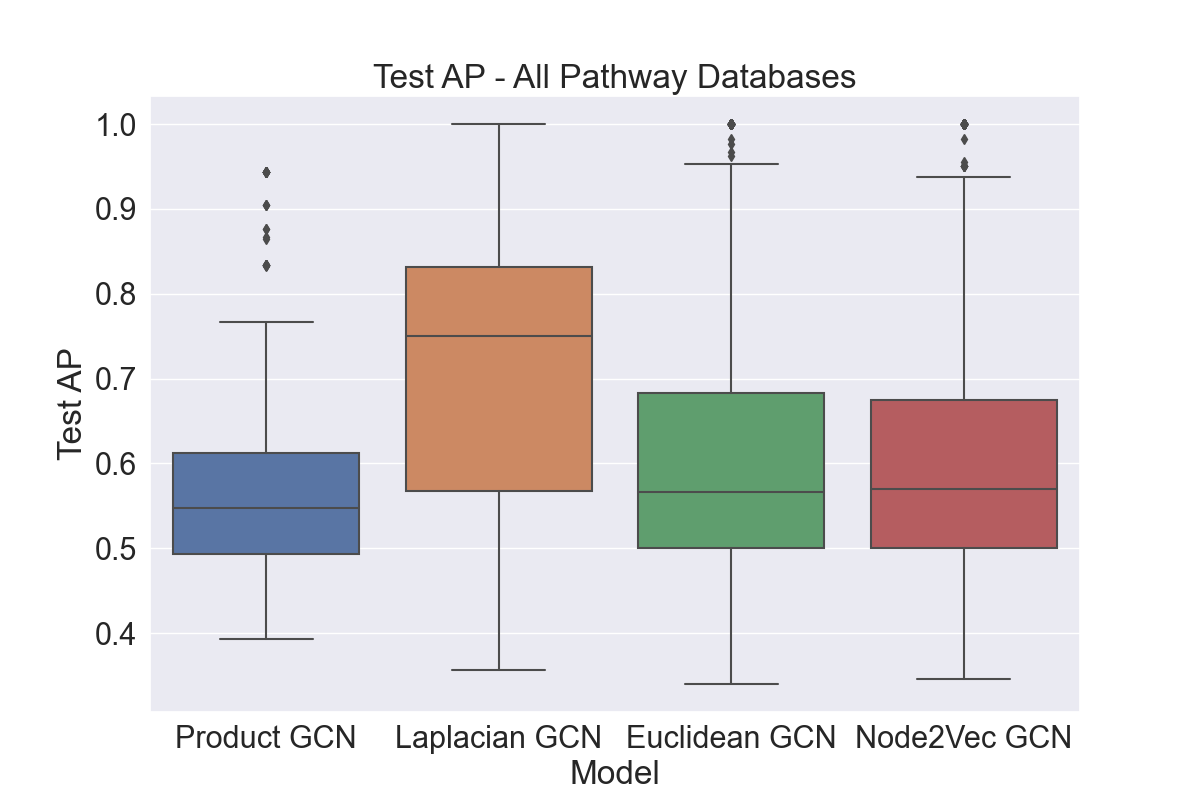}
\end{subfigure}%
\hfill
\begin{subfigure}[h]{.5\textwidth}
    \includegraphics[width=\textwidth]{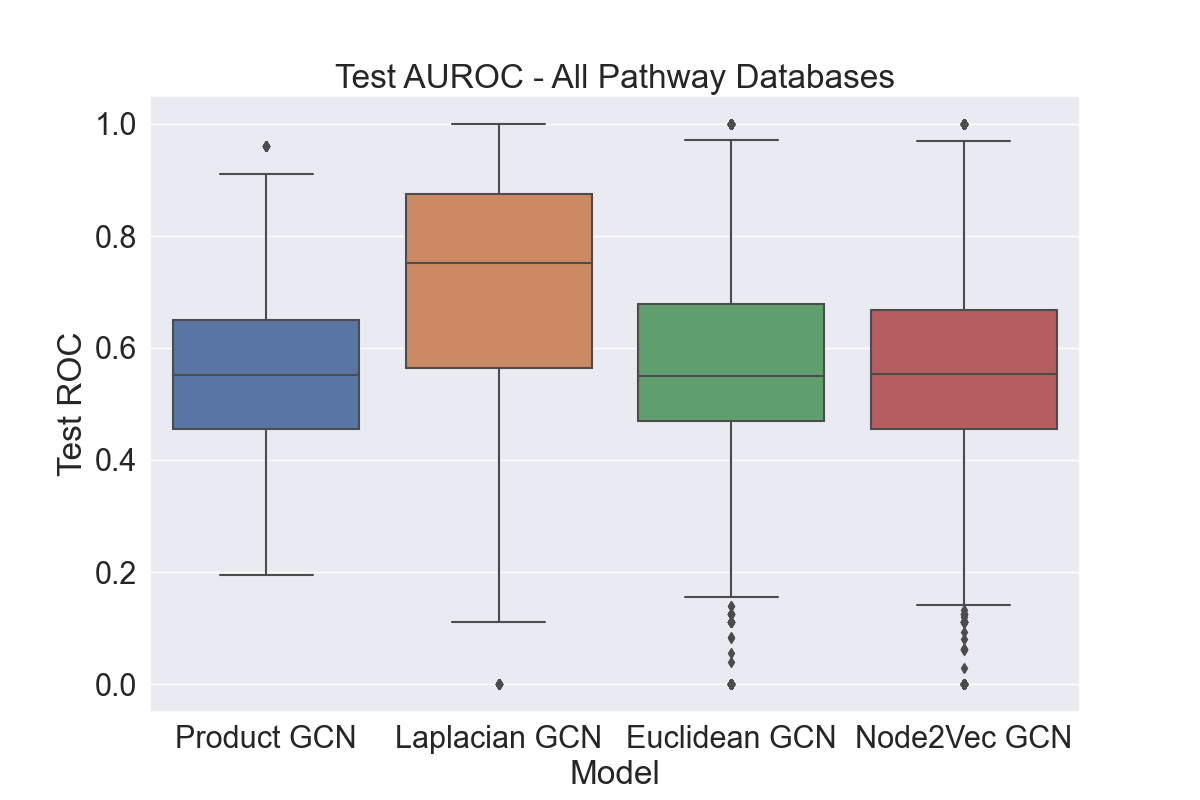}
\end{subfigure}
\caption{Edge prediction performance on all five pathway datasets for all models as given by four metrics: Validation Set AP, Validation Set AUROC, Test Set AP,
and Test Set AUROC.
Test set edges come from STRING.}
\label{fig:box-plots}
\end{figure*}

\begin{enumerate}
    \item A Euclidean GCN initialized with pretrained node2vec embeddings.
    \item A Euclidean GCN initialized with pretrained Laplacian embeddings.
    \item A Euclidean GCN initialized with pretrained Euclidean embeddings designed to minimize distortion in $\R^n$.
    \item A Product GCN architecture that matches the signature of the optimal non-Euclidean pathway embedding as measured by minimum distortion and is initialized with
    the pretrained mixed-curvature embeddings from the first task described in Section~\ref{sec:path_embed}.
\end{enumerate}

The pretrained Euclidean embeddings in method 3 are learned using the same Python package as
is used for learning the mixed-curvature embeddings \citep{Gu2018LearningMR} but use a Euclidean 
manifold as the target output space. For all methods, we keep the total dimensionality of the
embedding space fixed at 100 dimensions, with the exception of the 
Laplacian embedding method for which the embedding dimension was 
set to be the number of nodes in the graph, $|V|$. Since most pathway graphs have $|V| \leq 100$ (Figure~\ref{fig:histograms}(a)) the reduced dimensionality of the 
Laplacian embedding would be expected to be harmful, though it could provide 
an outsized benefit in representational capacity for
pathway graphs with a large number of nodes.
%We also perform a full hyperparameter sweep over GCN
%parameters during training to find the optimal model for each pathway and embedding type. More
%details about the hyperparameter sweep are given in the Appendix.
%% Commented these lines because we discuss hyperparameter sweeping at the start of the next section
% Consider moving some of the above to Methods

For the Product GCN, we match the signature of the best mixed-curvature embedding.
For example, if the best embedding for pathway 21 of the PathBank dataset has the signature
$\H^2 \times \S^3 \times \E^2$, then we use a Product GCN architecture with two hyperbolic
manifold layers, three spherical manifold layers, and two Euclidean manifold layers. % A figure of this architecture would be very helpful for the next version
We furthermore split each node embedding into its constituent pieces as determined by the product
manifold signature. In this example, the embedding 
would be split into $2 + 3 + 2 = 7$ slices, each of dimension 14. Each slice of the embedding
is fed into the GCN for its corresponding manifold. Scores are then generated
by each of the component GCNs and averaged to produce a final output score that is used to predict the presence of an edge between a given pair of nodes.

\subsubsection{Model Training}
For each graph on which we perform edge prediction with the Product GCN, we perform a sweep over 324 hyperparameter
combinations (Appendix) and choose the one with the highest average of validation set AUROC and AP for use in downstream analyses. For the baseline methods, we perform a sweep over 108 hyperparameter configurations (Appendix). The difference in the two sweeps arises from the fact that the Product GCN has one additional hyperparameter, the curvature of the 
hyperbolic and spherical embedding spaces. We sweep over three values for this parameter, yielding three times as many hyperparameter configurations.

\subsubsection{Summary of Results}
We find that mixed-curvature product manifold representations yield substantial
benefits in representational capacity for boosting downstream edge performance on the in-distribution validation edges (Figure~\ref{fig:box-plots}). The Product GCN
initialized with pretrained product manifold representations outperforms, on
average, all baselines for prediction of in-distribution validation set 
edges.
%Furthermore, the Product GCN exhibits low variance relative to the other methods.
% Commented out the above because when taken over all pathways it looks comparable to Laplacian
Among the baseline methods, the graph Laplacian-based Euclidean GCN outperforms the two other Euclidean GCNs.
For pathways with more than 100 nodes, this advantage is likely due to its larger embedding dimension.
In the Appendix, we provide detailed paired comparisons of the AP and AUROC for the Product GCN and each Euclidean GCN on individual graphs in each pathway database.

The Product GCN does not have the same advantage over the baselines for the prediction of out-of-distribution 
PPI edges from STRING. 
The Euclidean GCN with pretrained graph Laplacian embeddings is the clear best model, and all others have near random AUROC. % Add the random AP baseline and statistical testing to support these statements later
We surmise that the introduction of these
edges induces a substantial distribution shift in the graph topologies, which makes 
the initially learned product space embeddings no longer a good fit. This causes
the Product GCN to underperform relative to the Euclidean GCNs with their
respective pretrained embeddings. One conclusion we can draw is that common
Euclidean-style embeddings are more robust to distribution shifts of graph 
topology while product space embeddings, via their learned curvatures, impose
a strong inductive bias to tree-like and ring-like structures in the graphs.

\section{Discussion and Conclusion}
We find that performing representation learning in non-Euclidean and mixed-curvature spaces yields 
notable improvements in distortion and downstream in-distribution edge prediction performance.
In all cases, a mixed-curvature representation yields an embedding with lower distortion
than a simple Euclidean embedding. However, the exact decomposition of the product space into its mixture 
of components is highly dependent on the graph topology. Thus, it is beneficial to perform a hyperparameter sweep over the number and types of components, as well as their
dimensionalities, when learning a representation for a biological pathway graph. Such a sweep
need not be highly time intensive as biological pathway graphs are generally of a modest size.

For the out-of-distribution edge prediction task, the full structure of the graph is not known when the node embeddings are learned because the test set edges from STRING are not 
included in the training set.
This leads to a number of questions about how to further generalize our approach.
For example, we may ask how close the full graph structure must be to the observed structure in order to produce node embeddings that are usable for edge prediction via our method.
We can also explore whether there are alternative hyperparameter tuning strategies that produce slightly worse distortions or in-distribution edge prediction performance but generalized better to out-of-distribution edge prediction. % Is this is a good idea? Feasible?
% We could explore the sensitivity of the optimal mixed-curvature space for distortion and edge prediction. We had started to do that for PathBank.
The Laplacian embedding has the best out-of-distribution performance, consistent with its robust behavior in previous biological graph representation learning benchmarking \citep{song_benchmarking_2023}.
Currently, we only assess out-of-distribution edge predictions using STRING physical protein interactions.
An evaluation of 45 interaction networks \citep{wright_state_2025} could be used to prioritize other networks for evaluating these predictions and understand biases in the interaction networks.

Another future direction would be to consider additional downstream tasks, as edge prediction is not the only useful task for biological pathways that can be improved by non-Euclidean representation learning.
We plan to investigate how our pathway embeddings benefit other
problems such as node classification, for example, predicting the type (gene, protein, small molecule, metabolite, etc.) of a pathway node.

The lower performance of all models on the out-of-distribution edge prediction setting relative to the in-distribution setting may be partially attributable to biological context.
Edges in a pathway reflect a pair of proteins that interact in the context of conducting some specific biological process, potentially only in certain cells or tissue types.
The trained GCNs predict edges with that same context.
However, the STRING-based evaluation lacks that context.
There may be false negative edges in which there is a real PPI that is not relevant to the pathway.
Alternatively, there may be false positive edges that are cell- or tissue-specific and not yet identified by the experiments aggregated into STRING.

Non-Euclidean embedding models have not been applied to pathway graphs, perhaps
due to a lack of awareness among network biology researchers of their utility
in reducing distortion of graph distances and improving downstream predictive performance. We demonstrate that pathway graphs benefit
from the incorporation of non-Euclidean geometries into embeddings
and prediction models. We encourage researchers to consider making use of these
non-standard geometries when learning embeddings and making downstream 
predictions.

\section*{Software and Data availability}
We provide code at \url{https://github.com/mcneela/Mixed-Curvature-Pathways} and 
\url{https://github.com/mcneela/Mixed-Curvature-GCN}.

\section*{Acknowledgements}
We thank David Merrell, Christopher Magnano, and Sam Gelman for their help in processing 
Pathway Commons graphs and generating visualizations.
This research was supported by NIH award R01GM135631, the
Wisconsin Alumni Research Foundation, and computing resources provided by the \citet{chtc}.

\nocite{Lee2002IntroductionTS} % cite in text (intro?)
\nocite{TuManifolds}
\bibliography{example_paper}
\bibliographystyle{icml2023}

%%%%%%%%%%%%%%%%%%%%%%%%%%%%%%%%%%%%%%%%%%%%%%%%%%%%%%%%%%%%%%%%%%%%%%%%%%%%%%%
%%%%%%%%%%%%%%%%%%%%%%%%%%%%%%%%%%%%%%%%%%%%%%%%%%%%%%%%%%%%%%%%%%%%%%%%%%%%%%%
% APPENDIX
%%%%%%%%%%%%%%%%%%%%%%%%%%%%%%%%%%%%%%%%%%%%%%%%%%%%%%%%%%%%%%%%%%%%%%%%%%%%%%%
%%%%%%%%%%%%%%%%%%%%%%%%%%%%%%%%%%%%%%%%%%%%%%%%%%%%%%%%%%%%%%%%%%%%%%%%%%%%%%%
\newpage
\appendix
\onecolumn

% Same formatting as title take from icml2023.sty
{\Large\bf Appendix}

\section{Supplementary Methods}
\subsection{Pathway Commons Data}
We use the Pathway Commons v12 PathBank pathways provided in this file:
\url{https://www.pathwaycommons.org/archives/PC2/v12/PathwayCommons12.pathbank.hgnc.txt.gz}

Pathway Commons provides a number of different data formats, including text,
SIF, JSON, and BioPAX. We use the text format as it provides the simplest graph representation of the most important interactions in PathBank pathways.
%%%%%%%%%%%%%%%%%%%%%%%%%%%%%%%%%%%%%%%%%%%%%%%%%%%%%%%%%%%%%%%%%%%%%%%%%%%%%%%
%%%%%%%%%%%%%%%%%%%%%%%%%%%%%%%%%%%%%%%%%%%%%%%%%%%%%%%%%%%%%%%%%%%%%%%%%%%%%%%
%\section{Modeling}
\subsection{Initializing GCNs with Embeddings}
We save the embeddings learned by the embedding model in a PyTorch .pt file \cite{paszke_pytorch_2019}.
We then load these embeddings and use them to initialize the weights of the
embedding layer for the models in the edge prediction task, namely the Euclidean 
GCN and Product GCN. These embeddings are then further trained via backpropagation during
the edge prediction task.
\todo{Any other software to cite? Visualization packages like matplotlib and seaborn? We commented out the sklearn citation below?}

\section{Supplementary Tables}

%\subsection{Hyperparameters}
% \usepackage{graphicx}
\begin{table}[ht]
\centering
%\resizebox{\textwidth}{!}{%
\begin{tabular}{|l|l|l|l|}
\hline
\textbf{learning rate} & \textbf{hyperbolic components} & \textbf{Euclidean componentss} & \textbf{spherical components} \\ \hline
1e-3  & 0 & 0 & 0 \\ \hline
1e-2  & 1 & 1 & 1 \\ \hline
1e-1  & 2 & 2 & 2 \\ \hline
1     & 3 & 3 & 3 \\ \hline
\end{tabular}%
%}
\caption{Range of values for the pathway embedding hyperparameter sweep. Space dimensions were calculated automatically (to sum to 100) based on the number of components of each space. For example, if there were 2 hyperbolic components, 1 Euclidean component, and 3 spherical components, then there would be floor(100/6) = 16 dimensions assigned to each component.}
\label{tab:hyperparams}
\end{table}

% \usepackage{booktabs}
\begin{table}[hbtp]
\centering
\begin{tabular}{|l|l|l|l|l|l|}
\hline
\textbf{learning rate} & \textbf{hidden dim} & \textbf{\# of layers} & \textbf{dropout} & \textbf{use bias} & \textbf{curvature (for Product GCN)} \\ \hline
1e-4                   & 16                  & 2                     & 0.1              & true              & 0.5                                  \\ \hline
1e-3                   & 32                  & 3                     & 0.2              & false             & 1                                    \\ \hline
1e-2                   & 64                  & 4                     &                  &                   & 2                                    \\ \hline
\end{tabular}
\caption{Hyperparameters used in the sweep for
training the edge prediction GCNs.}
\label{tab:gcn-hyperparams}
\end{table}

\section{Supplementary Figures}
\begin{figure*}[hbtp]
\begin{subfigure}[h]{0.5\textwidth}
\includegraphics[width=\textwidth]{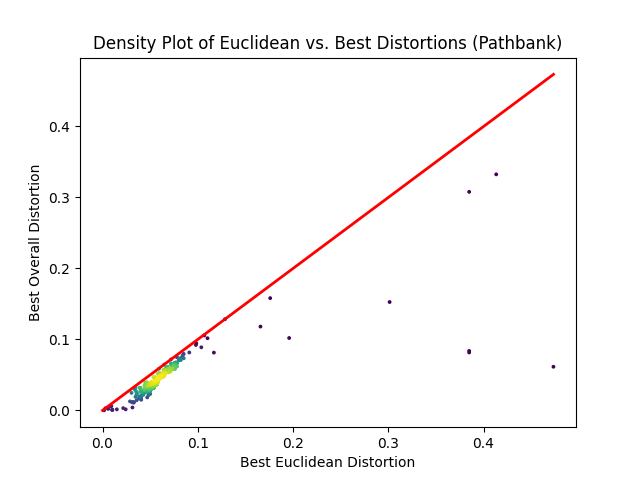}
\caption{PathBank dataset} % Change capitalization in figure if we regenerate it
\end{subfigure}
\hfill
\begin{subfigure}[h]{0.5\textwidth}
\includegraphics[width=\textwidth]{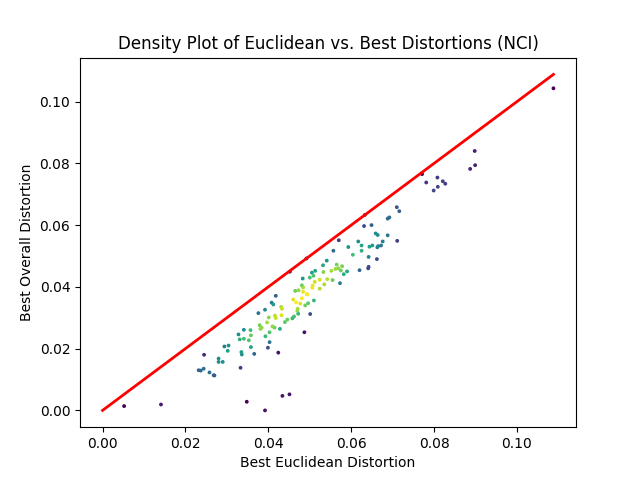}
\caption{NCI dataset}
\end{subfigure}
\begin{subfigure}[h]{0.5\textwidth}
\includegraphics[width=\textwidth]{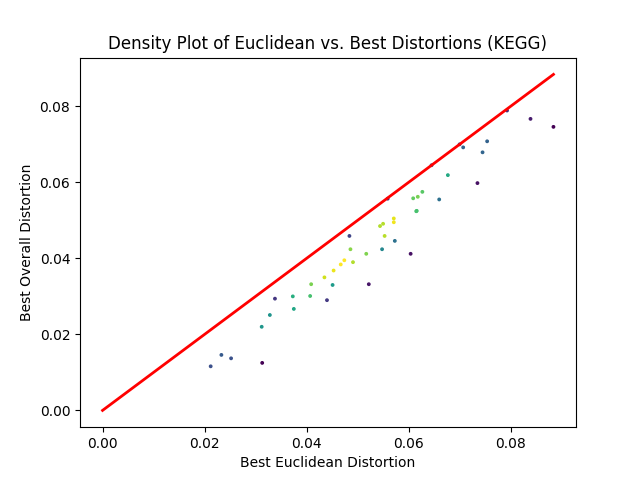}
\caption{KEGG dataset}
\end{subfigure}
\hfill
\begin{subfigure}[h]{0.5\textwidth}
\includegraphics[width=\textwidth]{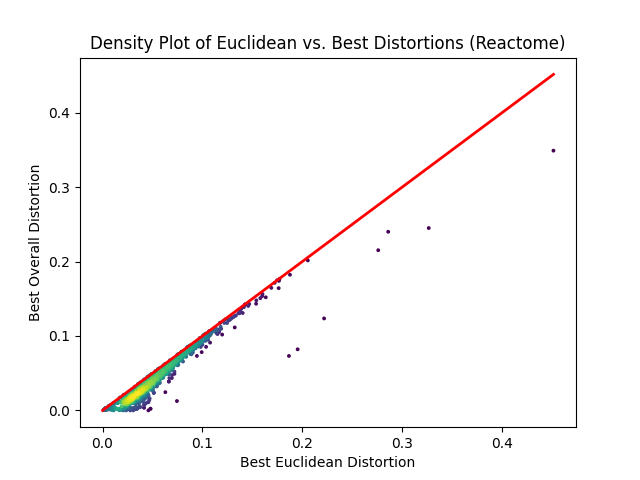}
\caption{Reactome dataset}
\end{subfigure}%
\hfill
\begin{subfigure}[h]{0.5\textwidth}
\includegraphics[width=\textwidth]{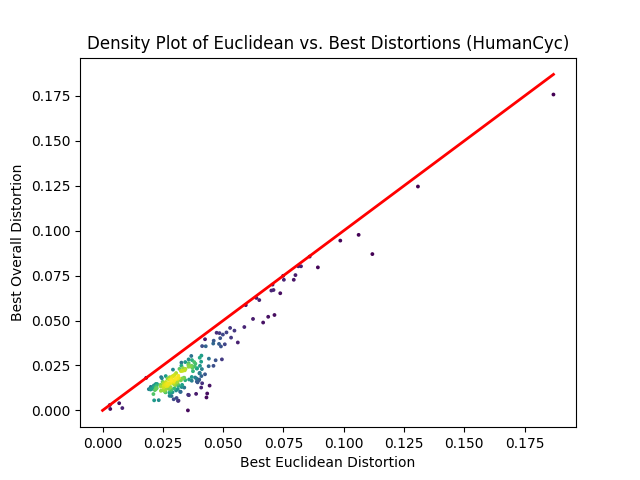}
\caption{HumanCyc}
\end{subfigure}
\hfill
\begin{subfigure}[h]{0.5\textwidth}
\includegraphics[width=\textwidth]{images/multi_dataset_scatterplot_density_zoomed.png}
\caption{All datasets}
\end{subfigure}%
\caption{Scatterplots of distortion in the Euclidean embedding versus distortion in
the mixed-curvature embedding for pathway datasets. Points are colored by local density, with yellow indicating the highest density. %All points are on or below the red diagonal because the mixed-curvature embeddings outperform the purely Euclidean embeddings in all cases.
}
\label{fig:distortion-scatter-apdx}
\end{figure*}
% --------------------- Pathbank ----------------------------
\newpage
\subsection{PathBank}
\begin{figure}[ht]
\centering
\begin{subfigure}{.5\textwidth}
    \centering
    \includegraphics[width=0.8\textwidth]{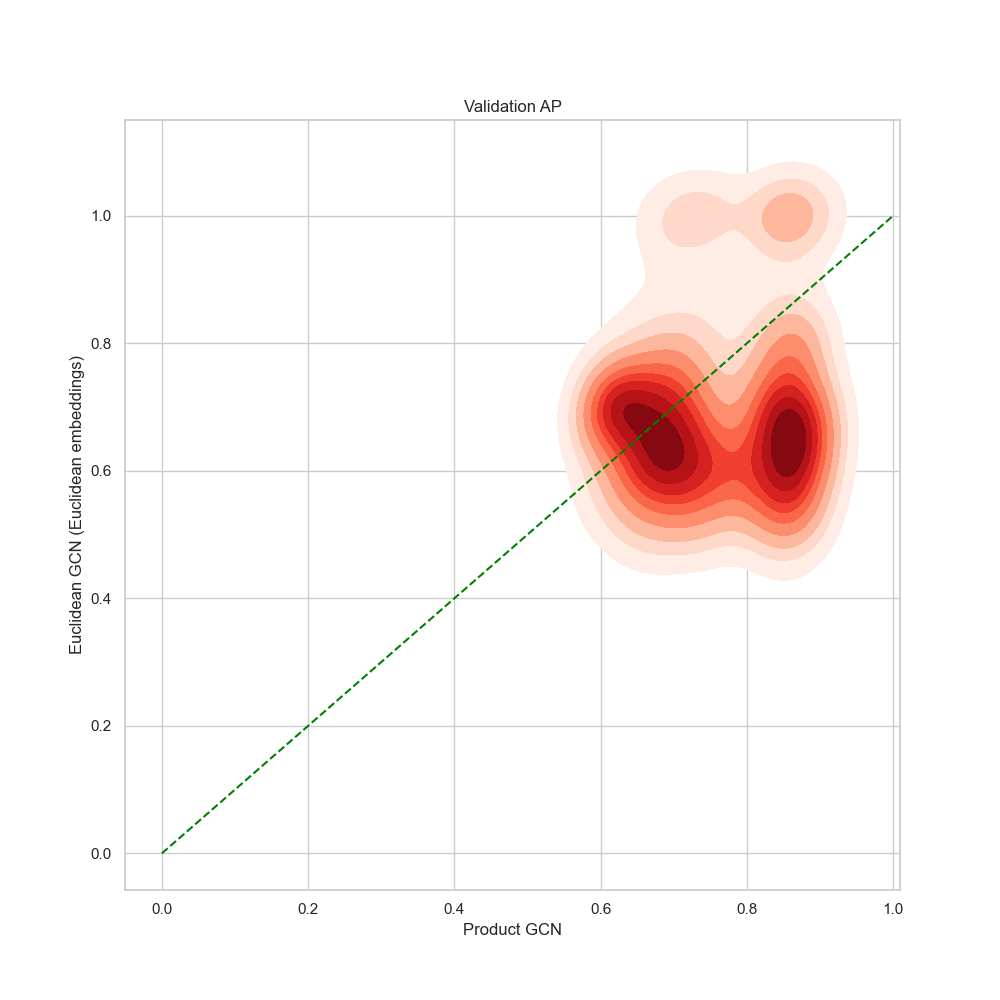}
\end{subfigure}%
\begin{subfigure}{.5\textwidth}
    \centering
    \includegraphics[width=0.8\textwidth]{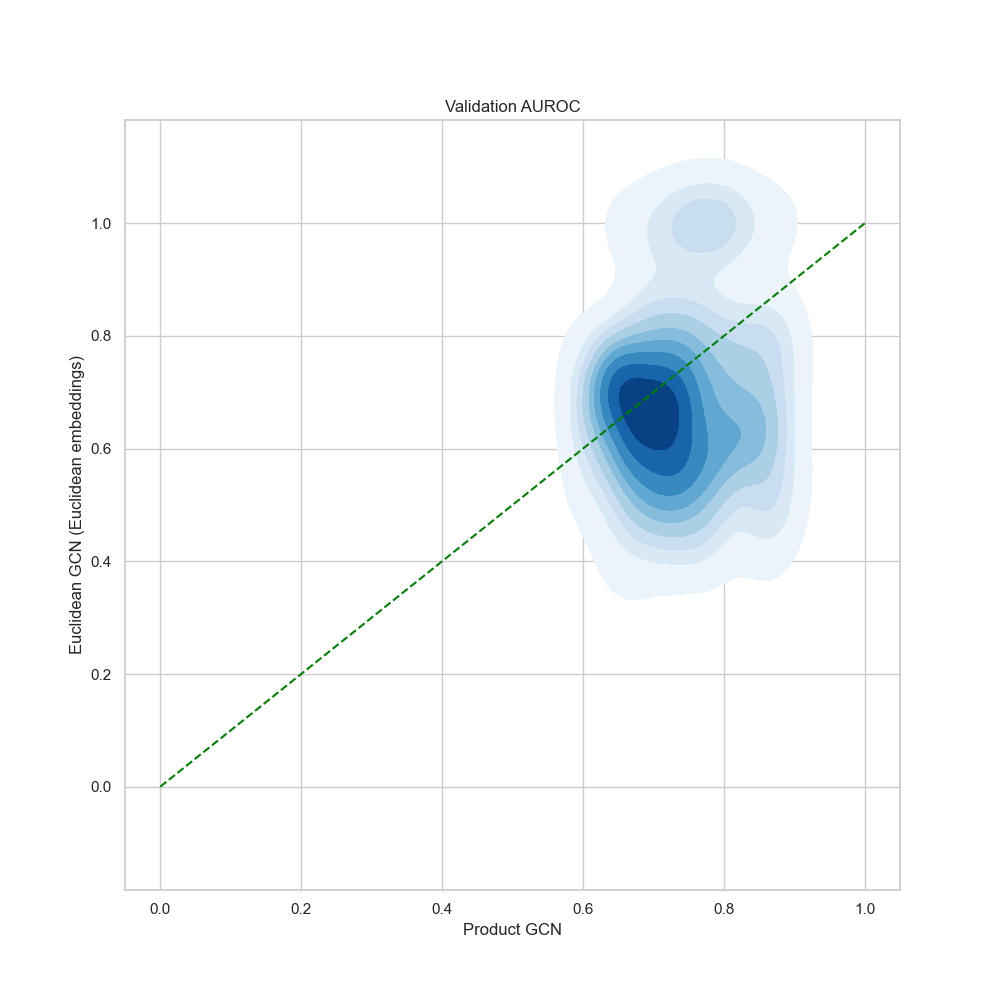}
\end{subfigure}
\begin{subfigure}{.5\textwidth}
    \centering
    \includegraphics[width=0.8\textwidth]{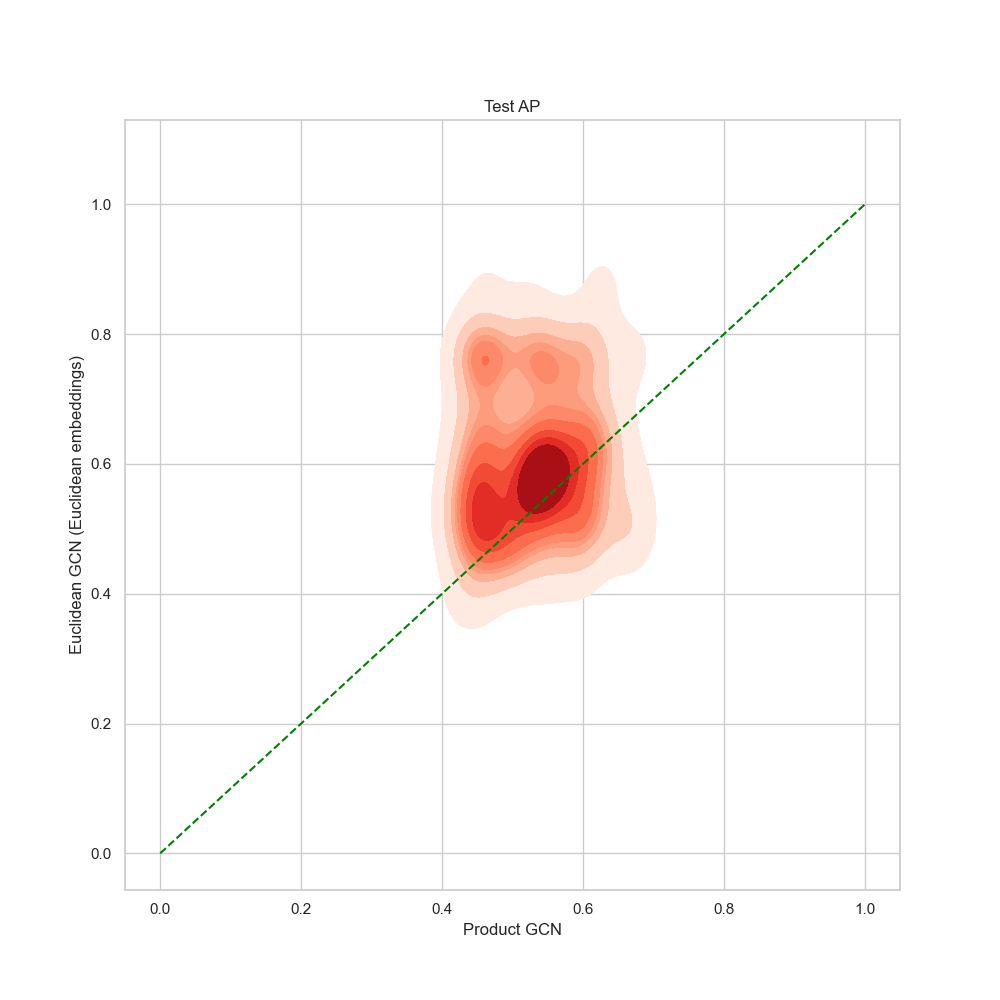}
\end{subfigure}%
\begin{subfigure}{.5\textwidth}
    \centering
    \includegraphics[width=0.8\textwidth]{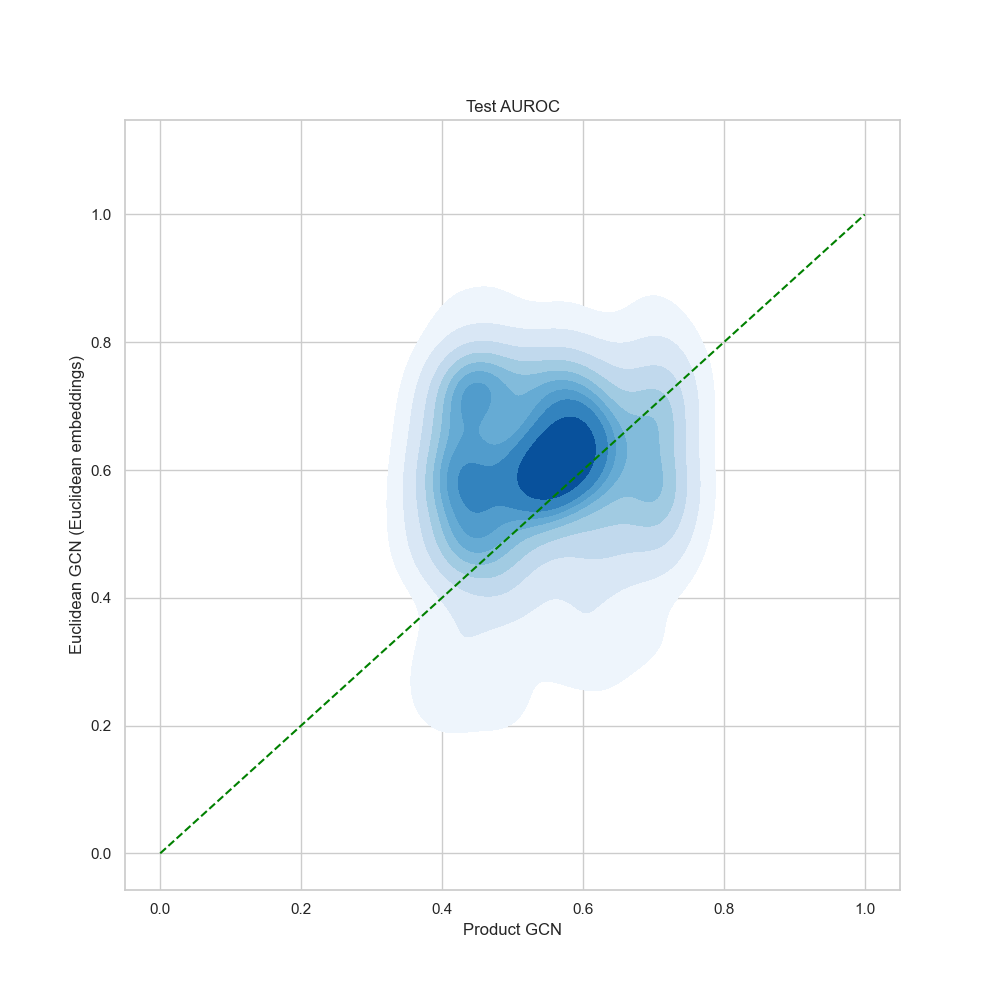}
\end{subfigure}

\caption[short]{Comparison of Euclidean GCN initialized with pretrained Euclidean embeddings and Product GCN performance on in-distribution 
validation set and out-of-distribution test set. Each density plot shows one of either AP or AUROC metrics taken across all graphs in the PathBank dataset.}
\end{figure}

\begin{figure}[ht]
\centering
\begin{subfigure}{.5\textwidth}
    \centering
    \includegraphics[width=0.8\textwidth]{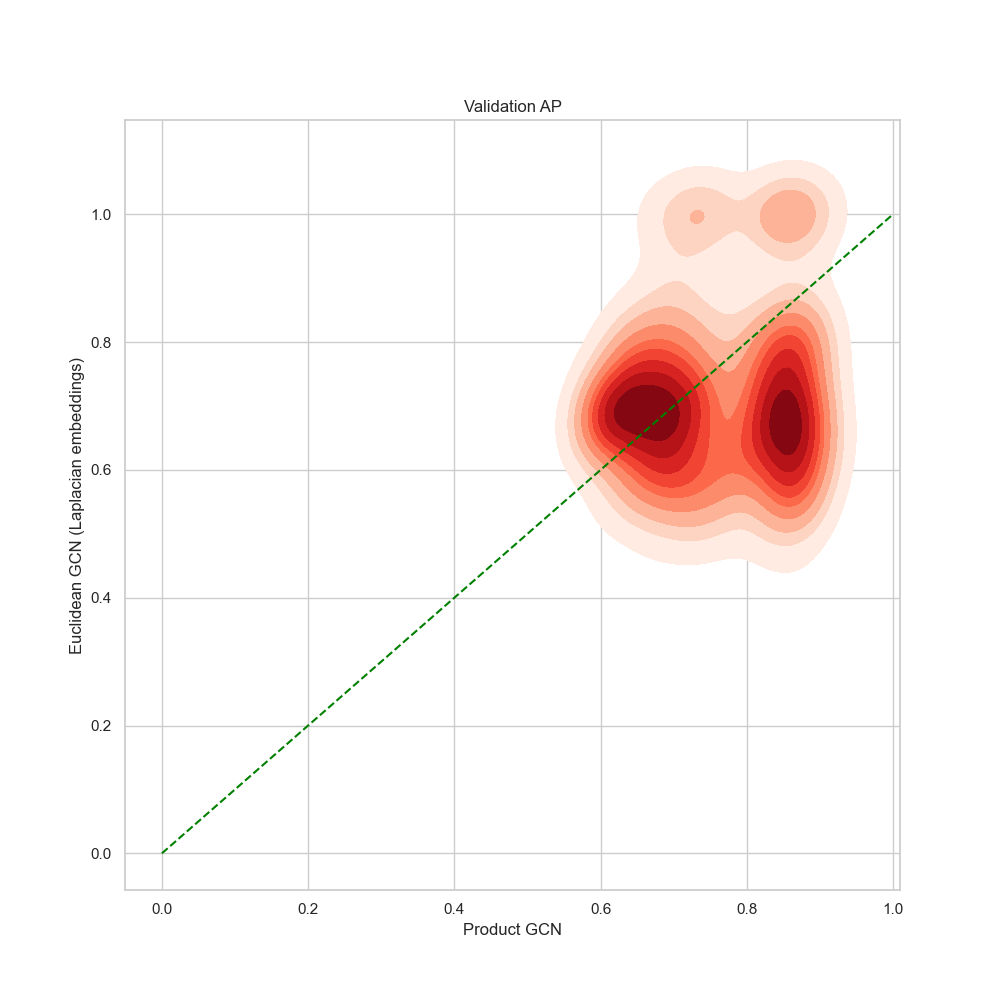}
\end{subfigure}%
\begin{subfigure}{.5\textwidth}
    \centering
    \includegraphics[width=0.8\textwidth]{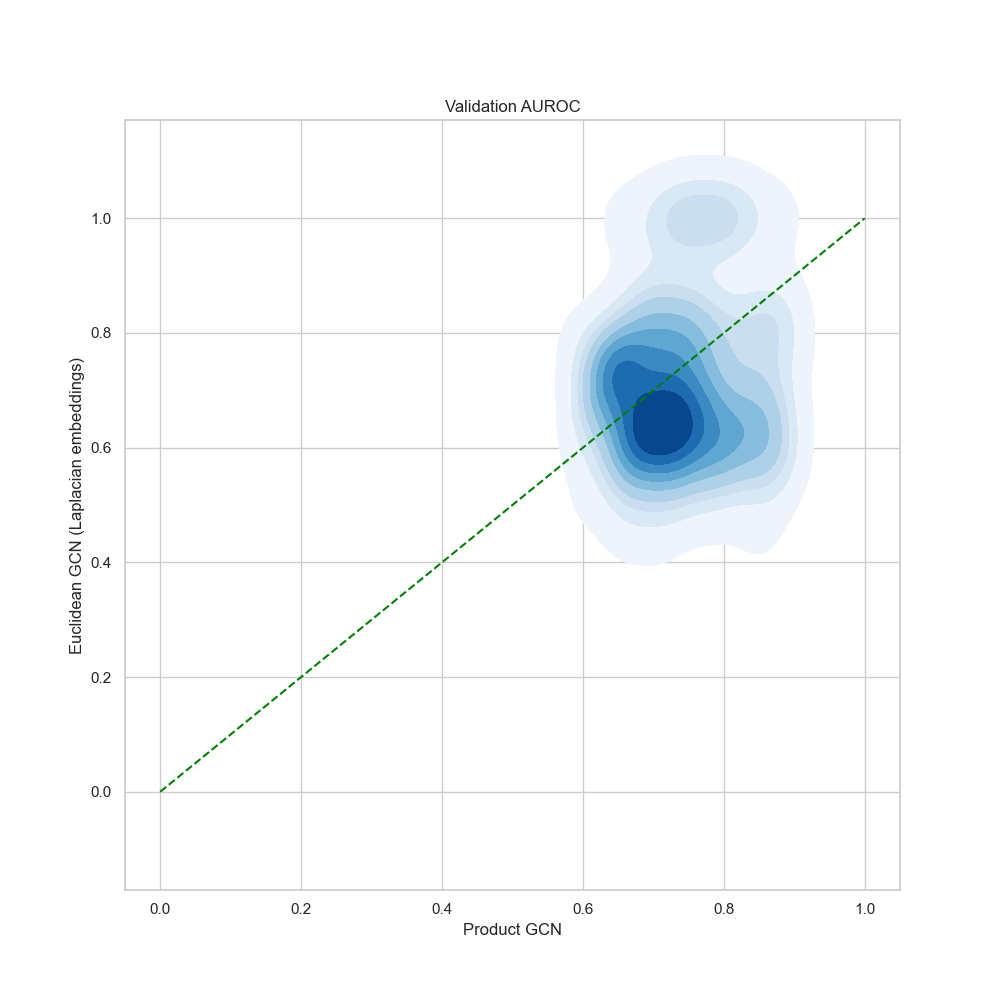}
\end{subfigure}
\begin{subfigure}{.5\textwidth}
    \centering
    \includegraphics[width=0.8\textwidth]{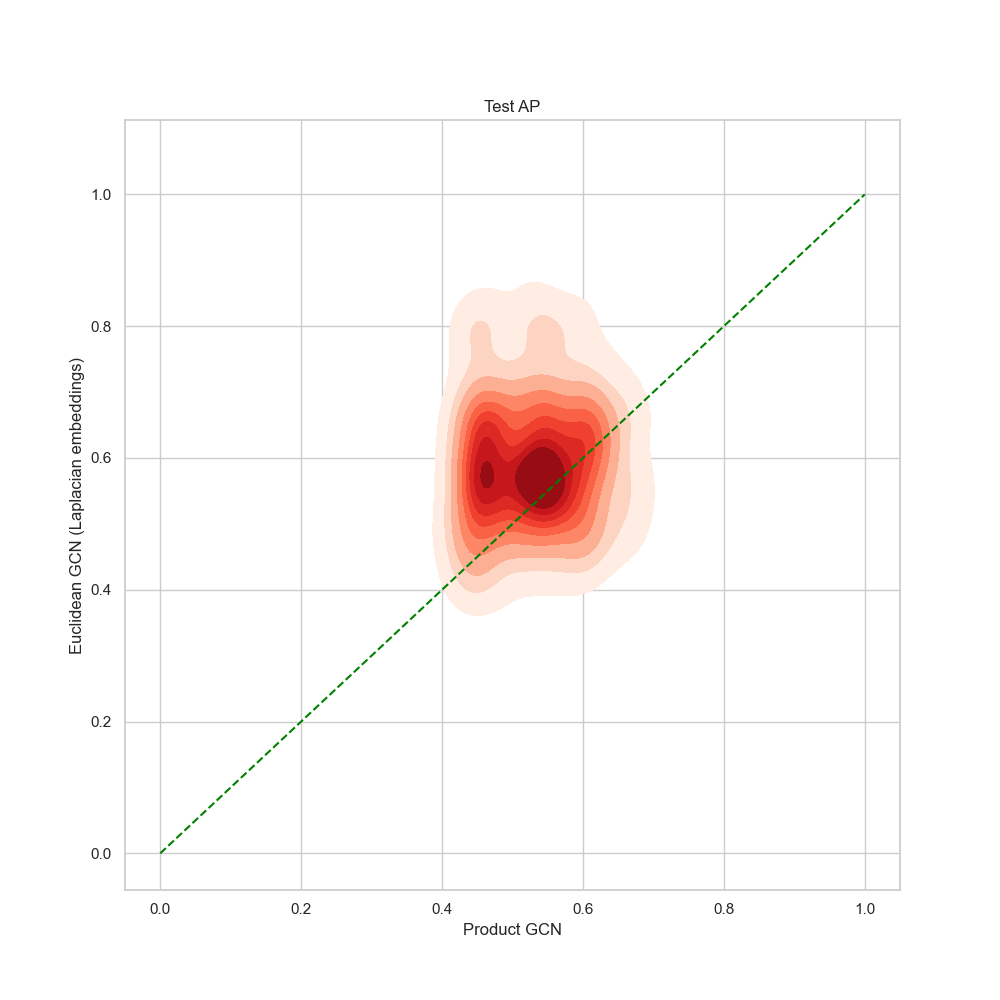}
\end{subfigure}%
\begin{subfigure}{.5\textwidth}
    \centering
    \includegraphics[width=0.8\textwidth]{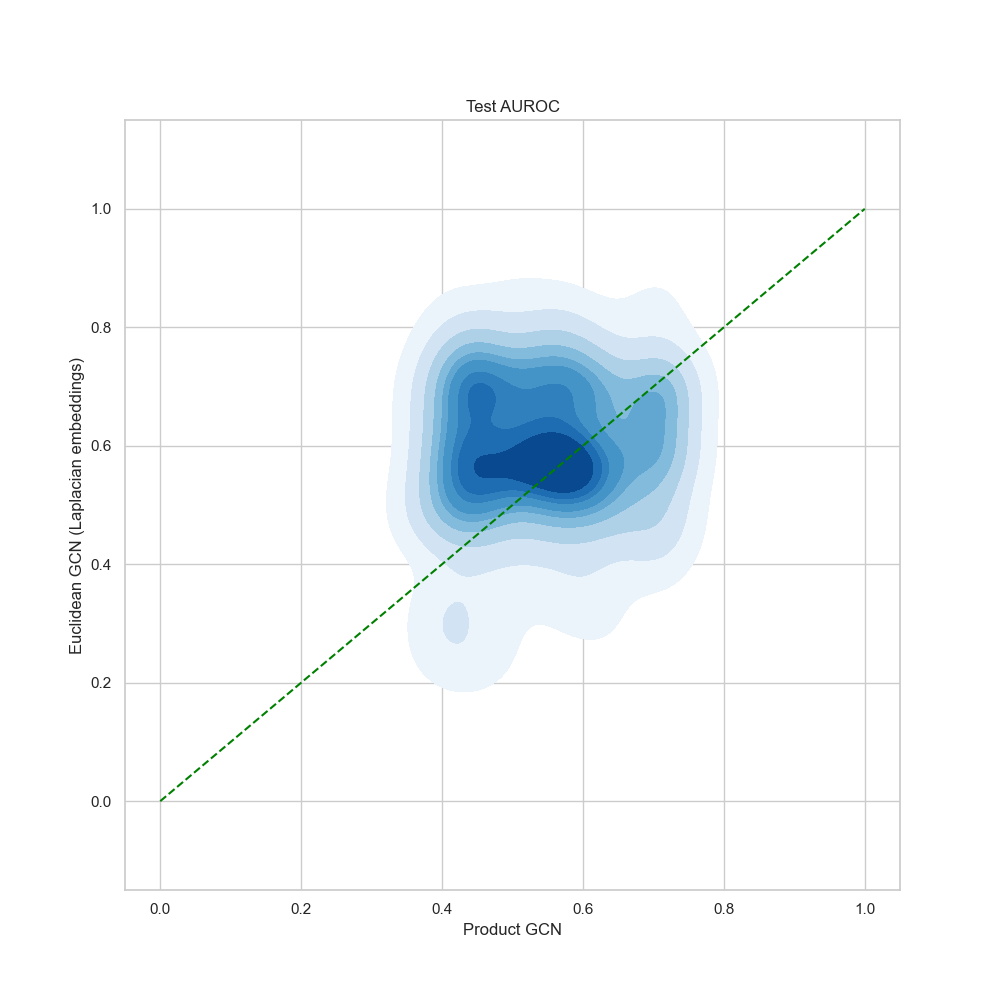}
\end{subfigure}
\caption[short]{Comparison of Euclidean GCN initialized with pretrained Laplacian embeddings and Product GCN performance on in-distribution 
validation set and out-of-distribution test set. Each density plot shows one of either AP or AUROC metrics taken across all graphs in the PathBank dataset.}
\end{figure}

\begin{figure}[ht]
\centering
\begin{subfigure}{.5\textwidth}
    \centering
    \includegraphics[width=0.8\textwidth]{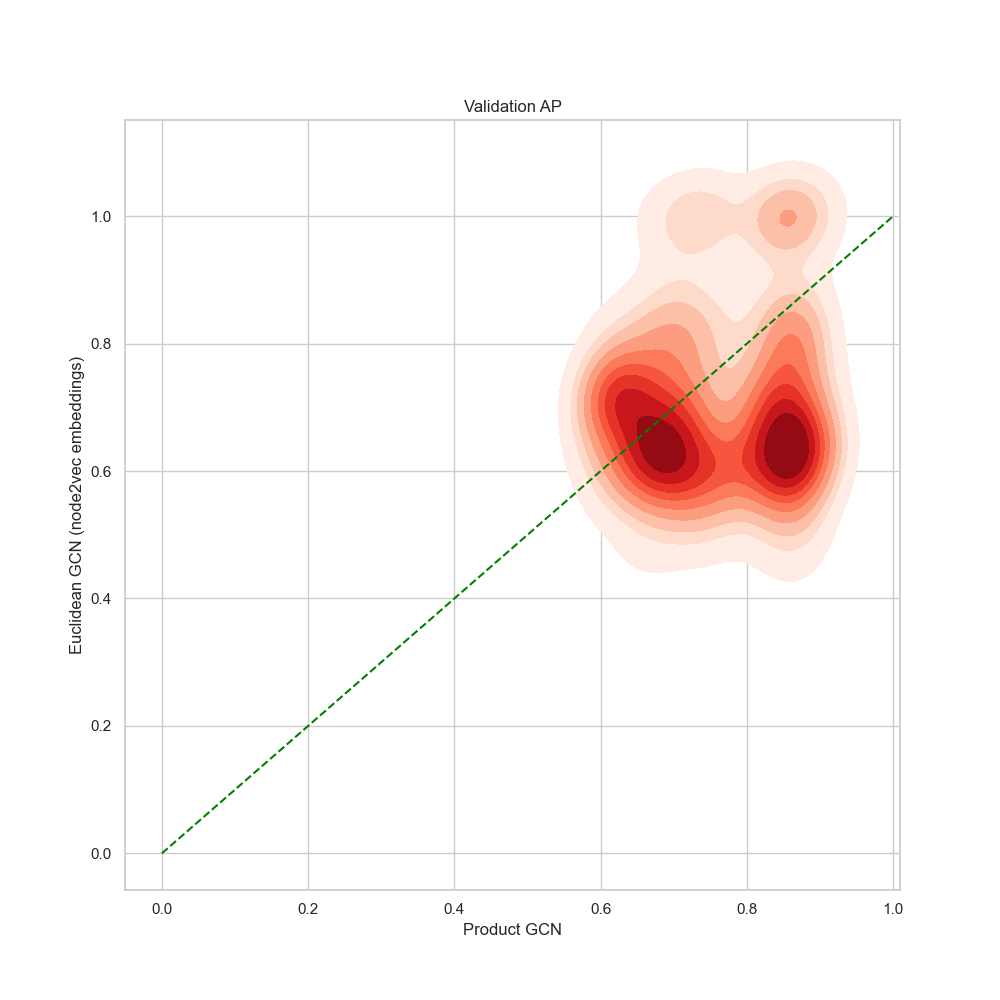}
\end{subfigure}%
\begin{subfigure}{.5\textwidth}
    \centering
    \includegraphics[width=0.8\textwidth]{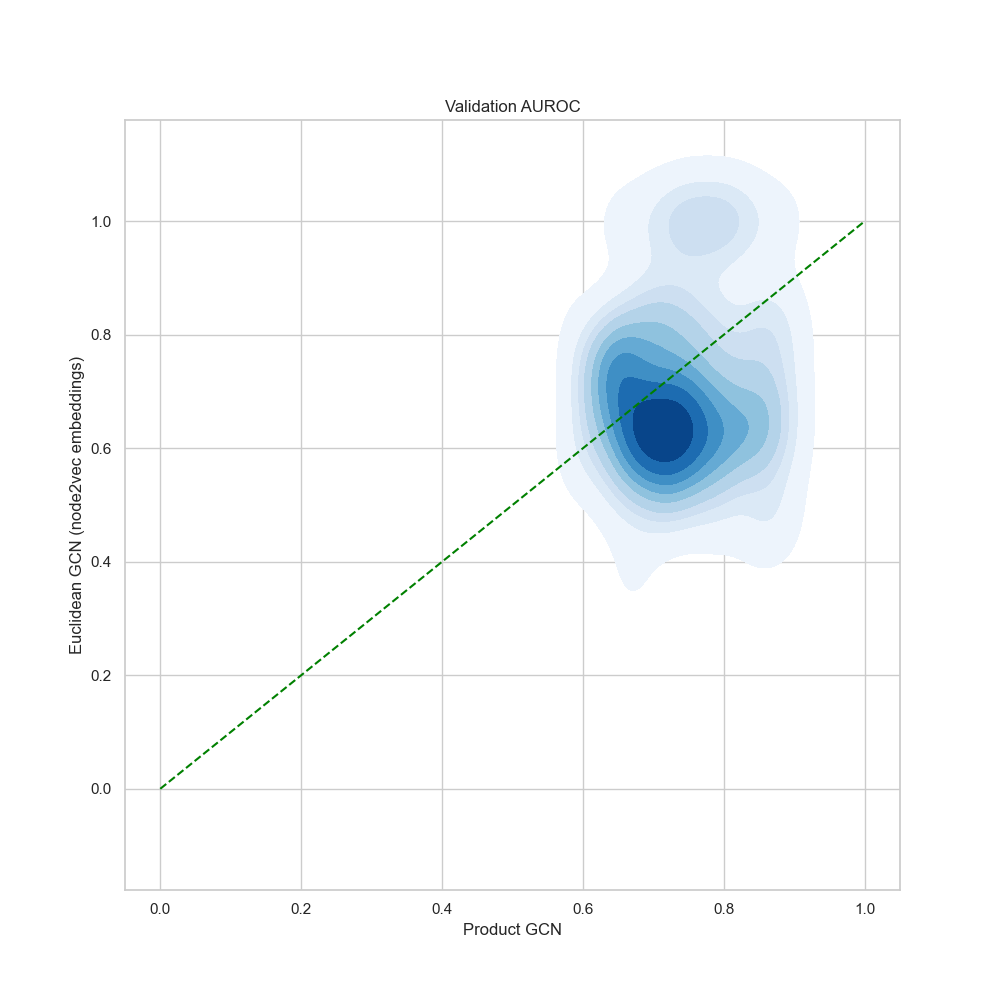}
\end{subfigure}
\begin{subfigure}{.5\textwidth}
    \centering
    \includegraphics[width=0.8\textwidth]{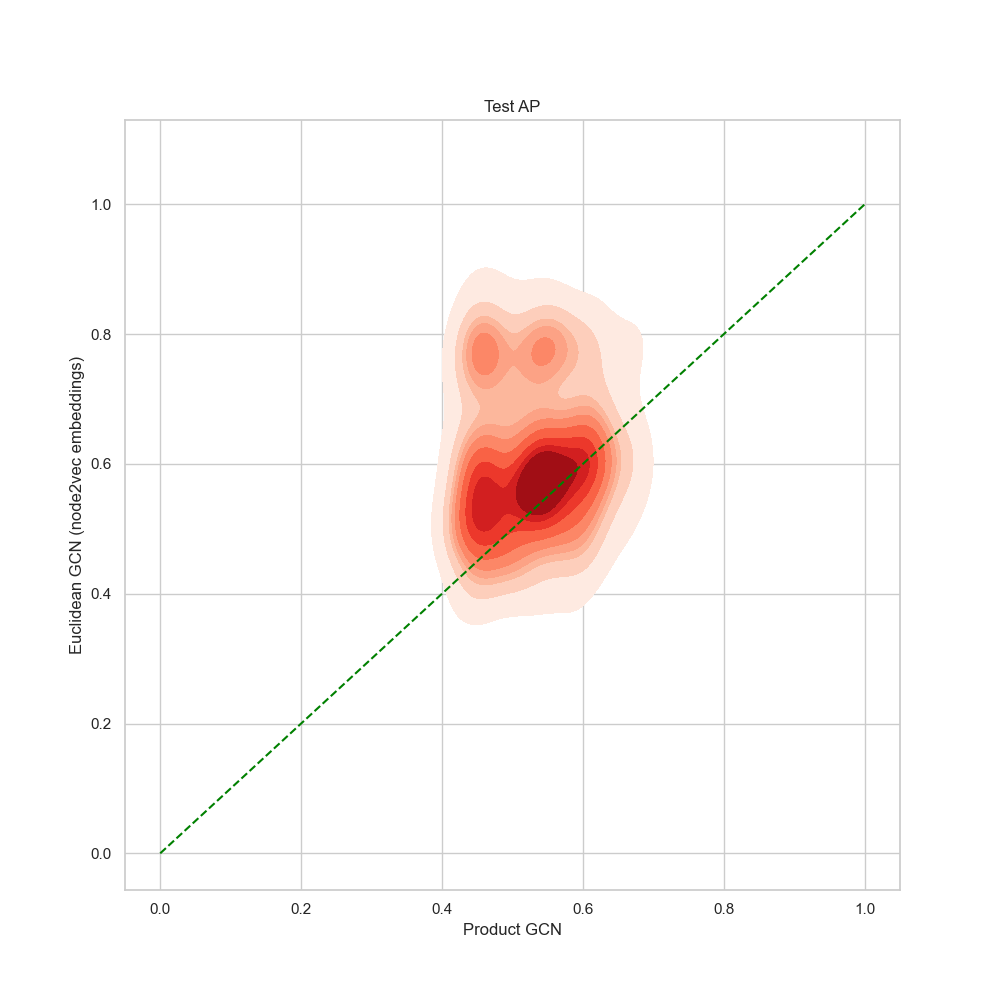}
\end{subfigure}%
\begin{subfigure}{.5\textwidth}
    \centering
    \includegraphics[width=0.8\textwidth]{images/pathbank_comparison_scatter_sweep_n2v_val_roc.png}
\end{subfigure}
\caption[short]{Comparison of Euclidean GCN initialized with pretrained node2vec embeddings and Product GCN performance on in-distribution 
validation set and out-of-distribution test set. Each density plot shows one of either AP or AUROC metrics taken across all graphs in the PathBank dataset.}
\end{figure}

\clearpage
% --------------------- Reactome ----------------------------
\subsection{Reactome}
\begin{figure}[ht]
\centering
\begin{subfigure}{.5\textwidth}
    \centering
    \includegraphics[width=0.8\textwidth]{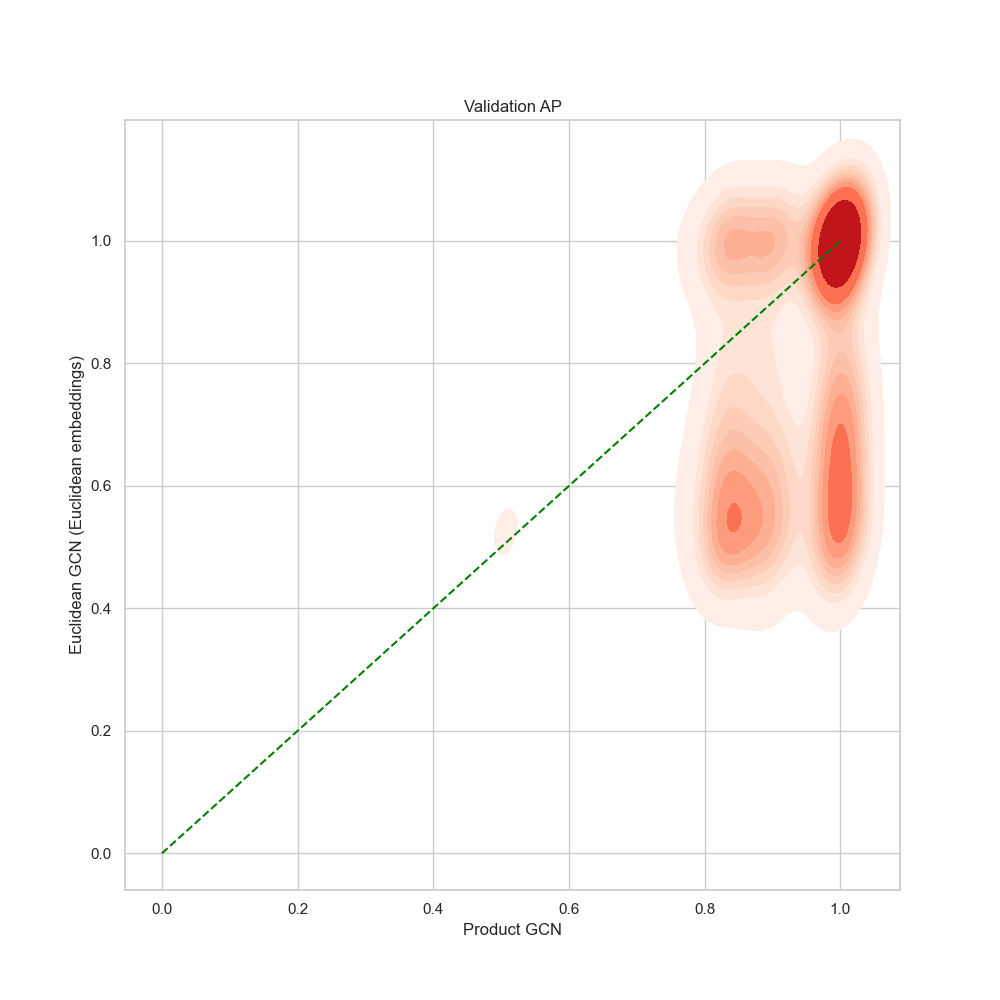}
\end{subfigure}%
\begin{subfigure}{.5\textwidth}
    \centering
    \includegraphics[width=0.8\textwidth]{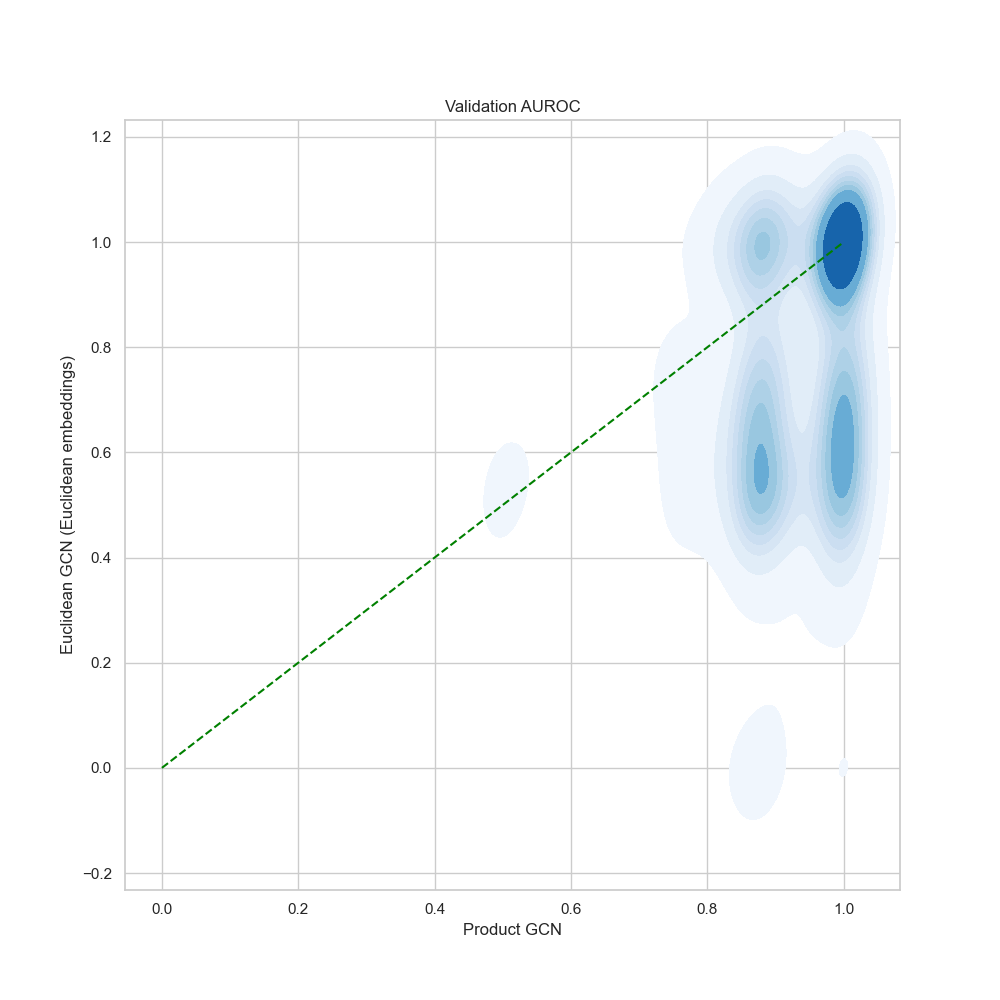}
\end{subfigure}
\begin{subfigure}{.5\textwidth}
    \centering
    \includegraphics[width=0.8\textwidth]{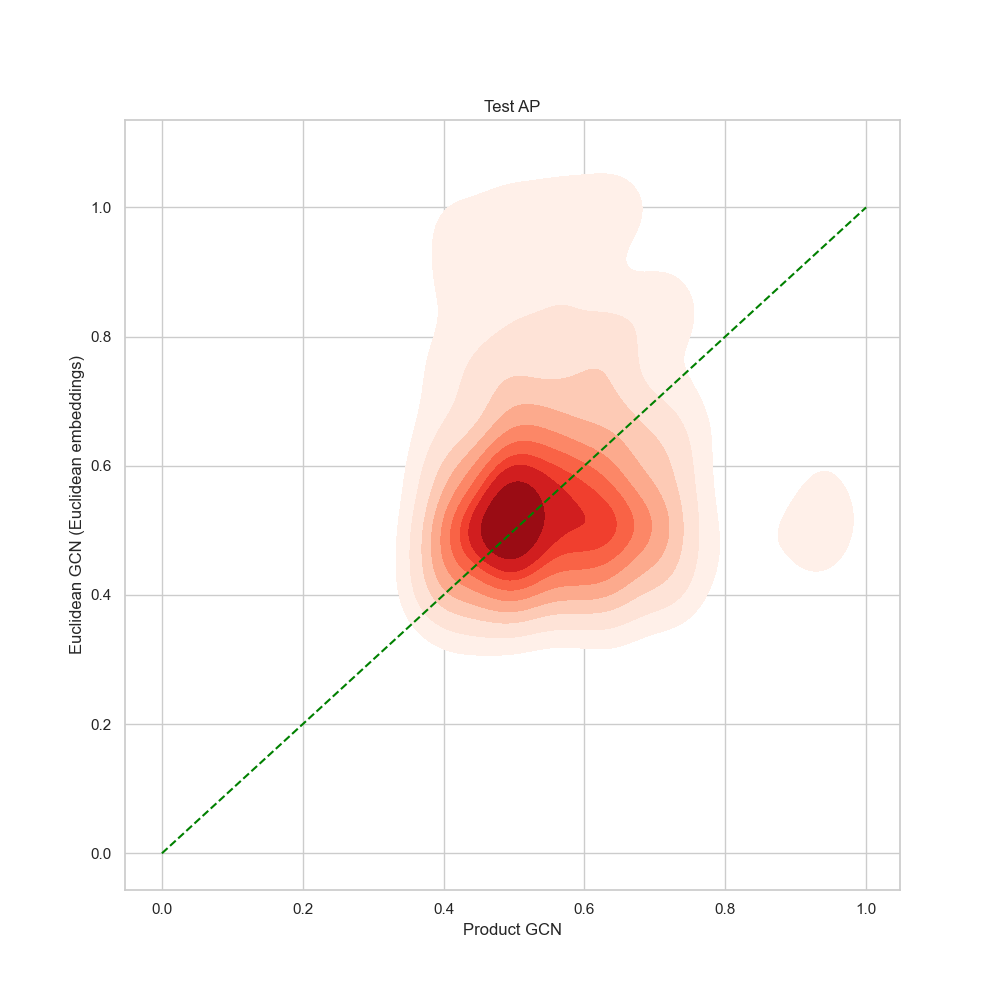}
\end{subfigure}%
\begin{subfigure}{.5\textwidth}
    \centering
    \includegraphics[width=0.8\textwidth]{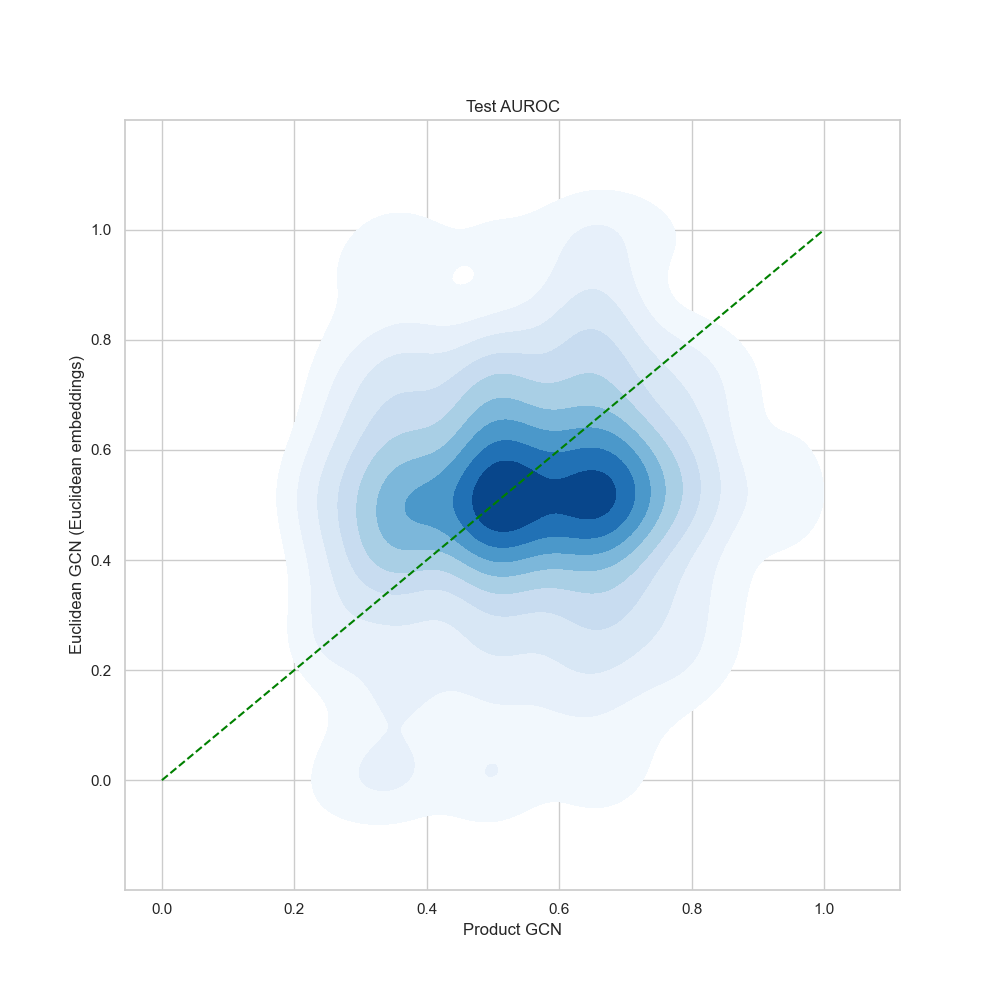}
\end{subfigure}

\caption[short]{Comparison of Euclidean GCN initialized with pretrained Euclidean embeddings and Product GCN performance on in-distribution 
validation set and out-of-distribution test set. Each density plot shows one of either AP or AUROC metrics taken across all graphs in the Reactome dataset.}
\end{figure}

\begin{figure}[ht]
\centering
\begin{subfigure}{.5\textwidth}
    \centering
    \includegraphics[width=0.8\textwidth]{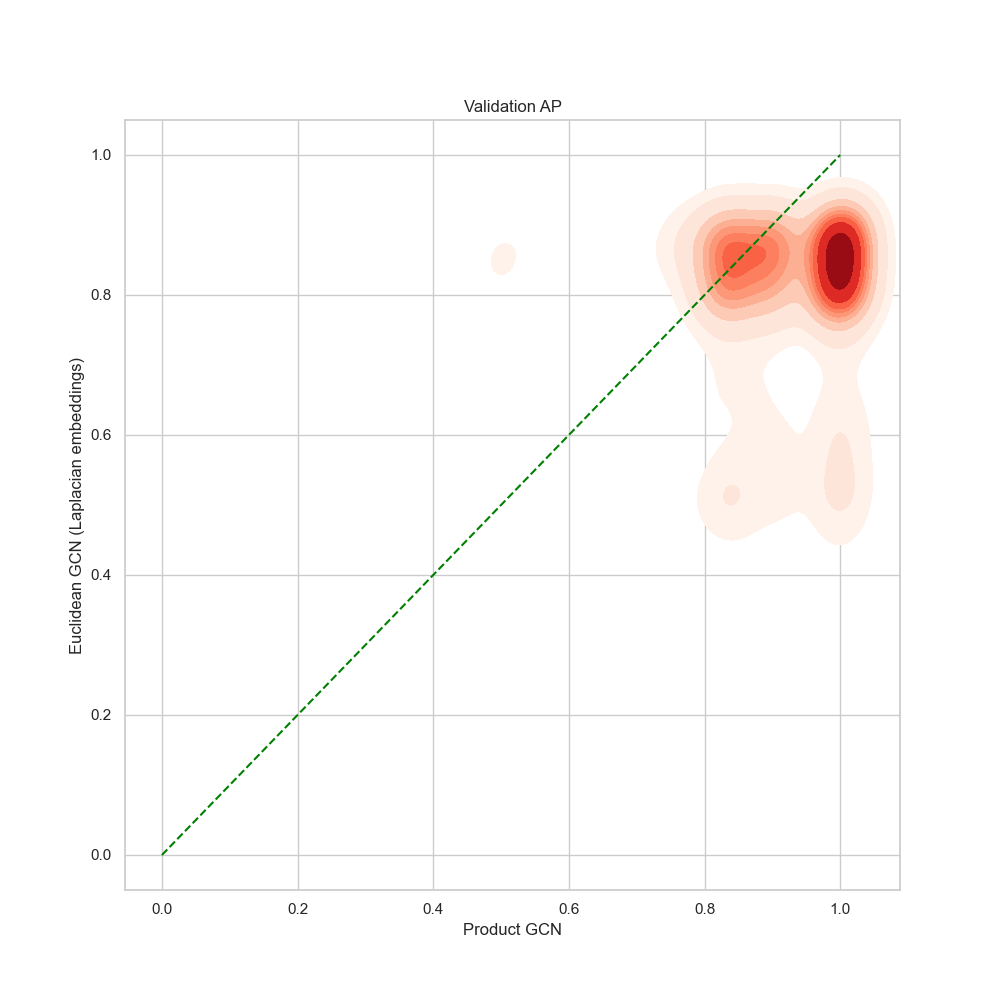}
\end{subfigure}%
\begin{subfigure}{.5\textwidth}
    \centering
    \includegraphics[width=0.8\textwidth]{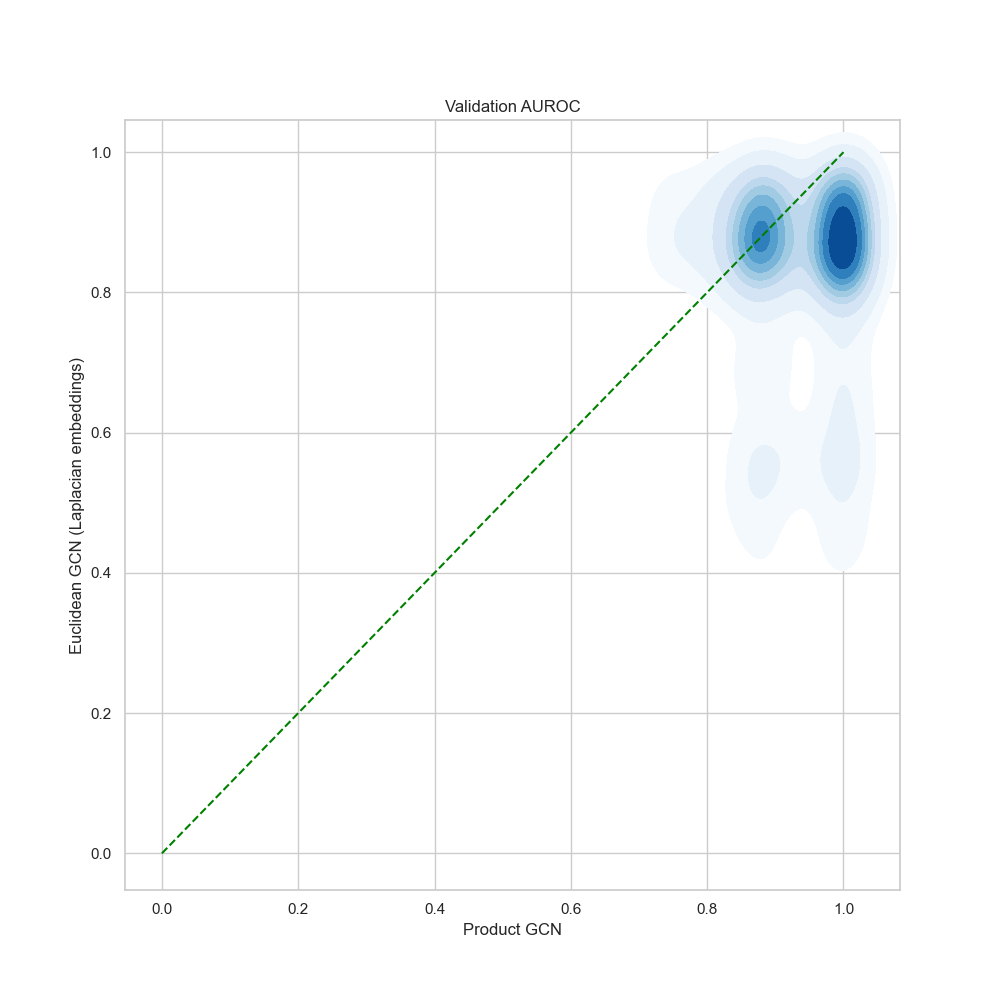}
\end{subfigure}
\begin{subfigure}{.5\textwidth}
    \centering
    \includegraphics[width=0.8\textwidth]{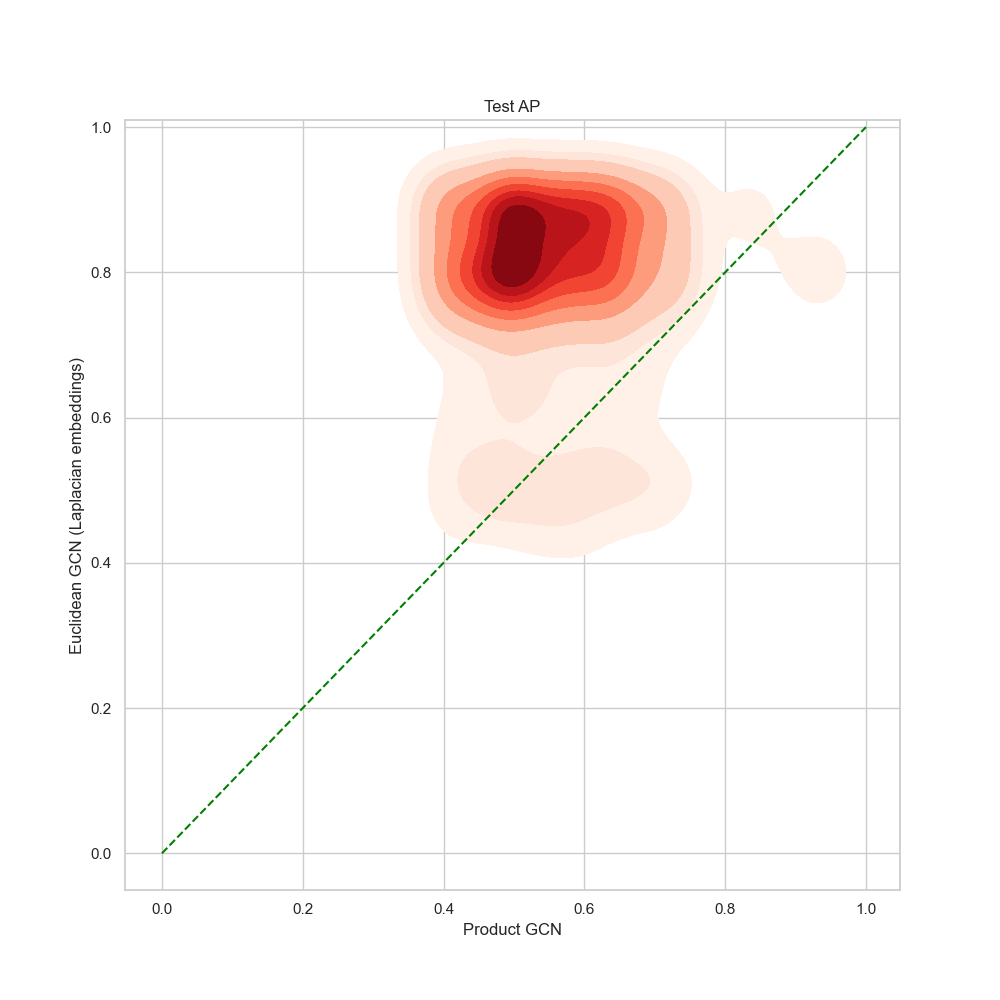}
\end{subfigure}%
\begin{subfigure}{.5\textwidth}
    \centering
    \includegraphics[width=0.8\textwidth]{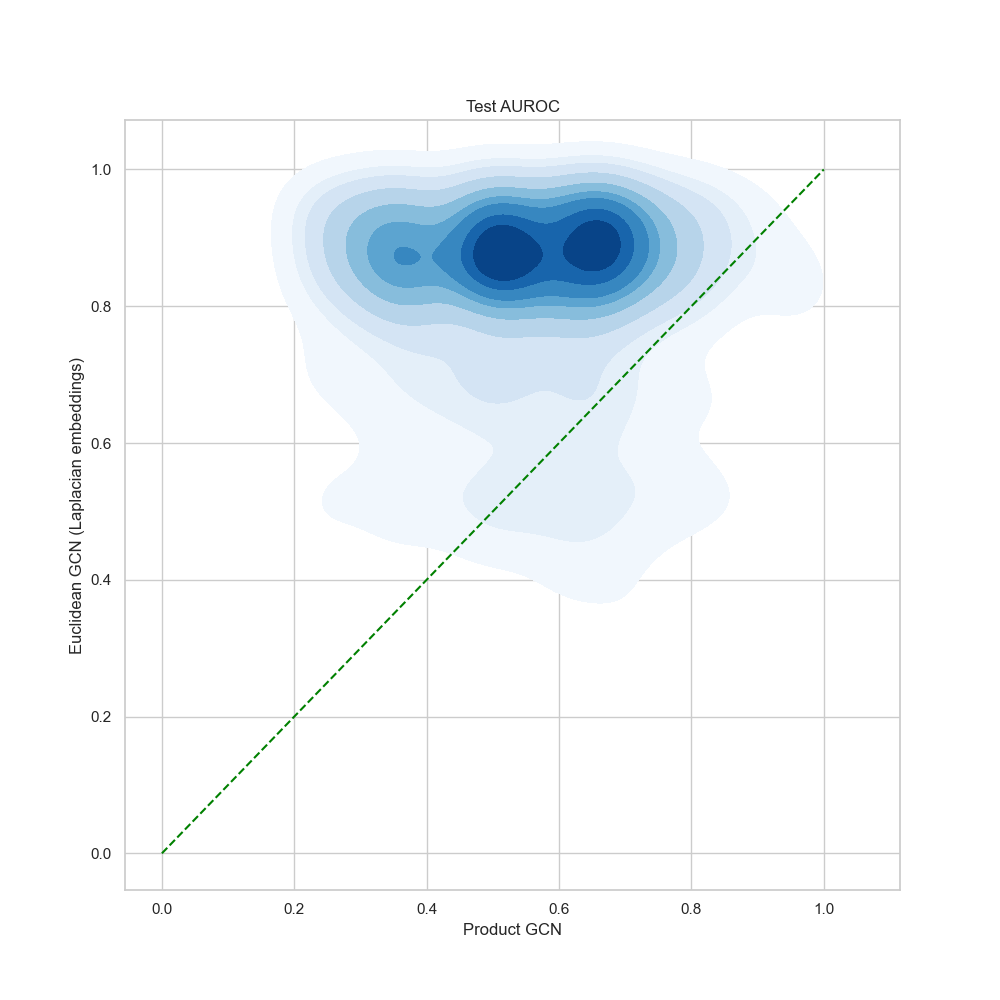}
\end{subfigure}
\caption[short]{Comparison of Euclidean GCN initialized with pretrained Laplacian embeddings and Product GCN performance on in-distribution 
validation set and out-of-distribution test set. Each density plot shows one of either AP or AUROC metrics taken across all graphs in the Reactome dataset.}
\end{figure}

\begin{figure}[ht]
\centering
\begin{subfigure}{.5\textwidth}
    \centering
    \includegraphics[width=0.8\textwidth]{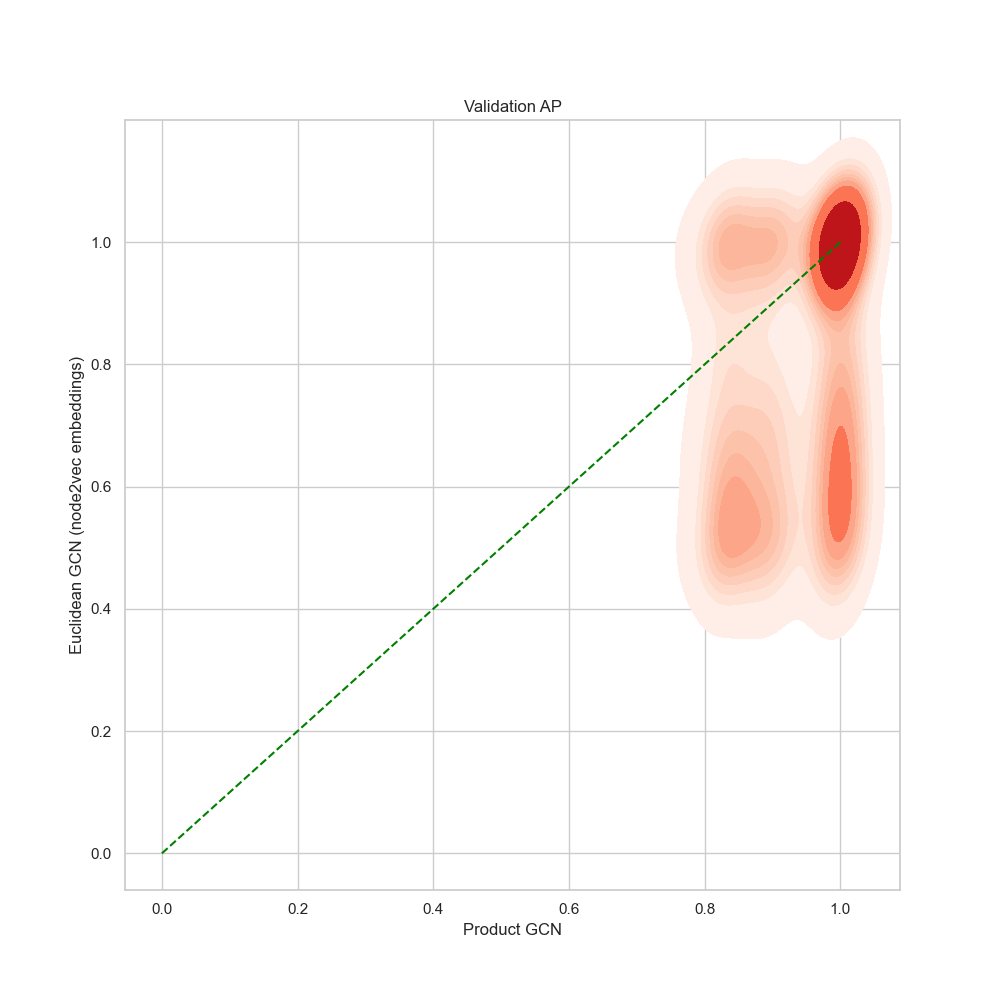}
\end{subfigure}%
\begin{subfigure}{.5\textwidth}
    \centering
    \includegraphics[width=0.8\textwidth]{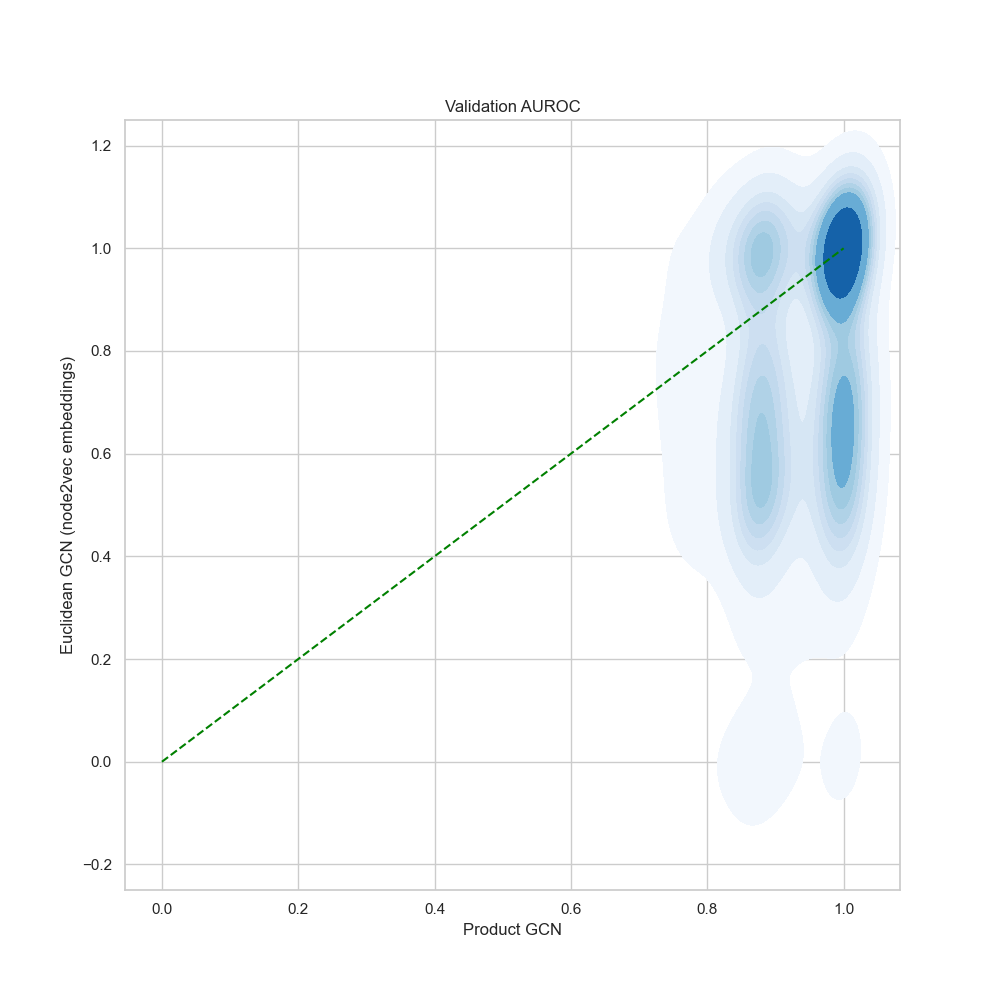}
\end{subfigure}
\begin{subfigure}{.5\textwidth}
    \centering
    \includegraphics[width=0.8\textwidth]{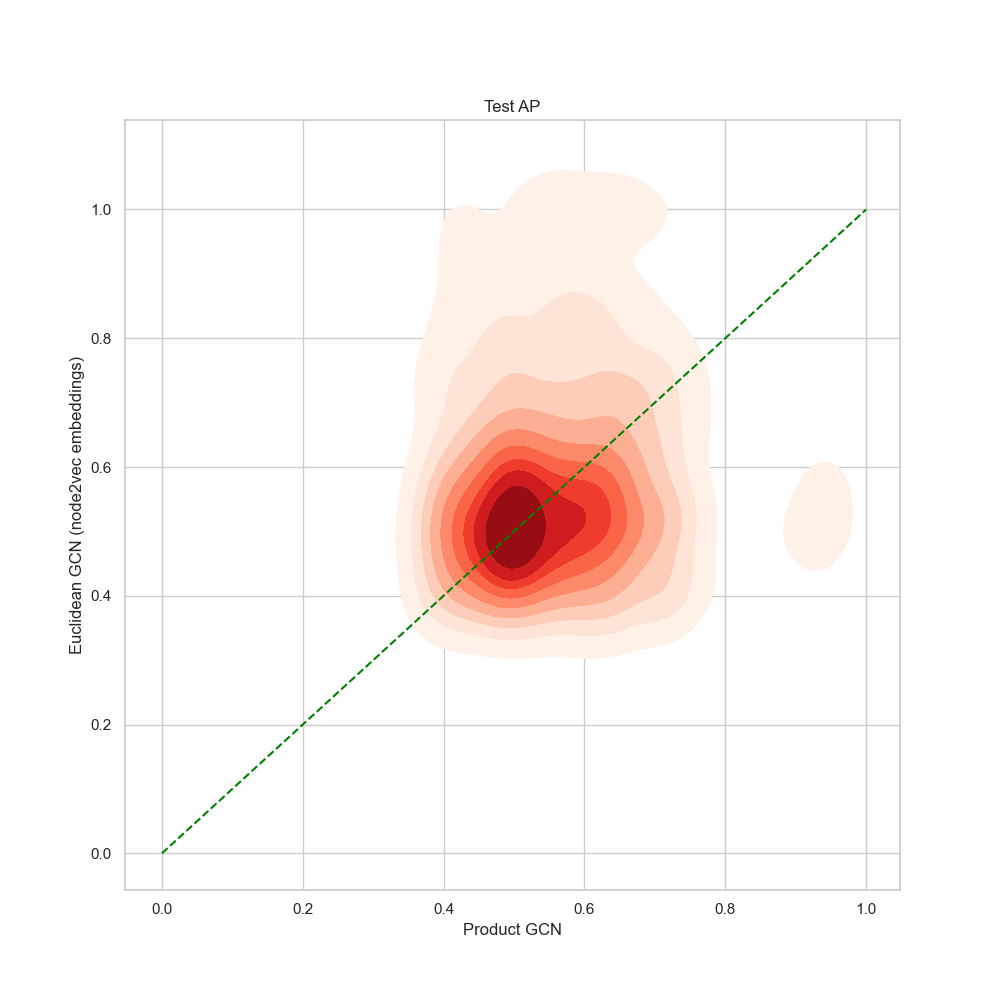}
\end{subfigure}%
\begin{subfigure}{.5\textwidth}
    \centering
    \includegraphics[width=0.8\textwidth]{images/reactome_comparison_scatter_sweep_n2v_val_roc.png}
\end{subfigure}
\caption[short]{Comparison of Euclidean GCN initialized with pretrained node2vec embeddings and Product GCN performance on in-distribution 
validation set and out-of-distribution test set. Each density plot shows one of either AP or AUROC metrics taken across all graphs in the Reactome dataset.}
\end{figure}

\clearpage
% --------------------- HumanCyc ----------------------------
\subsection{HumanCyc}
\begin{figure}[ht]
\centering
\begin{subfigure}{.5\textwidth}
    \centering
    \includegraphics[width=0.8\textwidth]{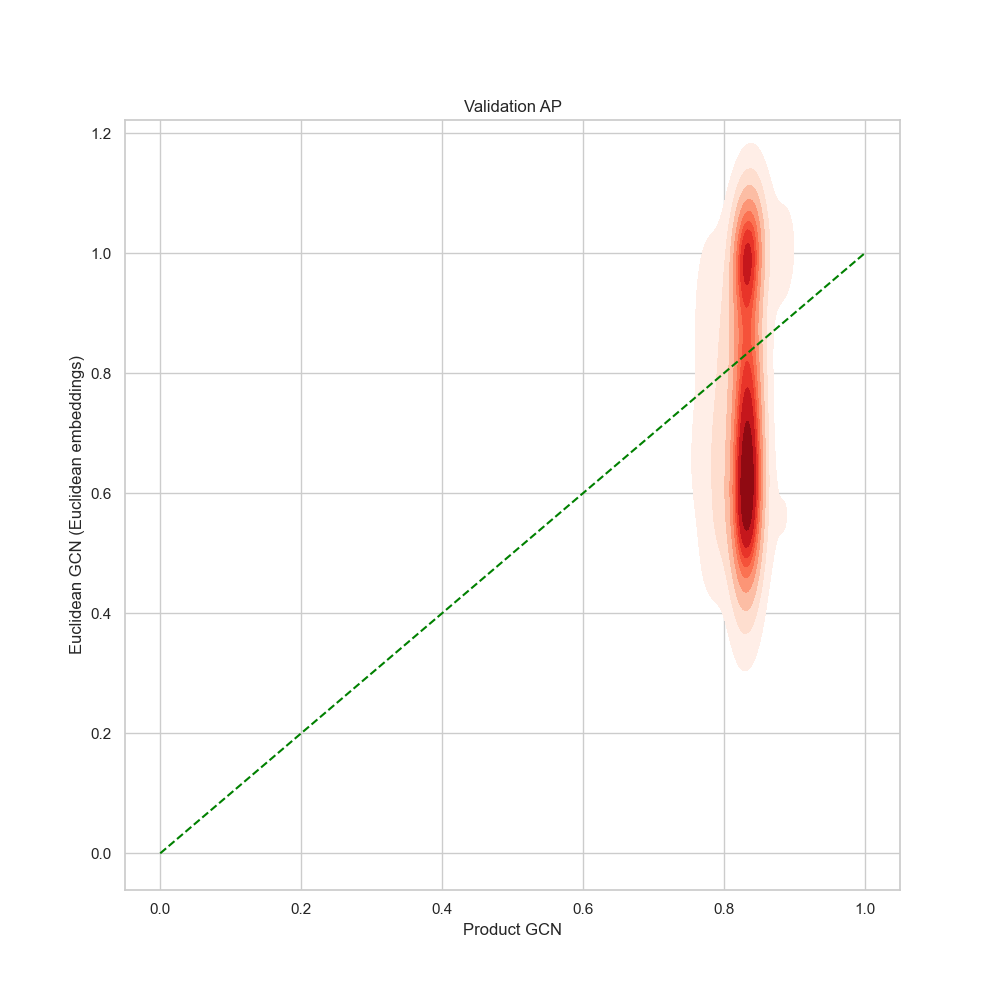}
\end{subfigure}%
\begin{subfigure}{.5\textwidth}
    \centering
    \includegraphics[width=0.8\textwidth]{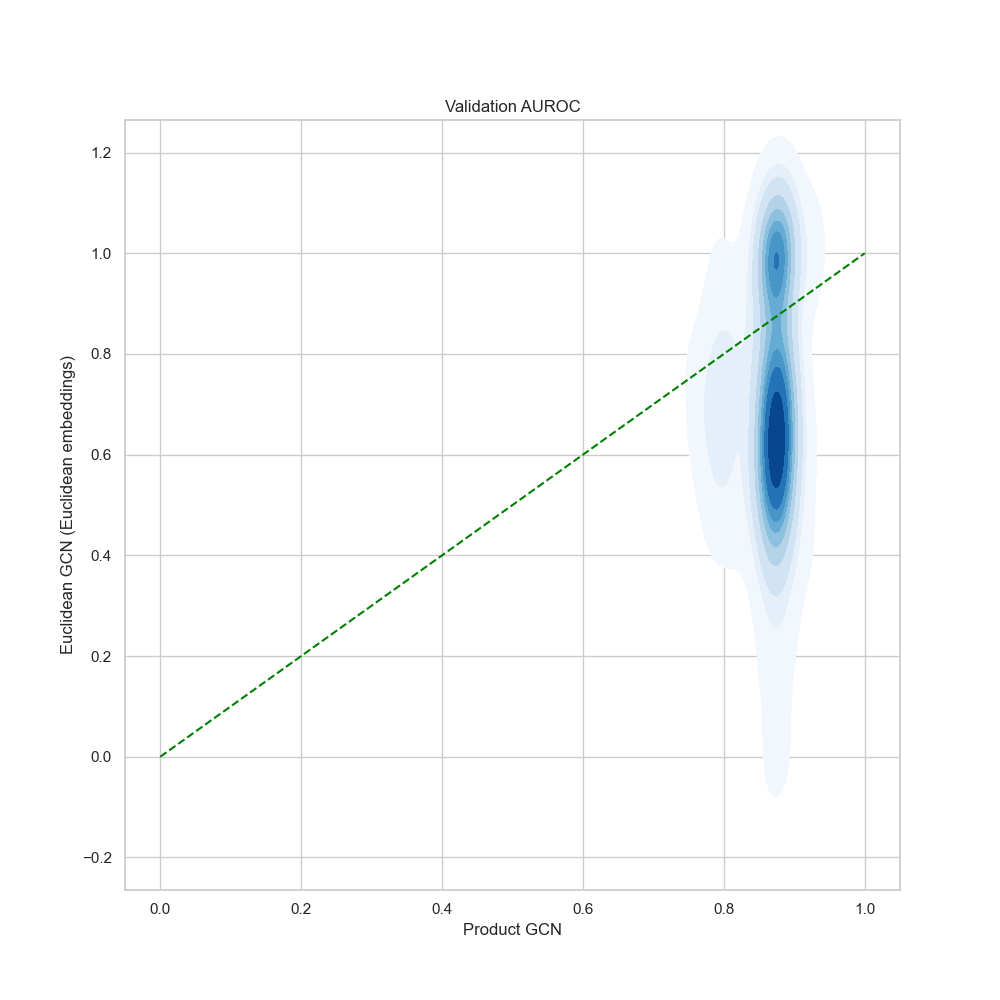}
\end{subfigure}
\begin{subfigure}{.5\textwidth}
    \centering
    \includegraphics[width=0.8\textwidth]{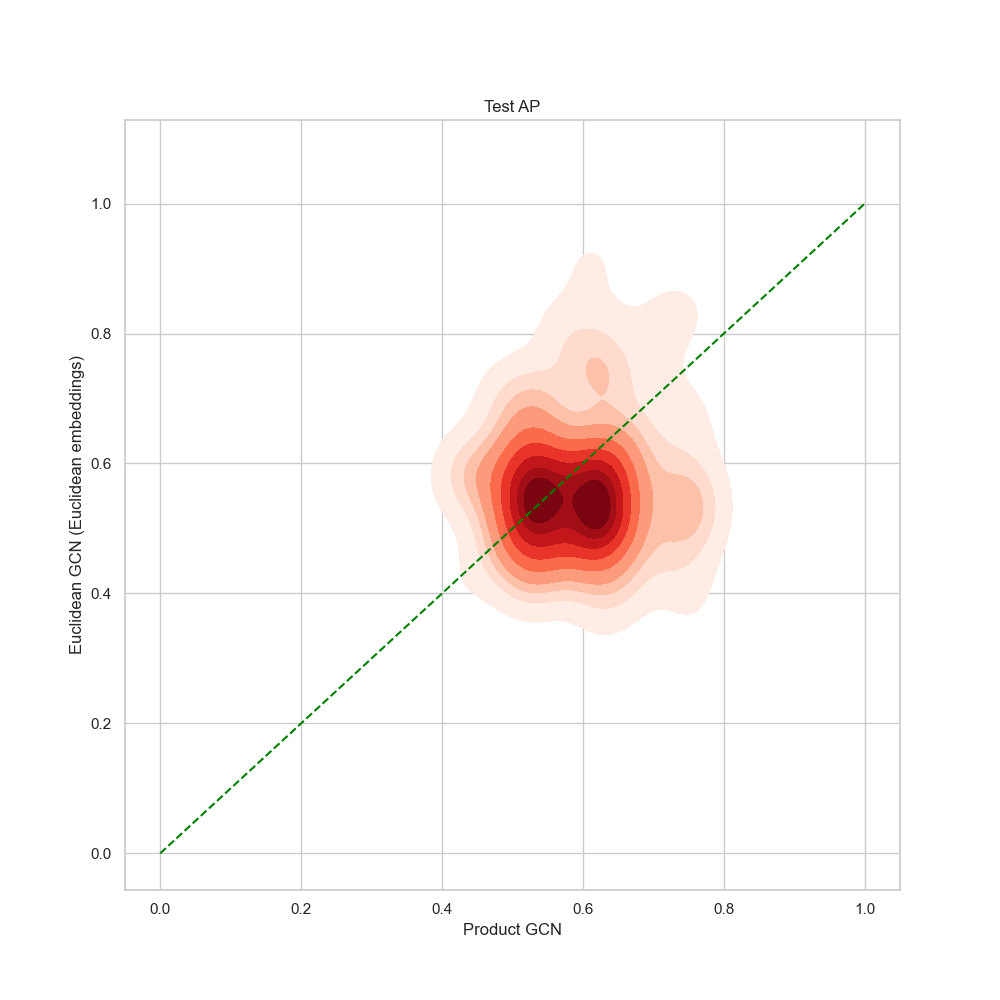}
\end{subfigure}%
\begin{subfigure}{.5\textwidth}
    \centering
    \includegraphics[width=0.8\textwidth]{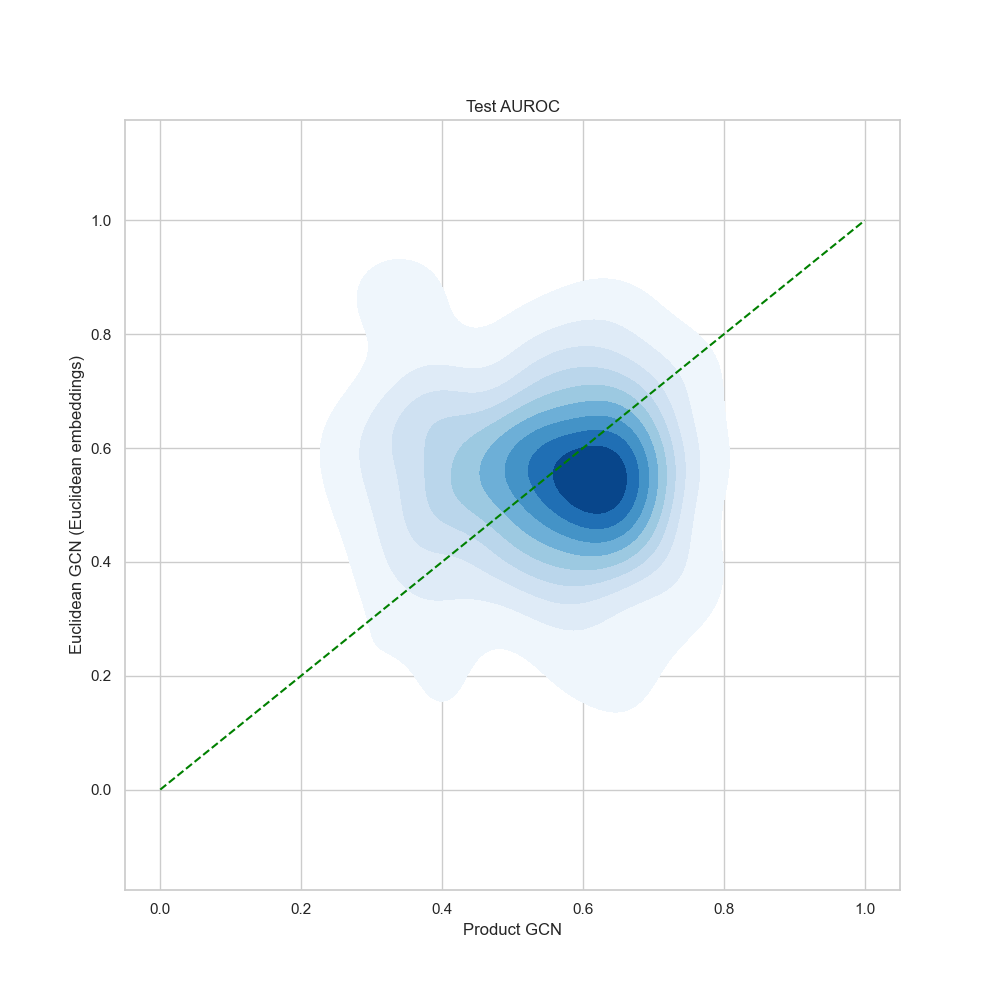}
\end{subfigure}

\caption[short]{Comparison of Euclidean GCN initialized with pretrained Euclidean embeddings and Product GCN performance on in-distribution 
validation set and out-of-distribution test set. Each density plot shows one of either AP or AUROC metrics taken across all graphs in the HumanCyc dataset.}
\end{figure}

\begin{figure}[ht]
\centering
\begin{subfigure}{.5\textwidth}
    \centering
    \includegraphics[width=0.8\textwidth]{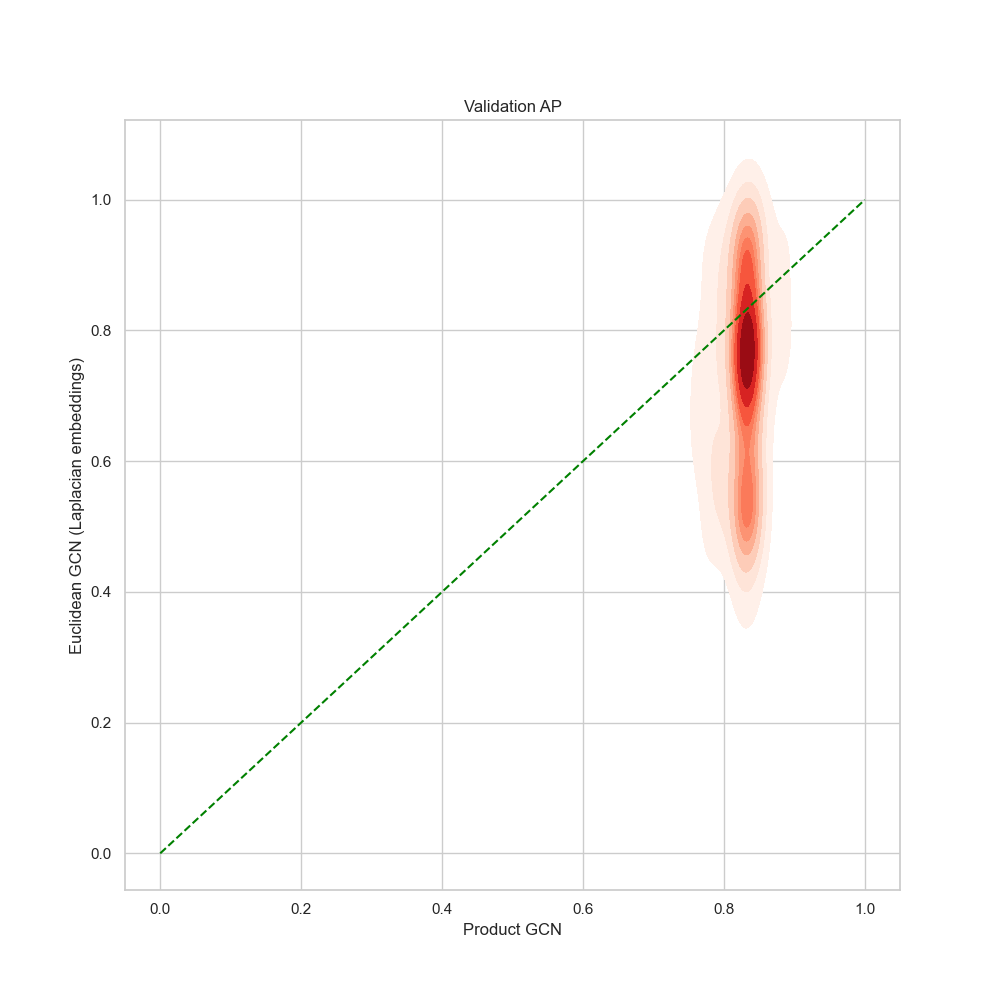}
\end{subfigure}%
\begin{subfigure}{.5\textwidth}
    \centering
    \includegraphics[width=0.8\textwidth]{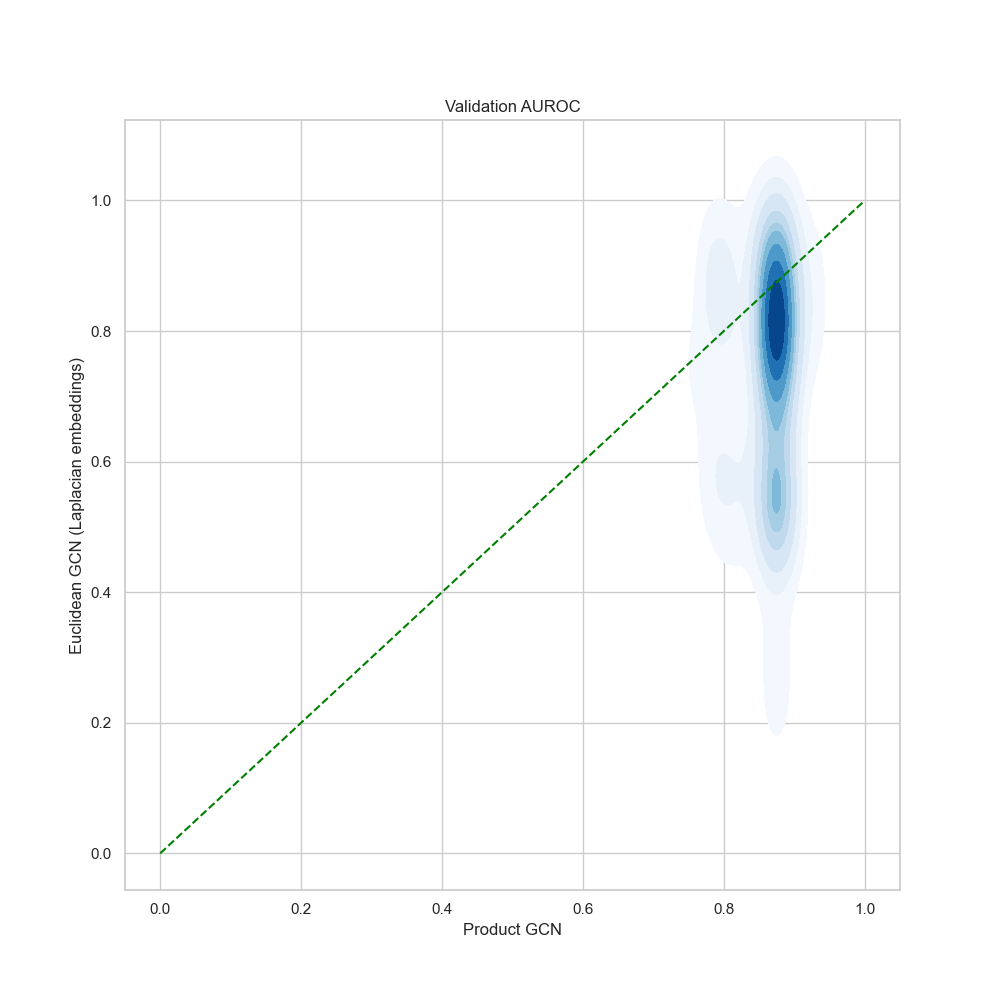}
\end{subfigure}
\begin{subfigure}{.5\textwidth}
    \centering
    \includegraphics[width=0.8\textwidth]{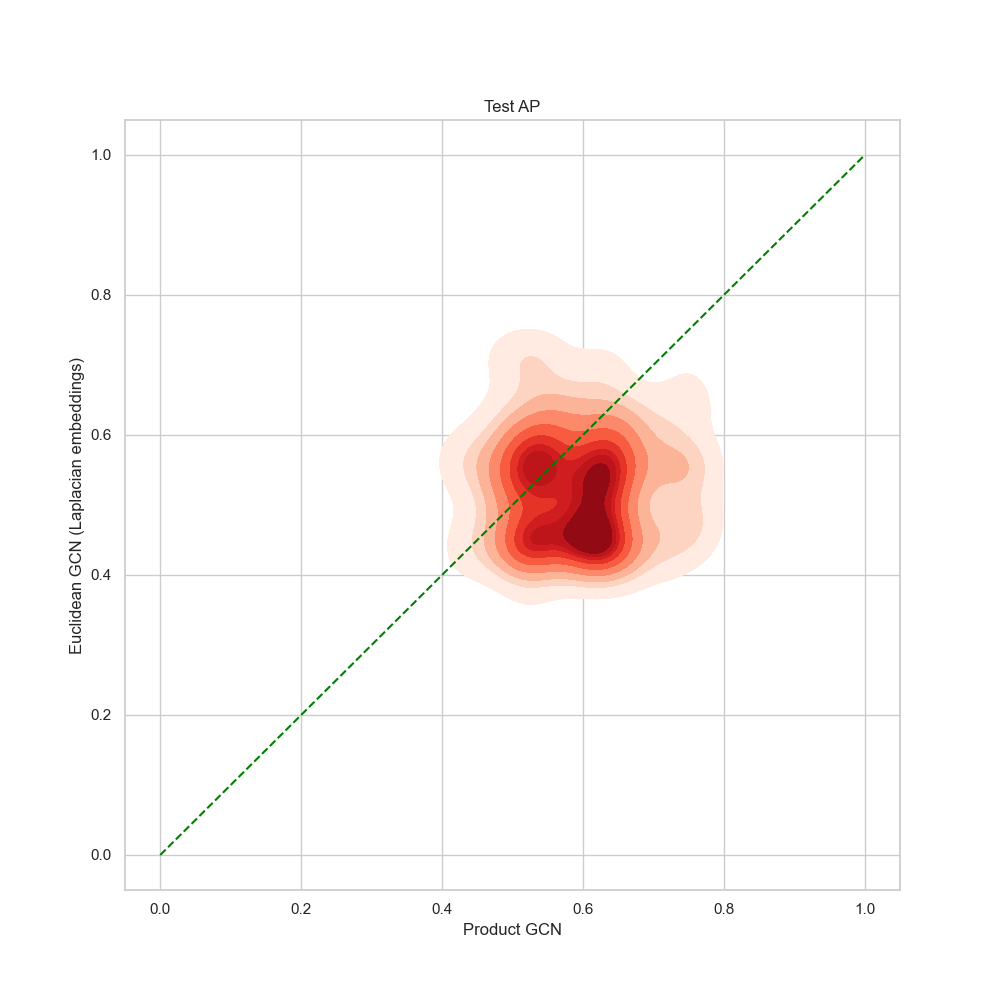}
\end{subfigure}%
\begin{subfigure}{.5\textwidth}
    \centering
    \includegraphics[width=0.8\textwidth]{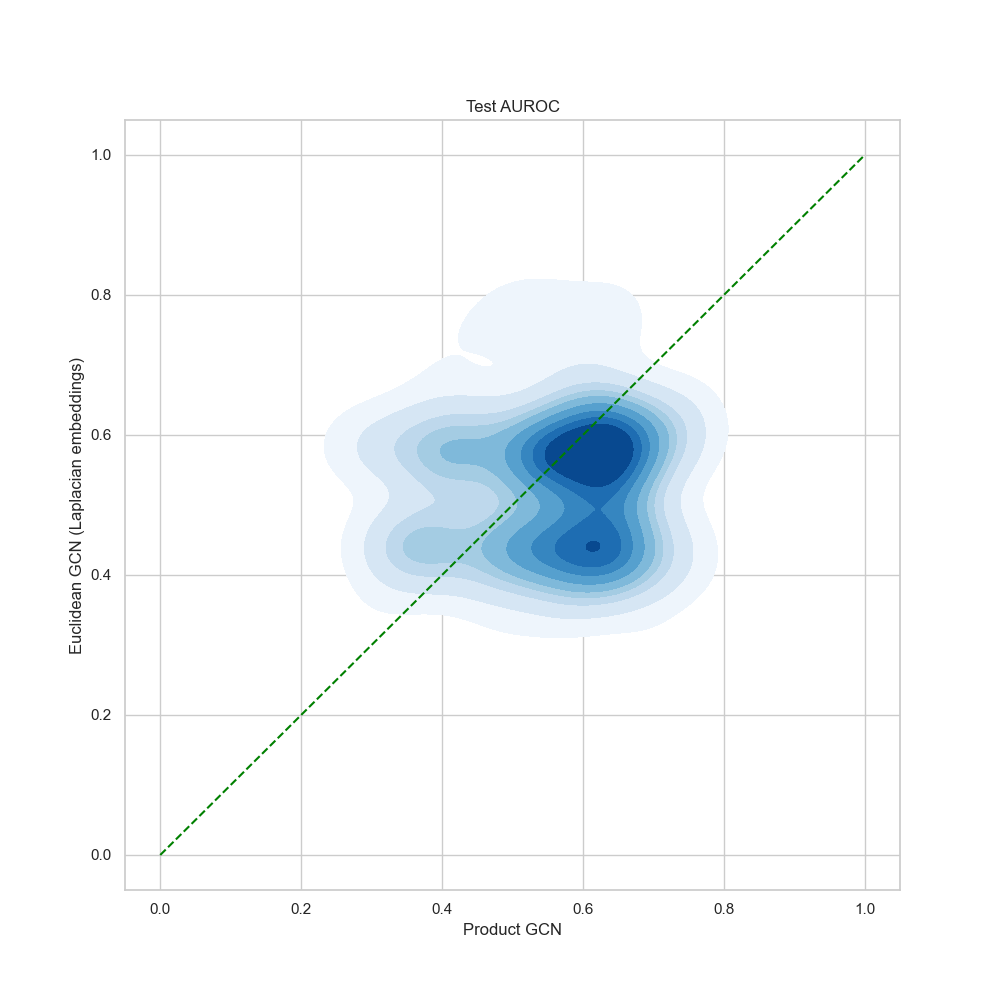}
\end{subfigure}
\caption[short]{Comparison of Euclidean GCN initialized with pretrained Laplacian embeddings and Product GCN performance on in-distribution 
validation set and out-of-distribution test set. Each density plot shows one of either AP or AUROC metrics taken across all graphs in the HumanCyc dataset.}
\end{figure}

\begin{figure}[ht]
\centering
\begin{subfigure}{.5\textwidth}
    \centering
    \includegraphics[width=0.8\textwidth]{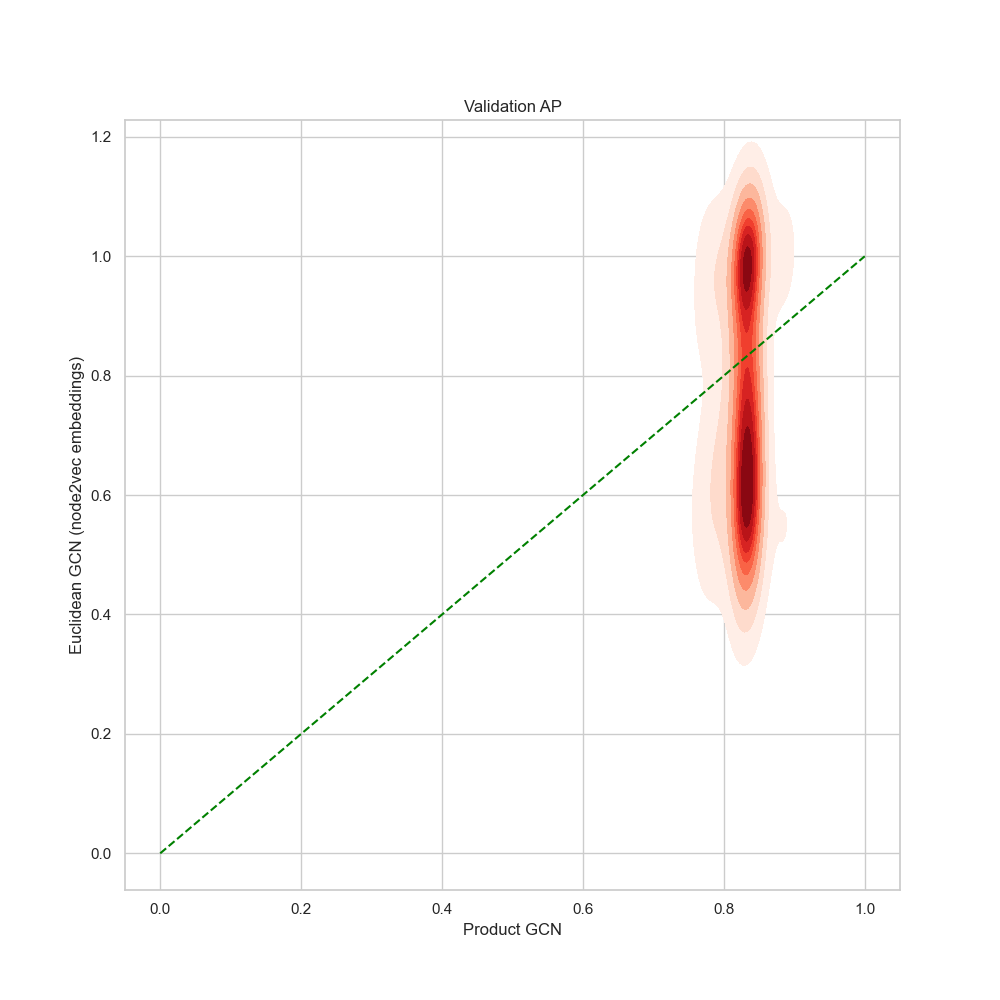}
\end{subfigure}%
\begin{subfigure}{.5\textwidth}
    \centering
    \includegraphics[width=0.8\textwidth]{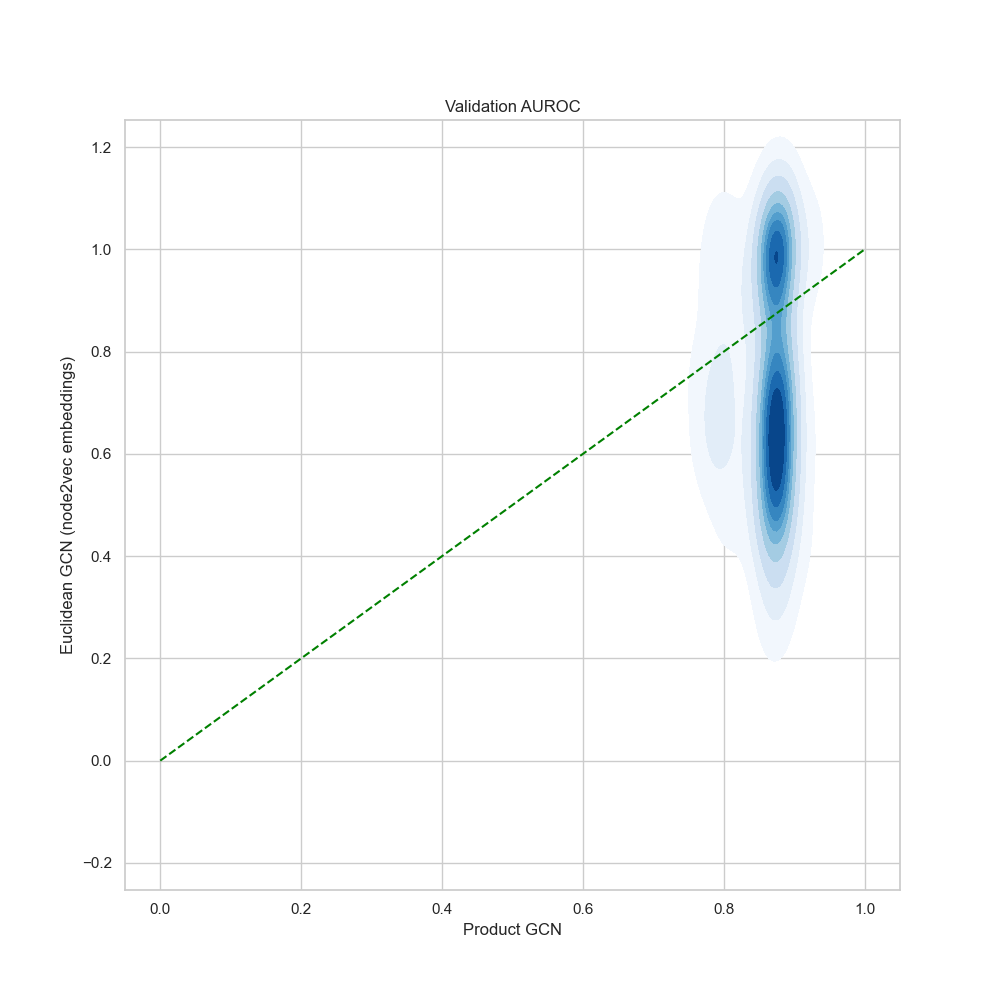}
\end{subfigure}
\begin{subfigure}{.5\textwidth}
    \centering
    \includegraphics[width=0.8\textwidth]{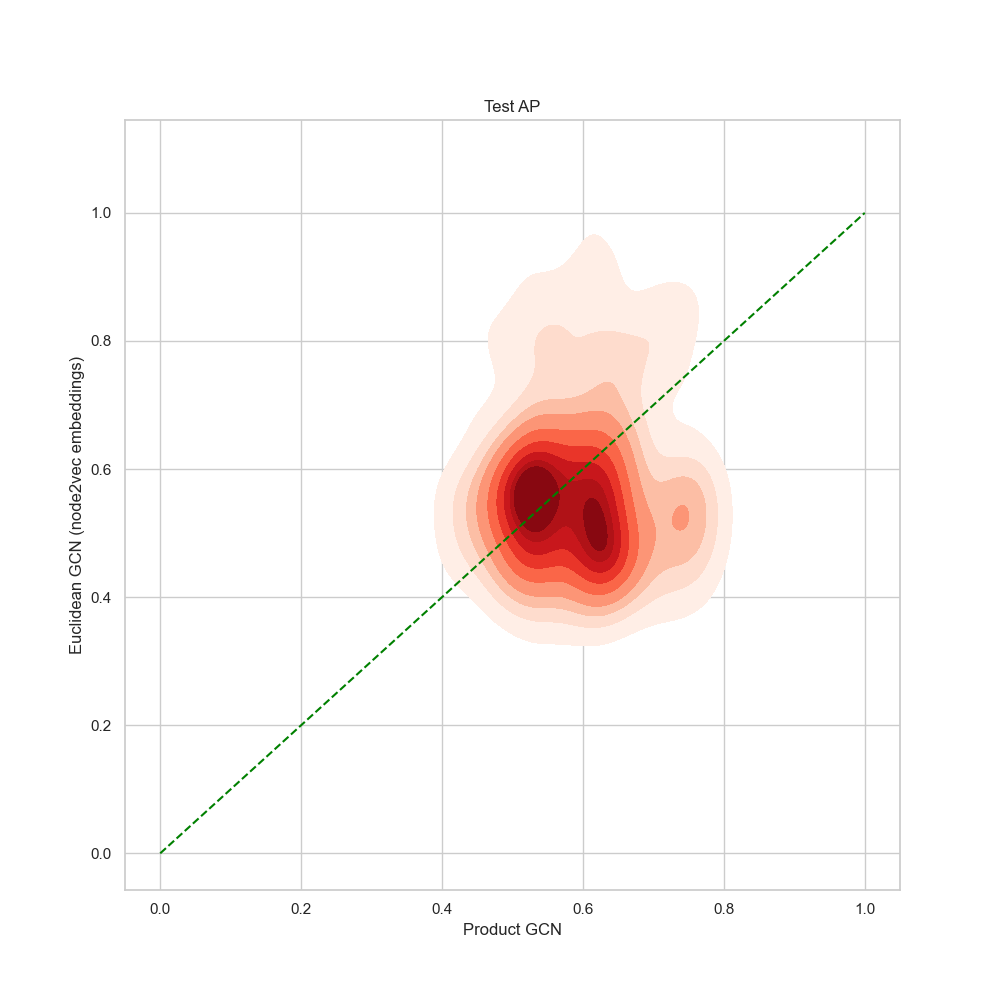}
\end{subfigure}%
\begin{subfigure}{.5\textwidth}
    \centering
    \includegraphics[width=0.8\textwidth]{images/humancyc_comparison_scatter_sweep_n2v_val_roc.png}
\end{subfigure}
\caption[short]{Comparison of Euclidean GCN initialized with pretrained node2vec embeddings and Product GCN performance on in-distribution 
validation set and out-of-distribution test set. Each density plot shows one of either AP or AUROC metrics taken across all graphs in the HumanCyc dataset.}
\end{figure}

\clearpage
% --------------------- NCI ----------------------------
\subsection{NCI}
\begin{figure}[ht]
\centering
\begin{subfigure}{.5\textwidth}
    \centering
    \includegraphics[width=0.8\textwidth]{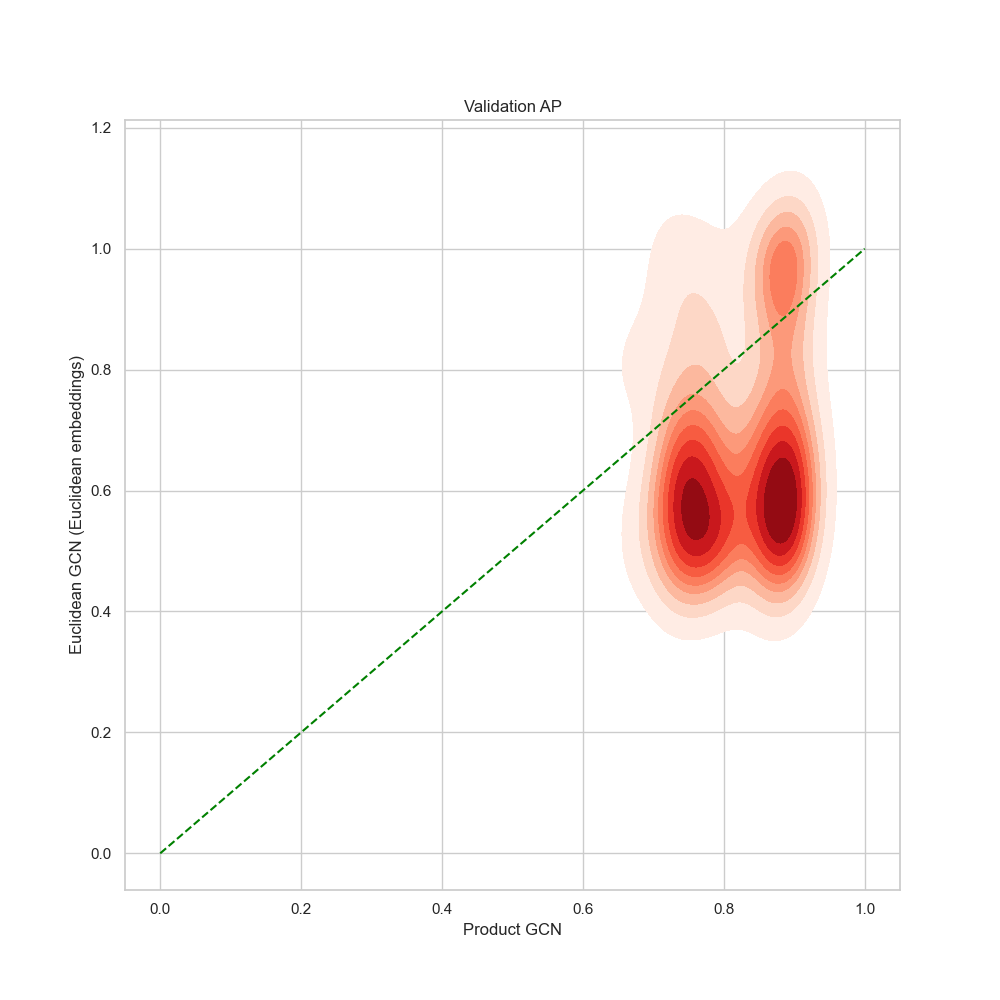}
\end{subfigure}%
\begin{subfigure}{.5\textwidth}
    \centering
    \includegraphics[width=0.8\textwidth]{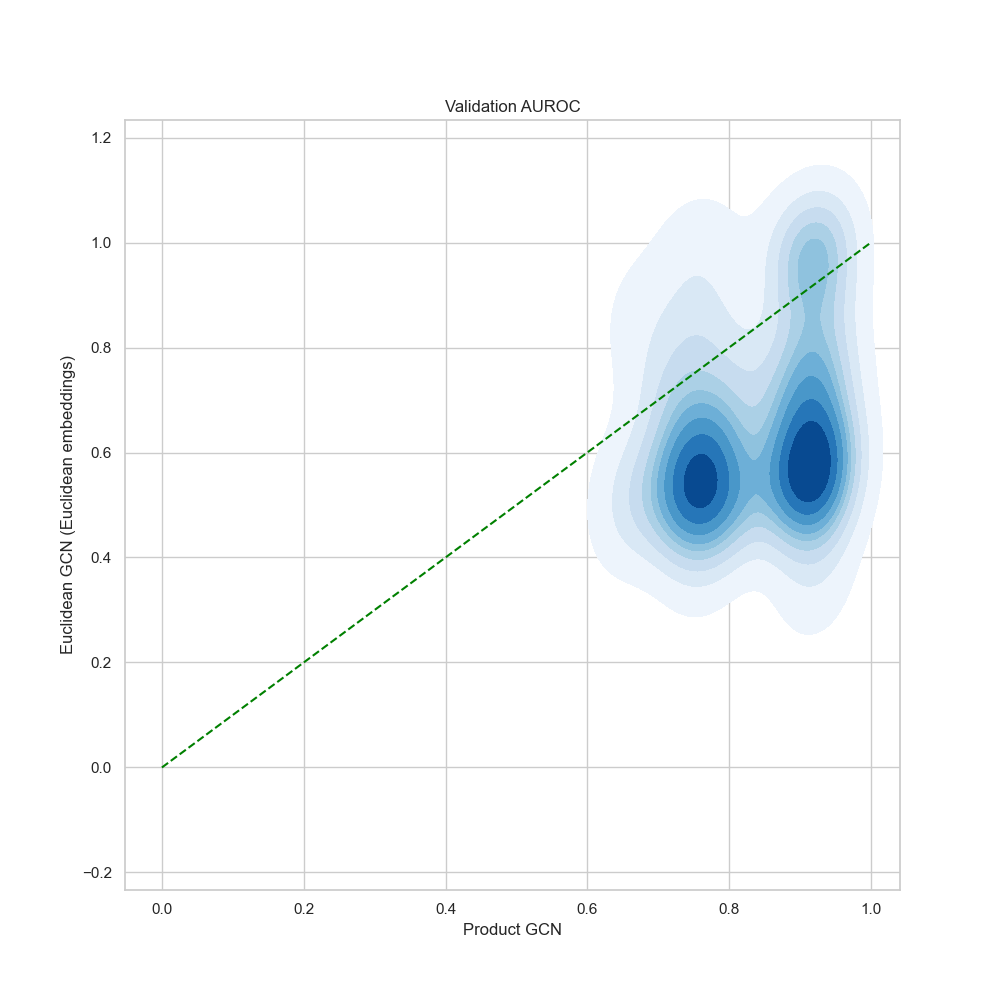}
\end{subfigure}
\begin{subfigure}{.5\textwidth}
    \centering
    \includegraphics[width=0.8\textwidth]{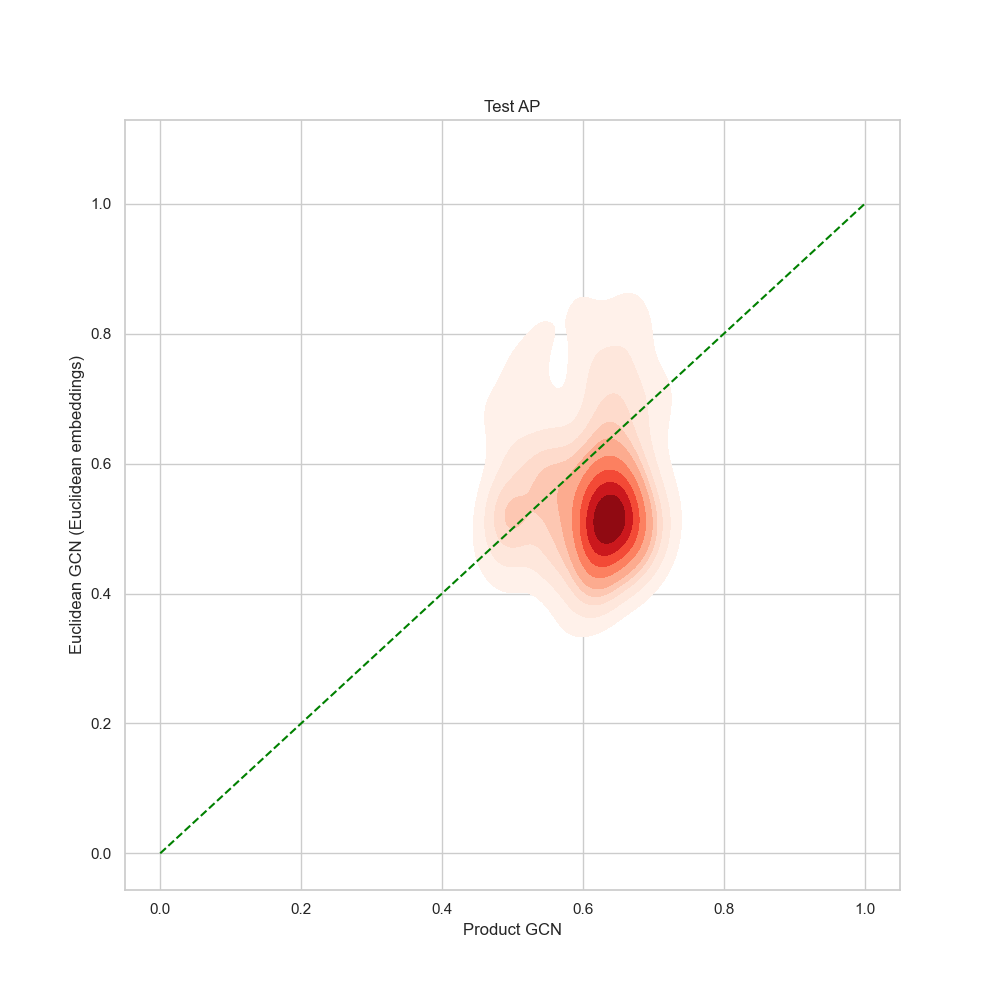}
\end{subfigure}%
\begin{subfigure}{.5\textwidth}
    \centering
    \includegraphics[width=0.8\textwidth]{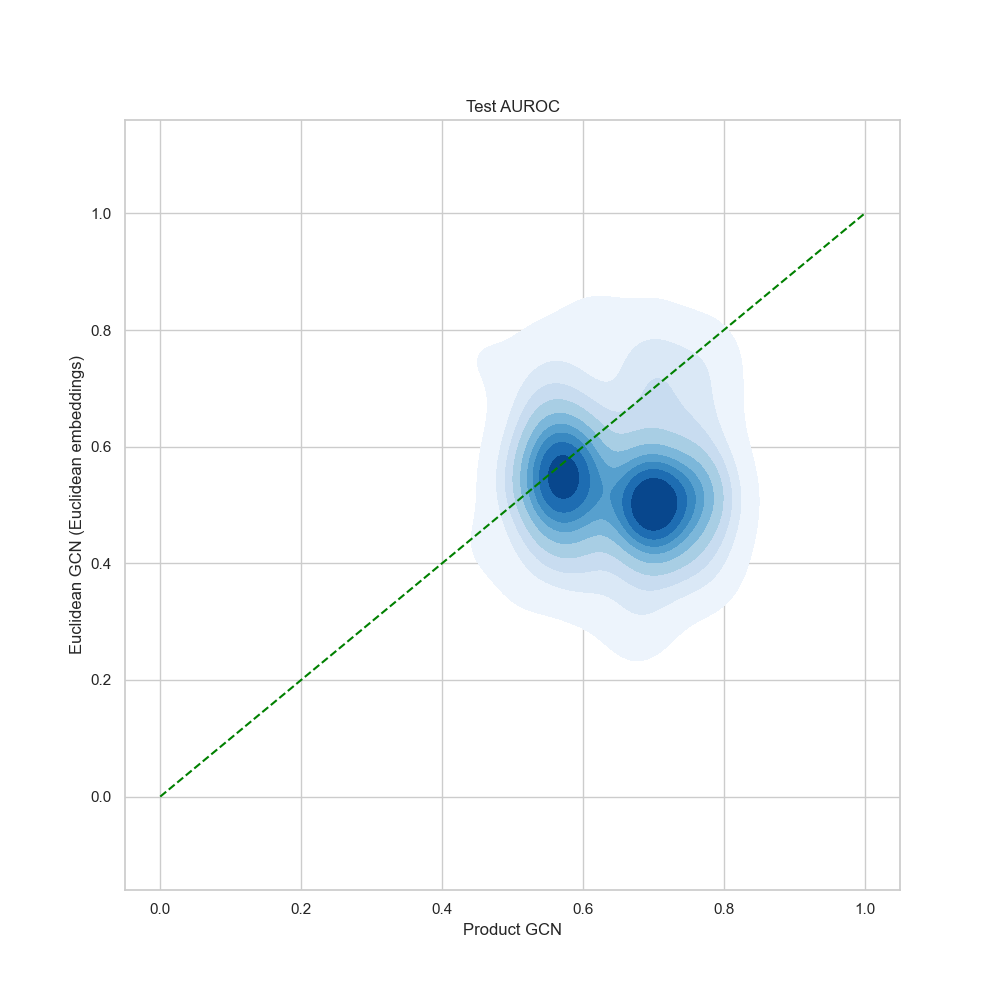}
\end{subfigure}

\caption[short]{Comparison of Euclidean GCN initialized with pretrained Euclidean embeddings and Product GCN performance on in-distribution 
validation set and out-of-distribution test set. Each density plot shows one of either AP or AUROC metrics taken across all graphs in the NCI dataset.}
\end{figure}

\begin{figure}[ht]
\centering
\begin{subfigure}{.5\textwidth}
    \centering
    \includegraphics[width=0.8\textwidth]{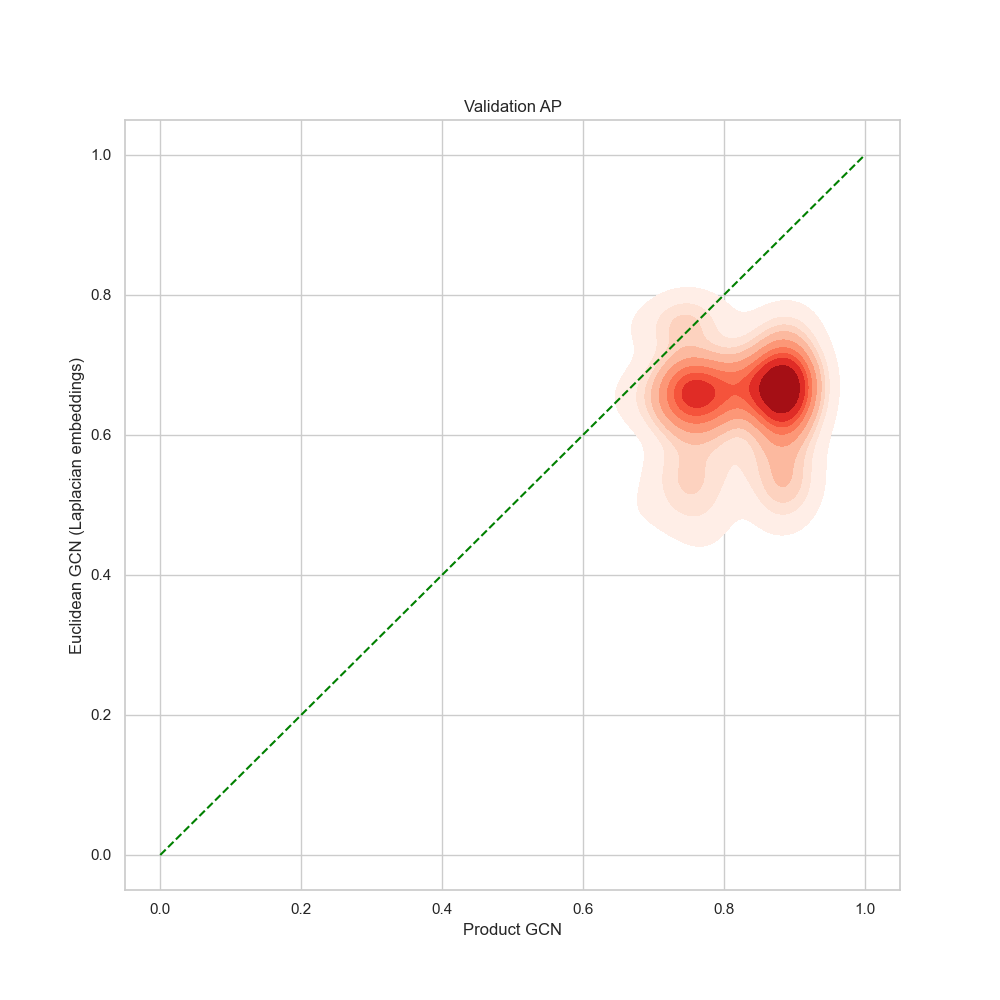}
\end{subfigure}%
\begin{subfigure}{.5\textwidth}
    \centering
    \includegraphics[width=0.8\textwidth]{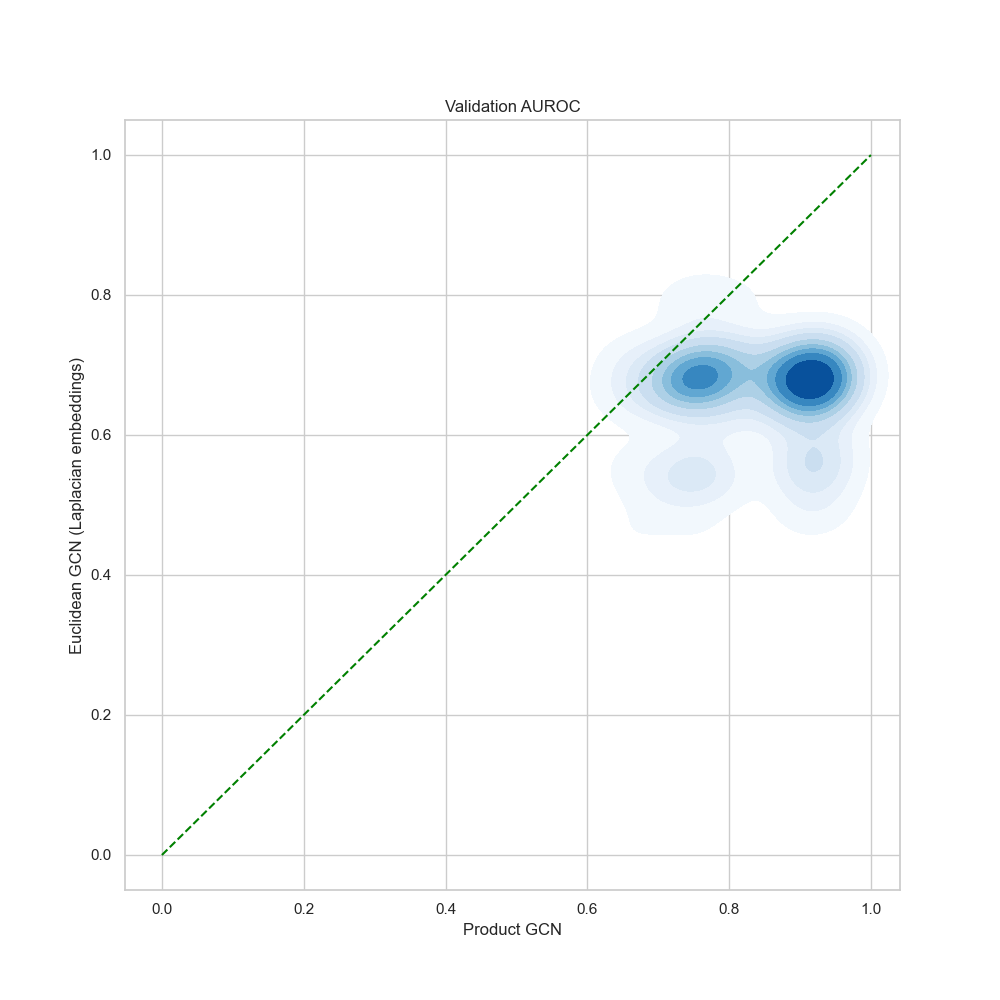}
\end{subfigure}
\begin{subfigure}{.5\textwidth}
    \centering
    \includegraphics[width=0.8\textwidth]{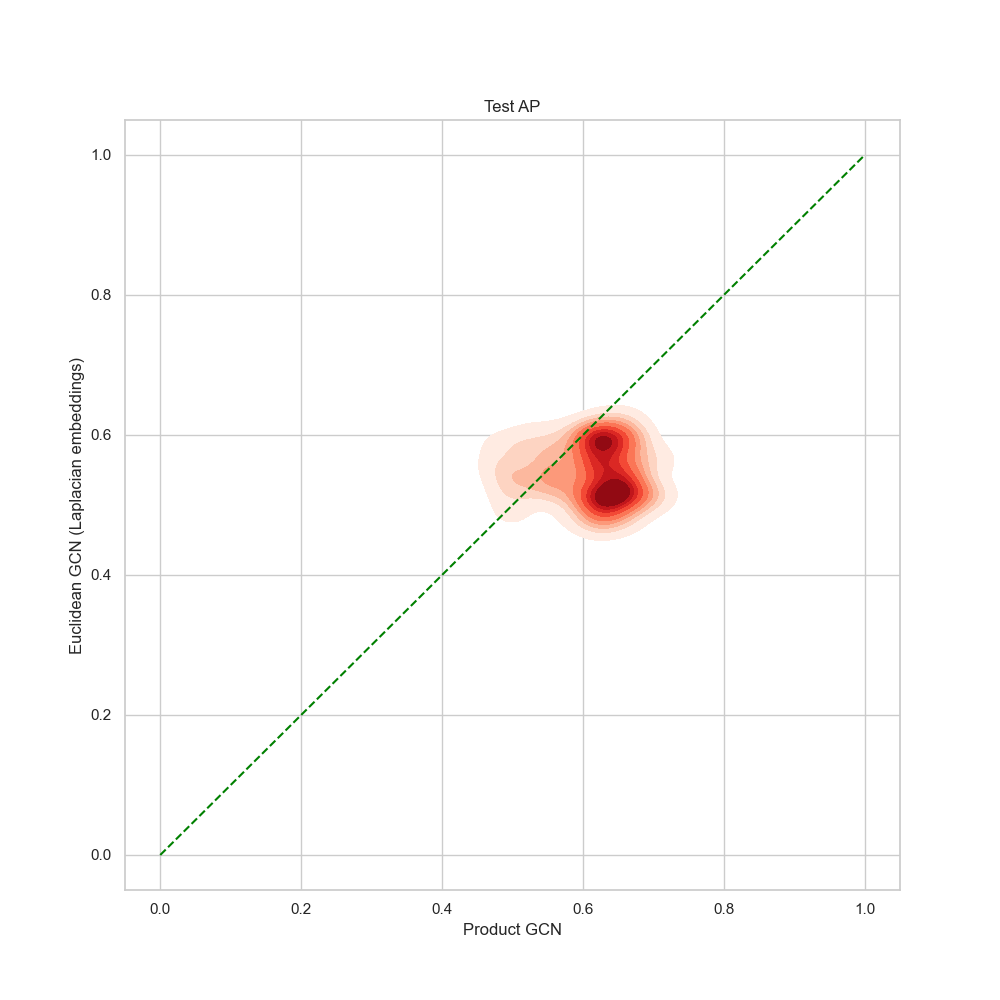}
\end{subfigure}%
\begin{subfigure}{.5\textwidth}
    \centering
    \includegraphics[width=0.8\textwidth]{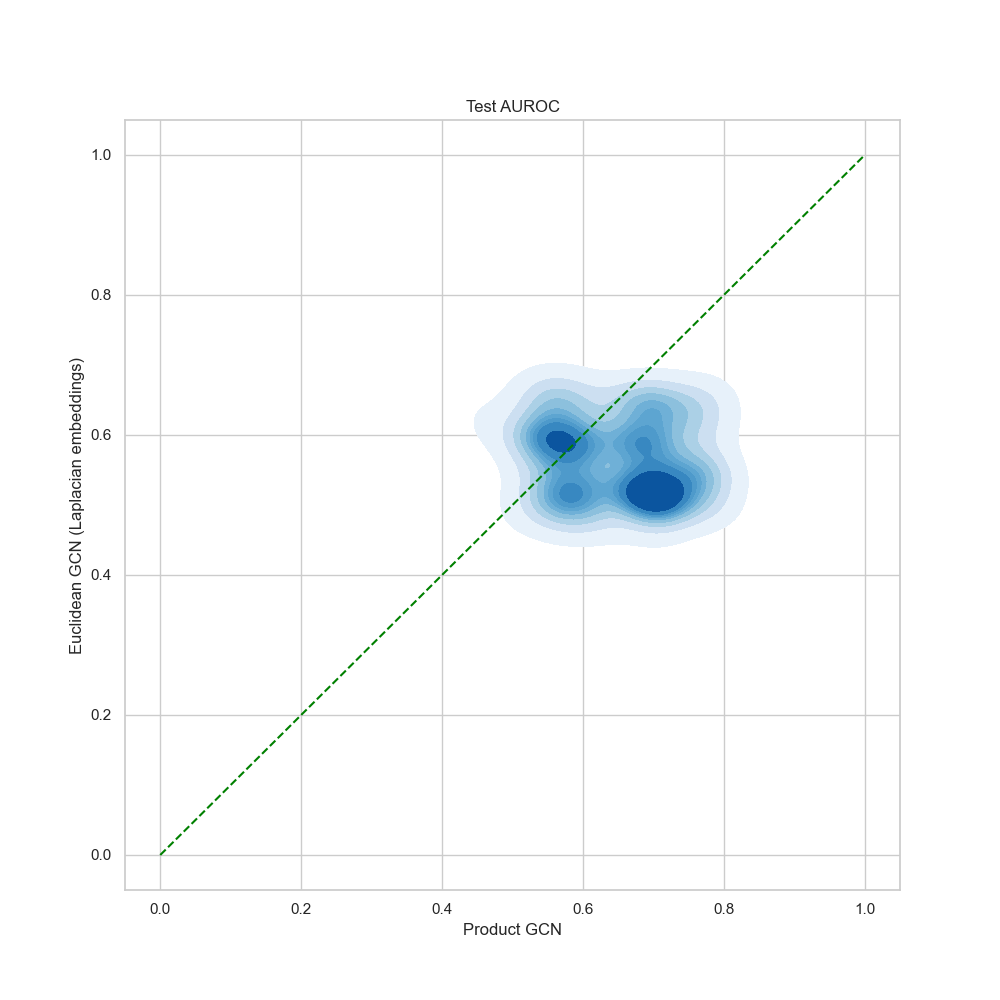}
\end{subfigure}
\caption[short]{Comparison of Euclidean GCN initialized with pretrained Laplacian embeddings and Product GCN performance on in-distribution 
validation set and out-of-distribution test set. Each density plot shows one of either AP or AUROC metrics taken across all graphs in the NCI dataset.}
\end{figure}

\begin{figure}[ht]
\centering
\begin{subfigure}{.5\textwidth}
    \centering
    \includegraphics[width=0.8\textwidth]{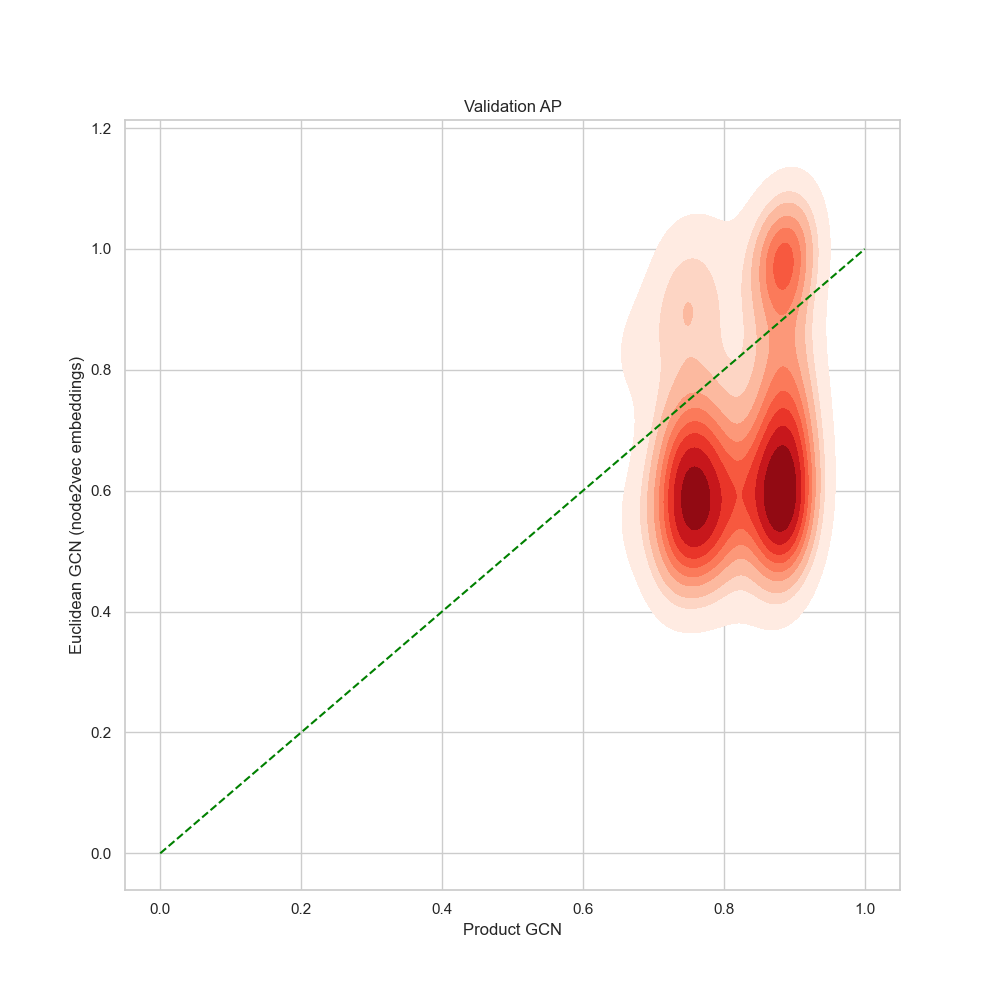}
\end{subfigure}%
\begin{subfigure}{.5\textwidth}
    \centering
    \includegraphics[width=0.8\textwidth]{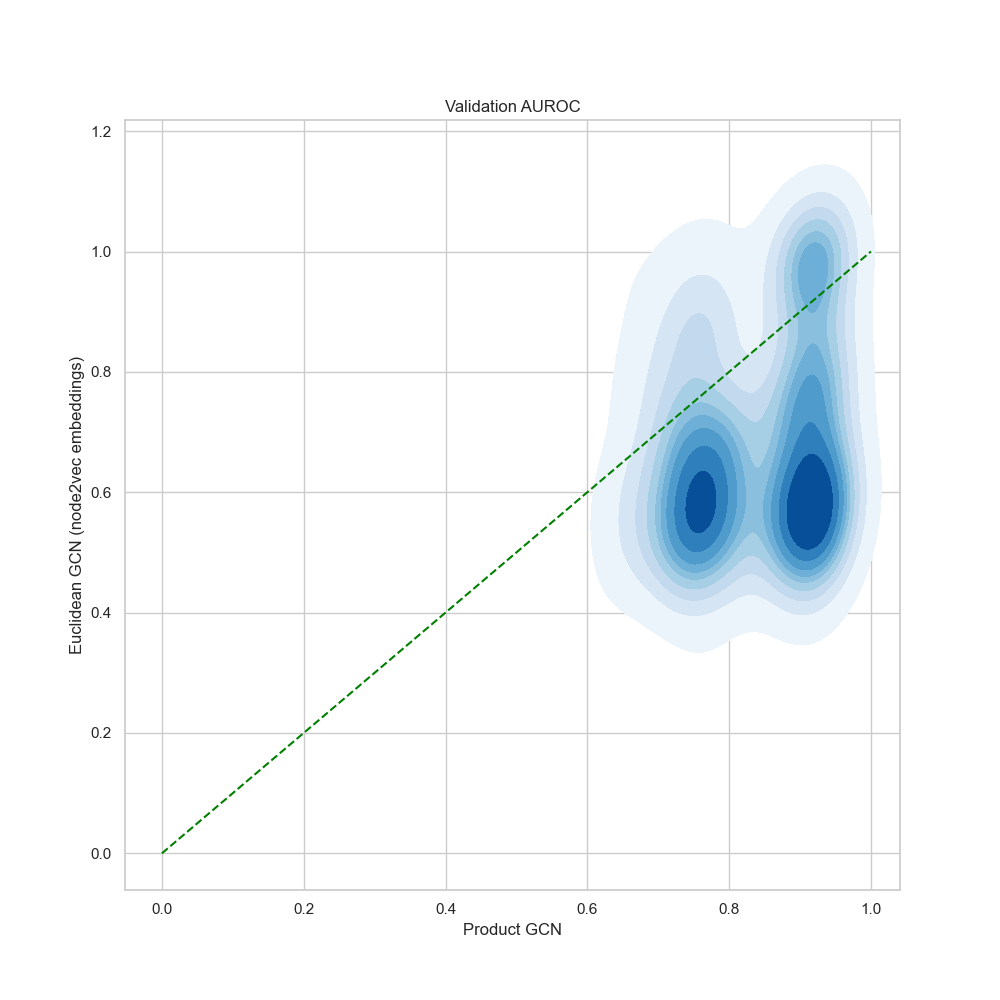}
\end{subfigure}
\begin{subfigure}{.5\textwidth}
    \centering
    \includegraphics[width=0.8\textwidth]{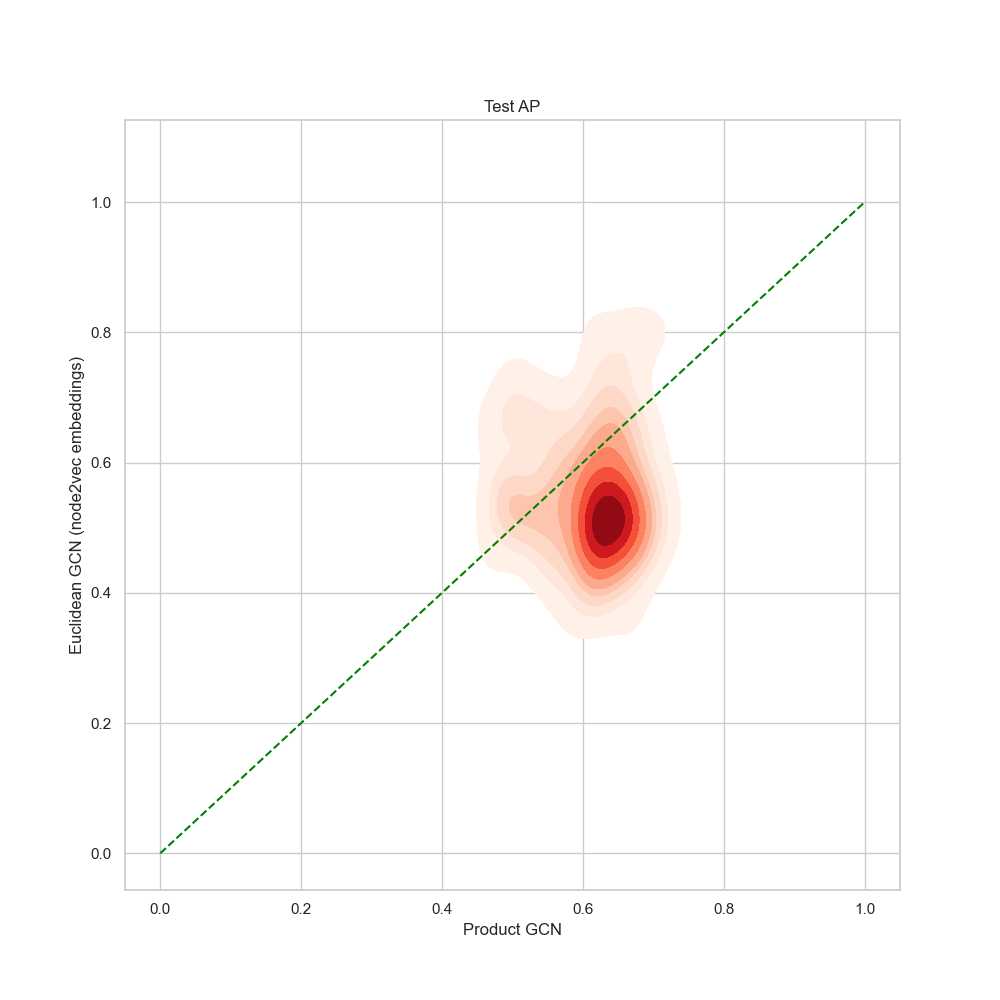}
\end{subfigure}%
\begin{subfigure}{.5\textwidth}
    \centering
    \includegraphics[width=0.8\textwidth]{images/ncbi_comparison_scatter_sweep_n2v_val_roc.png}
\end{subfigure}
\caption[short]{Comparison of Euclidean GCN initialized with pretrained node2vec embeddings and Product GCN performance on in-distribution 
validation set and out-of-distribution test set. Each density plot shows one of either AP or AUROC metrics taken across all graphs in the NCI dataset.}
\end{figure}

\clearpage
% --------------------- KEGG ----------------------------
\subsection{KEGG}
\begin{figure}[ht]
\centering
\begin{subfigure}{.5\textwidth}
    \centering
    \includegraphics[width=0.8\textwidth]{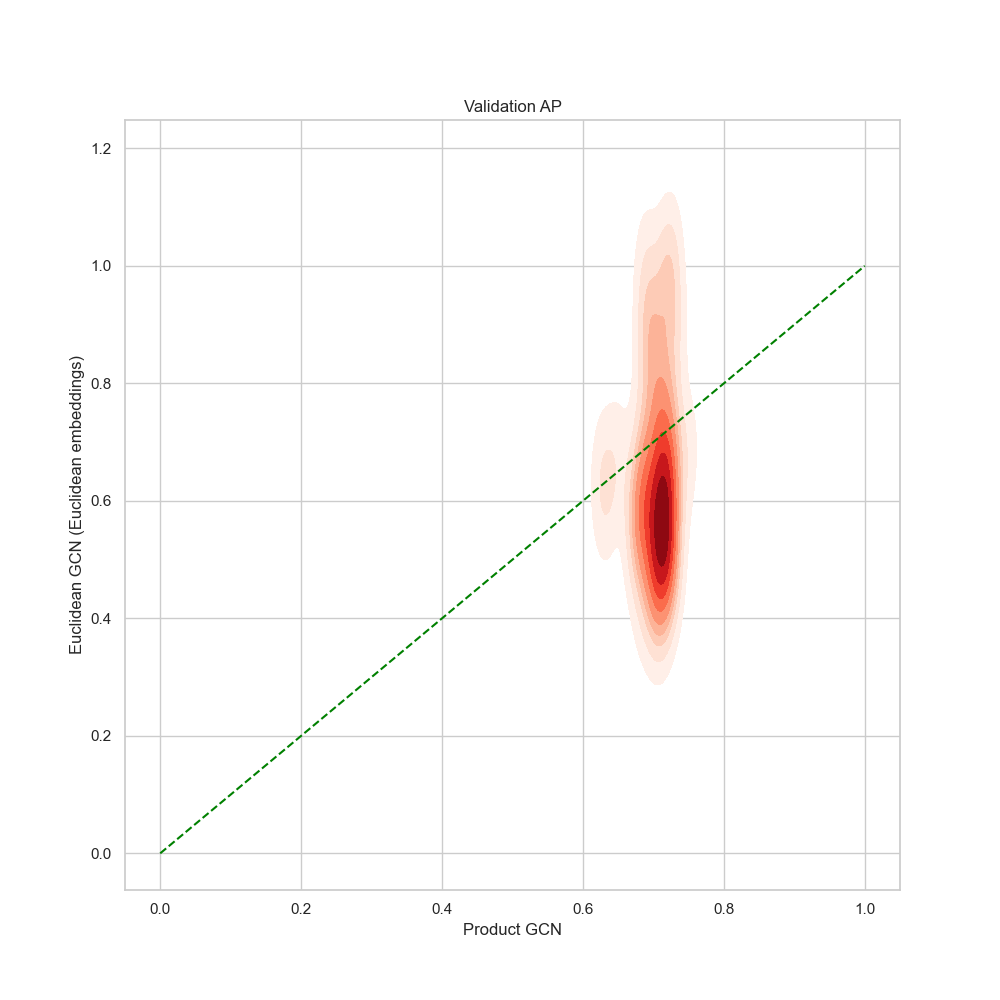}
\end{subfigure}%
\begin{subfigure}{.5\textwidth}
    \centering
    \includegraphics[width=0.8\textwidth]{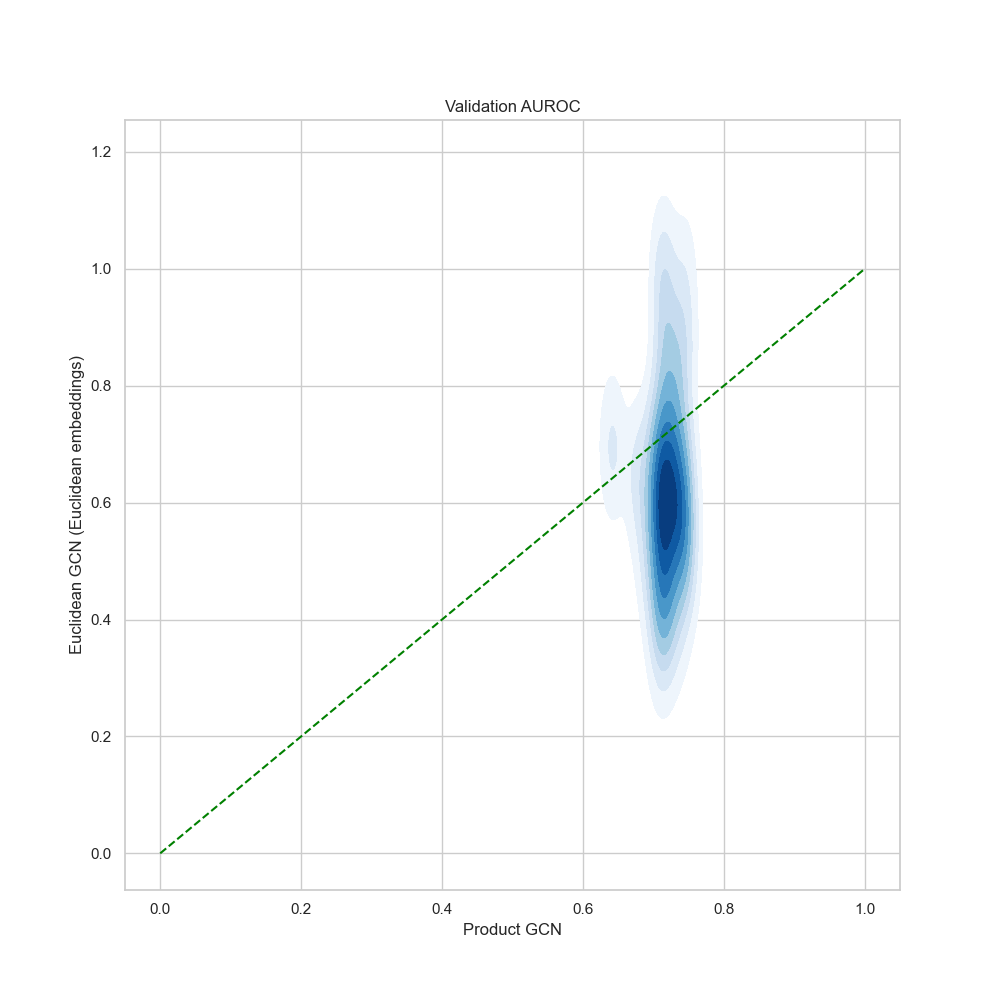}
\end{subfigure}
\begin{subfigure}{.5\textwidth}
    \centering
    \includegraphics[width=0.8\textwidth]{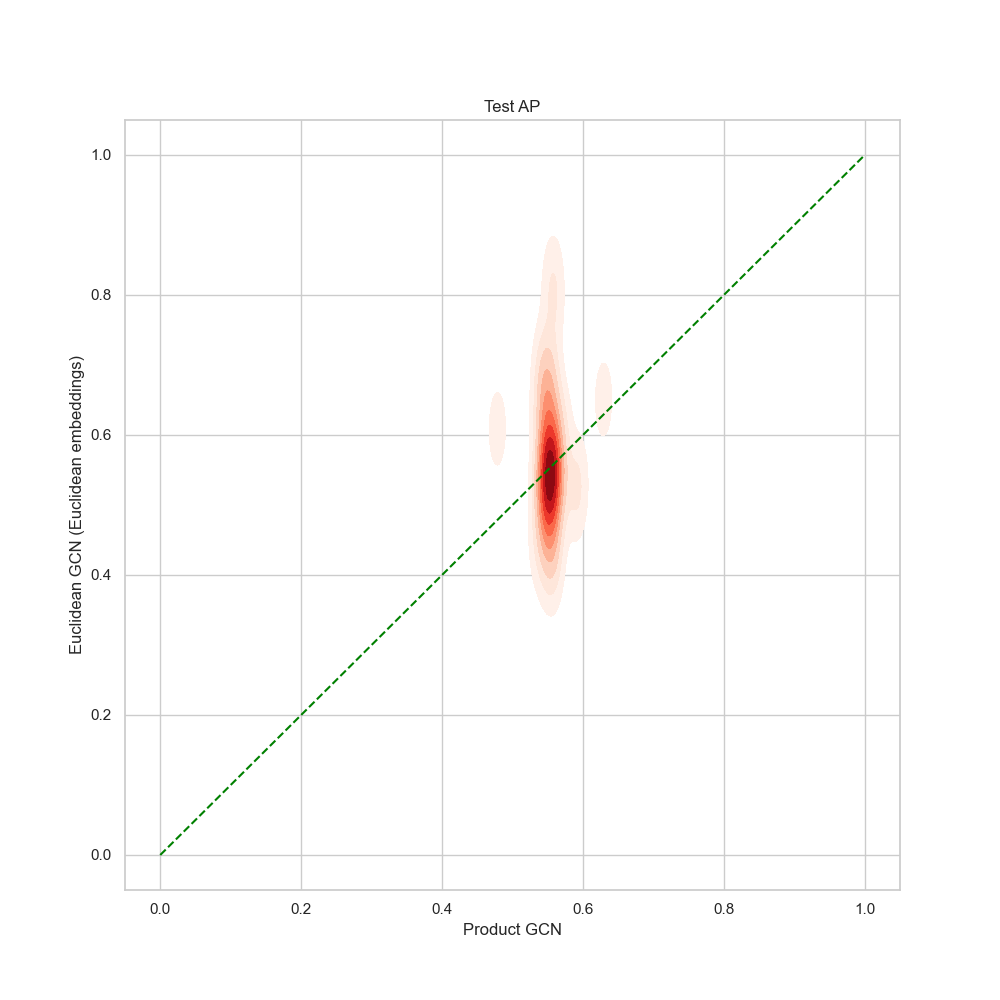}
\end{subfigure}%
\begin{subfigure}{.5\textwidth}
    \centering
    \includegraphics[width=0.8\textwidth]{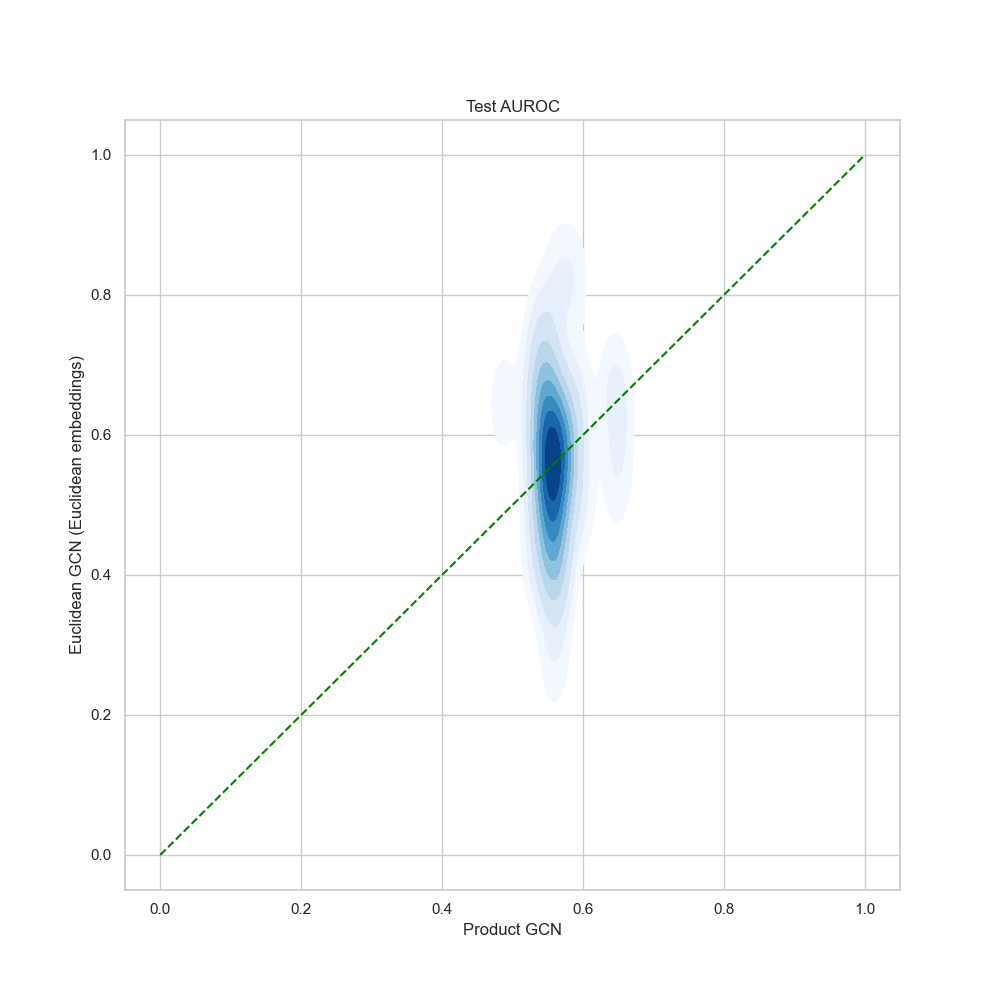}
\end{subfigure}

\caption[short]{Comparison of Euclidean GCN initialized with pretrained Euclidean embeddings and Product GCN performance on in-distribution 
validation set and out-of-distribution test set. Each density plot shows one of either AP or AUROC metrics taken across all graphs in the KEGG dataset.}
\end{figure}

\begin{figure}[ht]
\centering
\begin{subfigure}{.5\textwidth}
    \centering
    \includegraphics[width=0.8\textwidth]{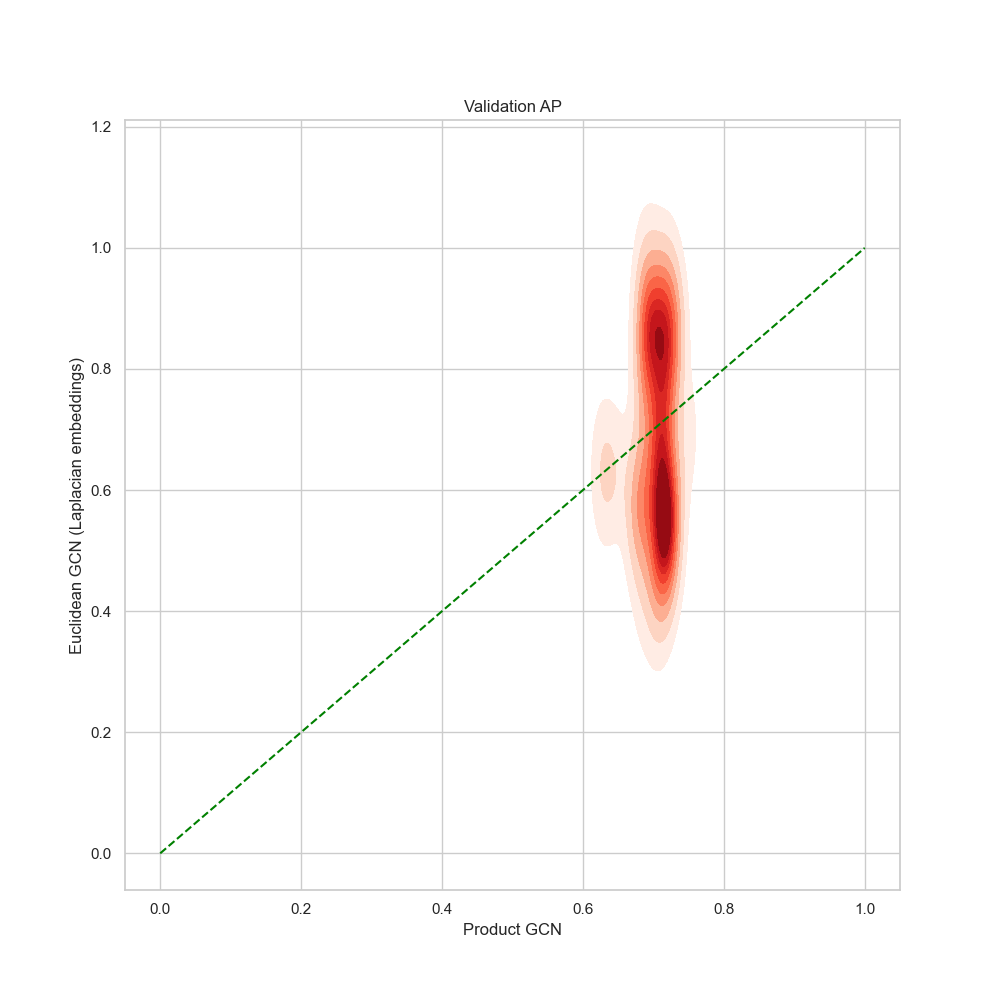}
\end{subfigure}%
\begin{subfigure}{.5\textwidth}
    \centering
    \includegraphics[width=0.8\textwidth]{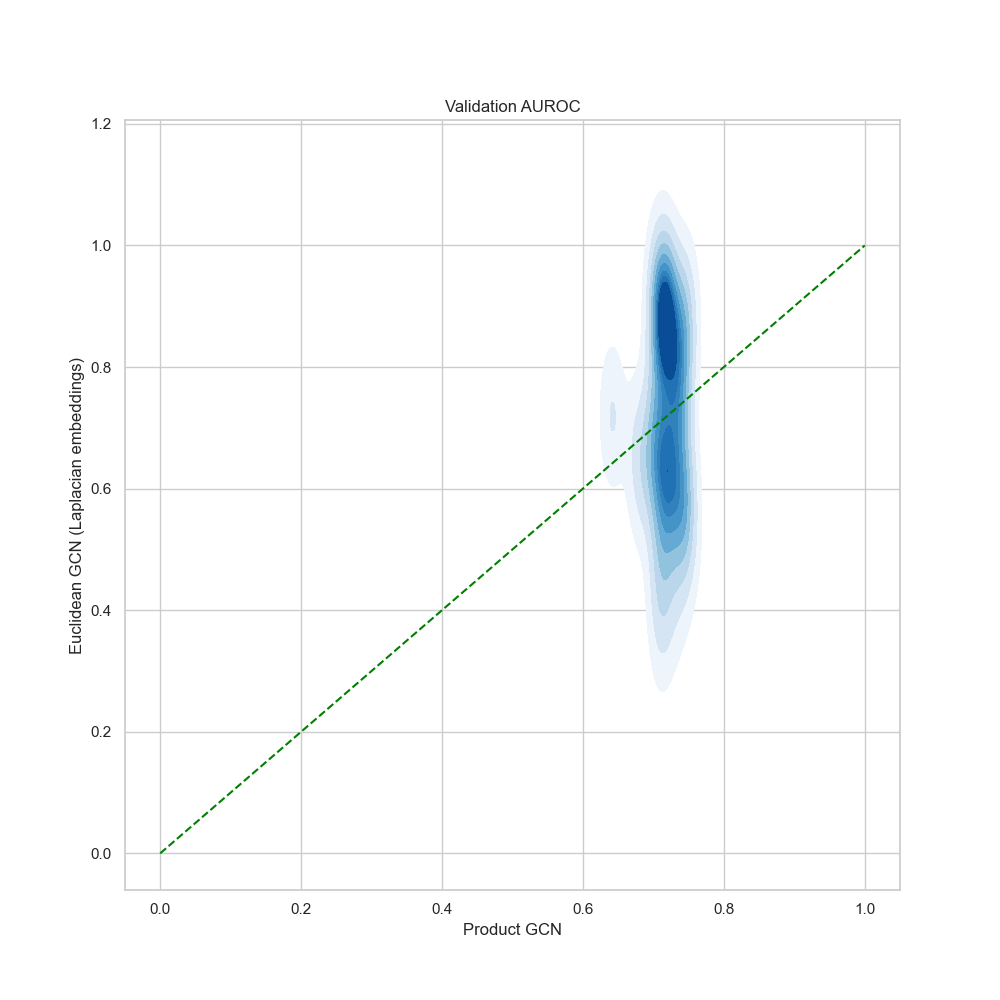}
\end{subfigure}
\begin{subfigure}{.5\textwidth}
    \centering
    \includegraphics[width=0.8\textwidth]{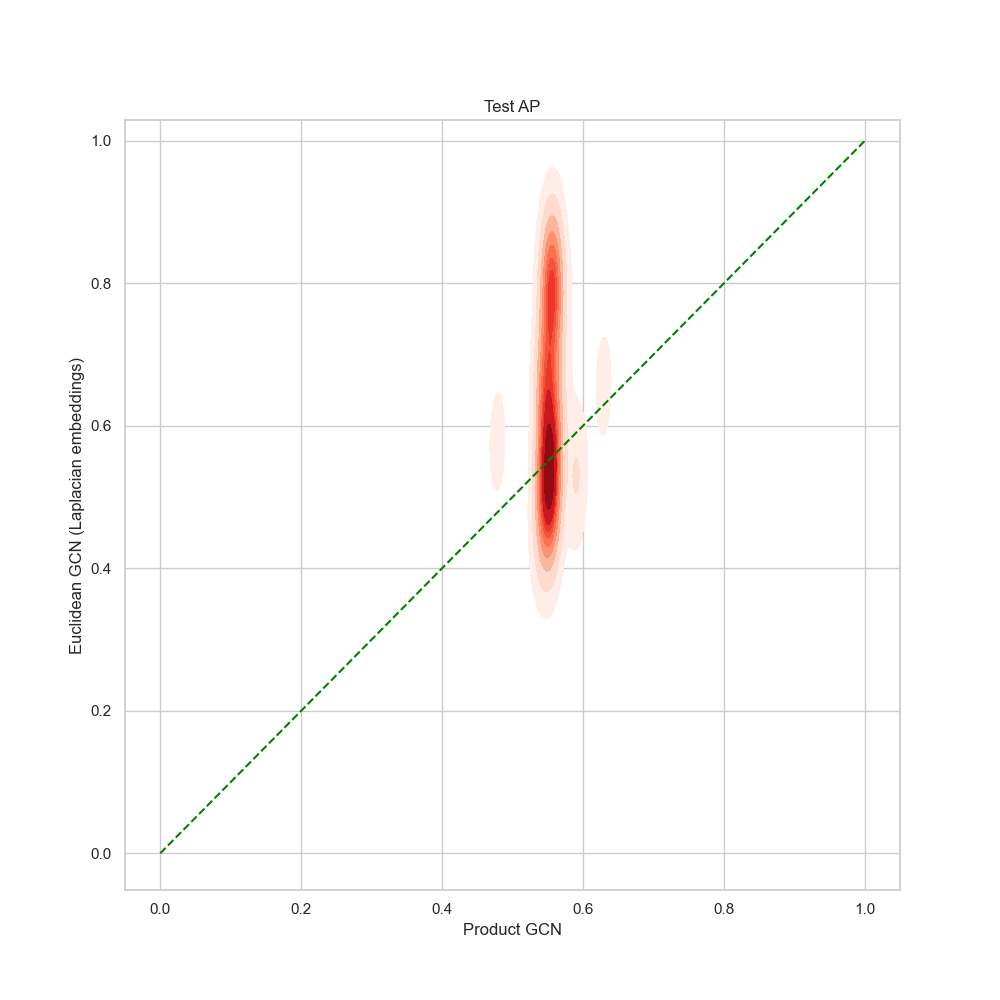}
\end{subfigure}%
\begin{subfigure}{.5\textwidth}
    \centering
    \includegraphics[width=0.8\textwidth]{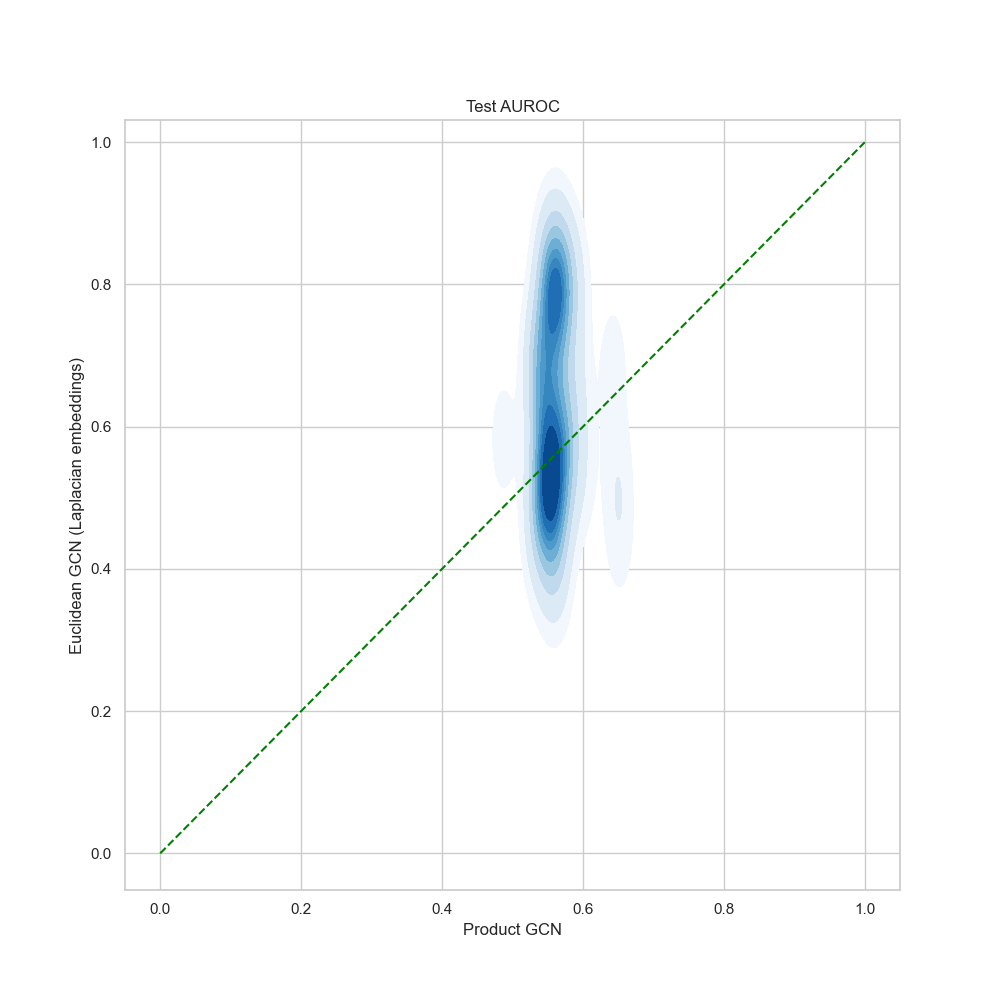}
\end{subfigure}
\caption[short]{Comparison of Euclidean GCN initialized with pretrained Laplacian embeddings and Product GCN performance on in-distribution 
validation set and out-of-distribution test set. Each density plot shows one of either AP or AUROC metrics taken across all graphs in the KEGG dataset.}
\end{figure}

\begin{figure}[ht]
\centering
\begin{subfigure}{.5\textwidth}
    \centering
    \includegraphics[width=0.8\textwidth]{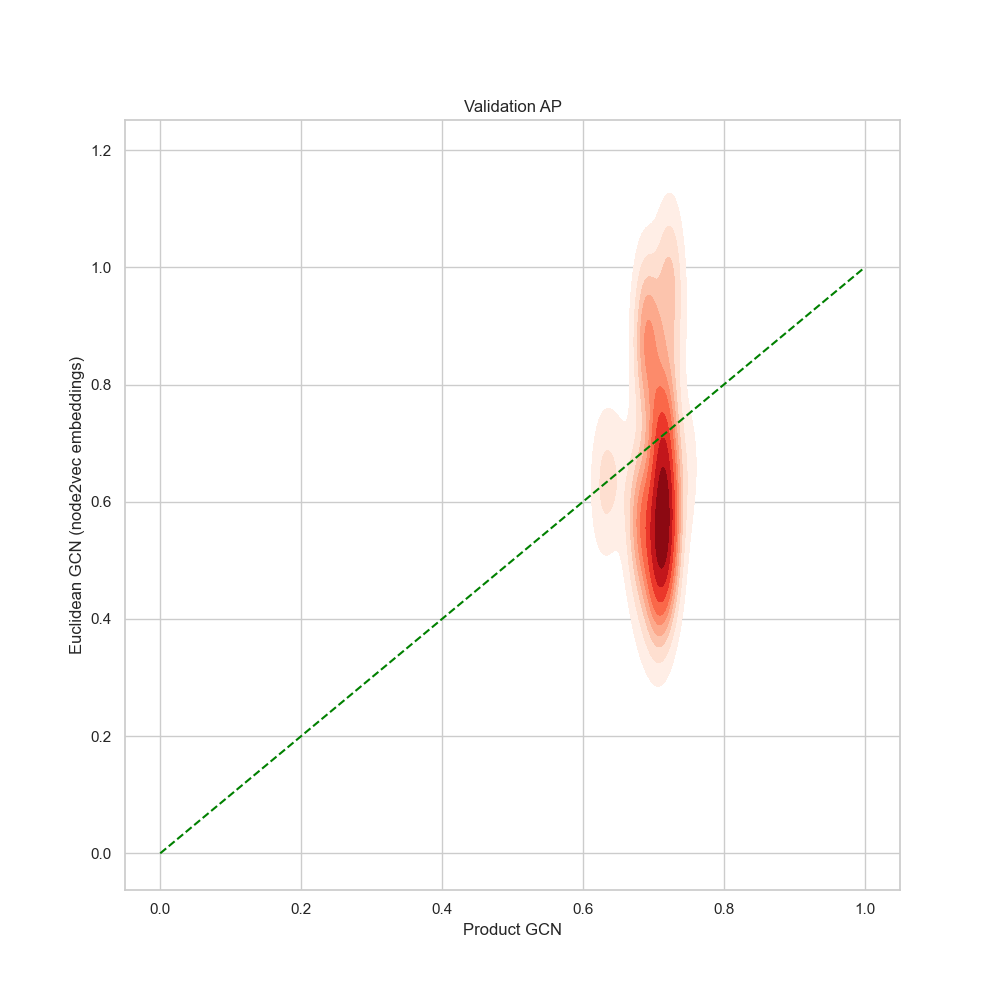}
\end{subfigure}%
\begin{subfigure}{.5\textwidth}
    \centering
    \includegraphics[width=0.8\textwidth]{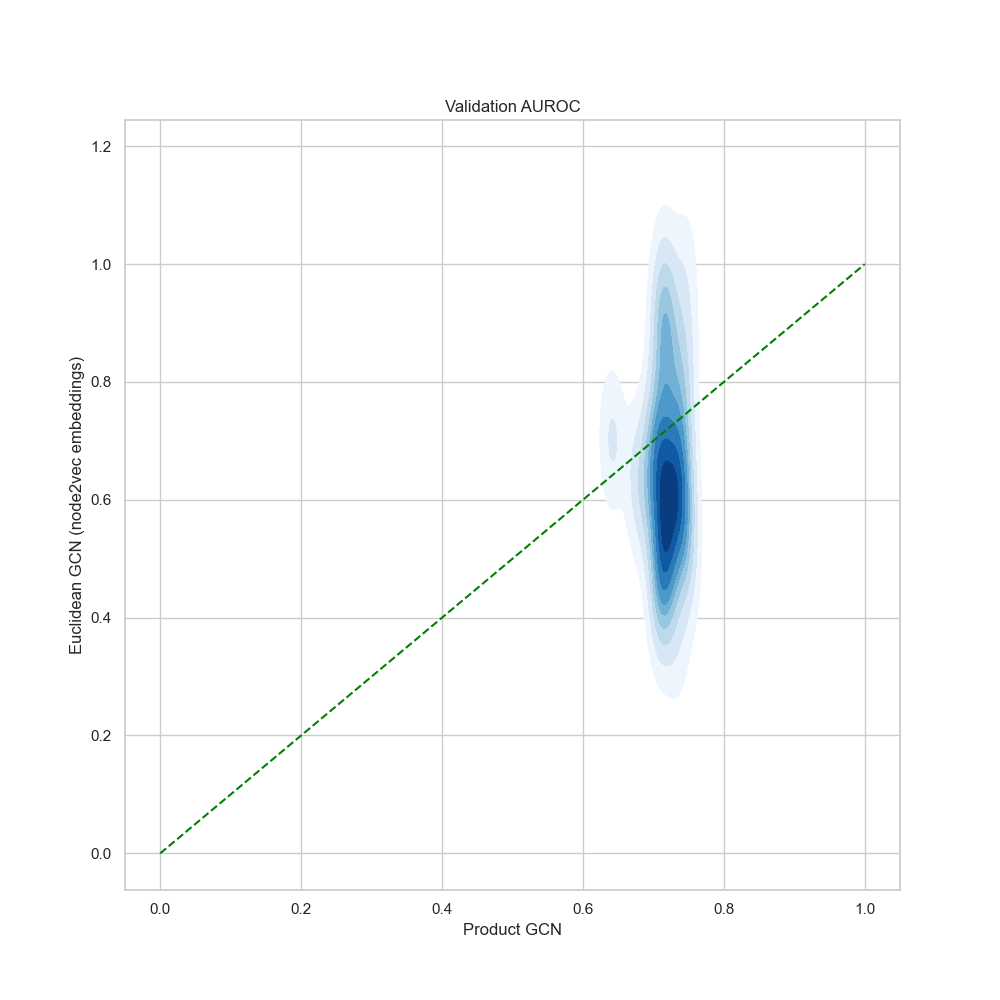}
\end{subfigure}
\begin{subfigure}{.5\textwidth}
    \centering
    \includegraphics[width=0.8\textwidth]{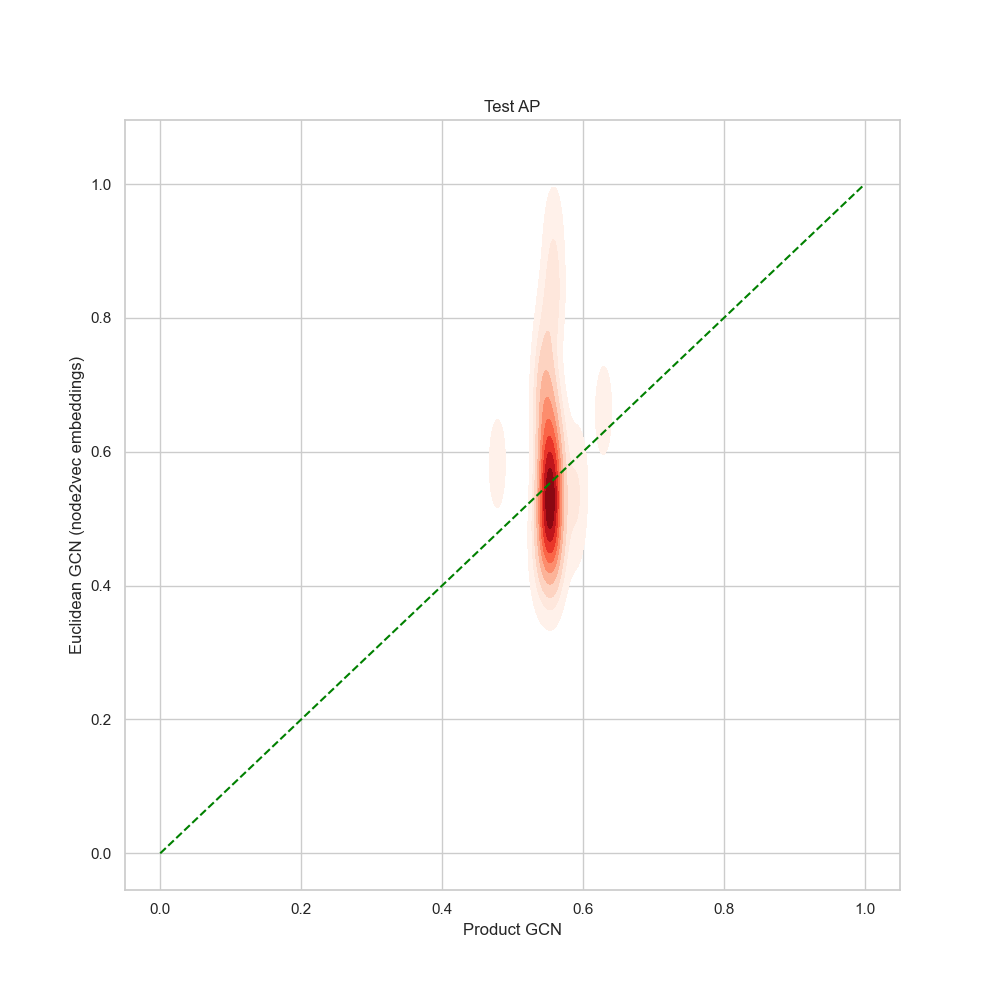}
\end{subfigure}%
\begin{subfigure}{.5\textwidth}
    \centering
    \includegraphics[width=0.8\textwidth]{images/kegg_comparison_scatter_sweep_n2v_val_roc.png}
\end{subfigure}
\caption[short]{Comparison of Euclidean GCN initialized with pretrained node2vec embeddings and Product GCN performance on in-distribution 
validation set and out-of-distribution test set. Each density plot shows one of either AP or AUROC metrics taken across all graphs in the KEGG dataset.}
\end{figure}

\end{document}